\documentclass[journal]{IEEEtran}
\usepackage{hyperref}
\usepackage{xcolor}
\hypersetup{
    colorlinks,
    linkcolor={black},
    citecolor={black},
    urlcolor={black}
}\usepackage{subfigure}
\newcommand{\squig}{{\scriptstyle\sim\mkern-3.9mu}}

\newcommand{\m}[1]{\mbox{$#1$}}

\usepackage{microtype}
\ifCLASSINFOpdf
\else
\fi
\usepackage{relsize}
\IEEEoverridecommandlockouts

\usepackage{setspace}

\usepackage{cite}
\usepackage{mathrsfs}
\usepackage{bbm}
\usepackage{pifont}
\usepackage{amsfonts}
\usepackage{amsmath, amsthm}

\let\labelindent\relax
\usepackage{enumitem}
\usepackage{tabularx}
\usepackage{listings}
\usepackage{graphicx}
\usepackage{float}
\usepackage{amssymb}
\usepackage{array}

\usepackage{fancyhdr}
\usepackage{cases}
 \usepackage{supertabular}
\usepackage{url}
\usepackage{fancyhdr}
\usepackage{color}
\usepackage{bbding}
\newcommand{\bcap} {\hspace{2pt} \mathlarger{\cap}
\hspace{2pt}}

\newcommand{\f}{it follows that }

\newcommand{\bcup} {\hspace{2pt} \mathlarger{\cup}
\hspace{2pt}}

\newcommand{\fsquare}{\vrule height6pt width7pt depth1pt}   
\newcommand{\pfe}{\hfill\fsquare }             

\def\centerhack#1{\hbox to 0pt{\hss\footnotesize #1\hss}}
\def\centerhackn#1{\hbox to 0pt{\hss #1\hss}}
\def\dchack#1{\vbox to 0pt{\vss{\hbox to 0pt{\hss#1\hss}}\vss}}

\newtheorem{lem}{Lemma}
\newtheorem{thm}{Theorem}

\newtheorem{rem}{Remark}

\newtheorem*{proposition1.1}{Proposition 1.1}
\newtheorem*{proposition1.2}{Proposition 1.2}
\newtheorem*{proposition1.3}{Proposition 1.3}
\newtheorem*{proposition2.1}{Proposition 2.1}
\newtheorem*{proposition2.2}{Proposition 2.2}
\begin{document}

\title{Transitional Behavior of $q$-Composite Random Key Graphs with Applications to Networked Control}

\author{Jun~Zhao,~\IEEEmembership{Member,~IEEE}\thanks{The author Jun Zhao obtained his PhD from Carnegie Mellon University, Pittsburgh, PA 15213, USA, where he was with the Cybersecurity Lab (CyLab). He was a postdoctoral scholar with Arizona State University, Tempe, AZ 85281, USA. He is now a research fellow at Nanyang Technological University in Singapore.
 Email: \texttt{junzhao@alumni.cmu.edu} \newline \indent
The
materials in this paper were presented in part at the 2015 Allerton Conference on Communication, Control, and Computing \cite{2015-Allerton-RKG,2015-Allerton-s-intersection}.
}}

\maketitle


{\color{black}
 \begin{abstract}

Random key graphs have received considerable attention and been used in various applications including secure sensor networks, social networks, the study of epidemics, cryptanalysis, and recommender systems. In this paper, we investigate a $q$-composite random key graph, whose construction on $n$ nodes is as follows: each node independently selects a set of $K_n$ different keys uniformly at random from the same pool of $P_n$ distinct keys, and two nodes establish an undirected edge in between if and only if they share at least $q$ key(s). Such graph denoted by $G_q(n,K_n,P_n)$ models a secure sensor network employing the well-known $q$-composite key predistribution. For $G_q(n,K_n,P_n)$, we analyze the probabilities of
 $G_q(n,K_n,P_n)$ having $k$-connectivity, $k$-robustness,  a Hamilton cycle and  a perfect matching, respectively. Our studies of these four properties are motivated by a detailed discussion of their applications to \textit{networked control}. Our results reveal that $G_q(n,K_n,P_n)$ exhibits a sharp transition for each property: as $K_n$ increases,  the probability that $G_q(n,K_n,P_n)$ has the property sharply increases from $0$ to $1$. These results provide fundamental guidelines to design secure sensor networks for different control-related applications: \textit{distributed in-network parameter estimation}, \textit{fault-tolerant consensus}, and \textit{resilient data backup}.

\end{abstract}
%
%
}


 \section{Introduction} \label{sec-Introduction}
\begin{figure*}[!t]
  \centering
 \includegraphics[width=.97\textwidth]{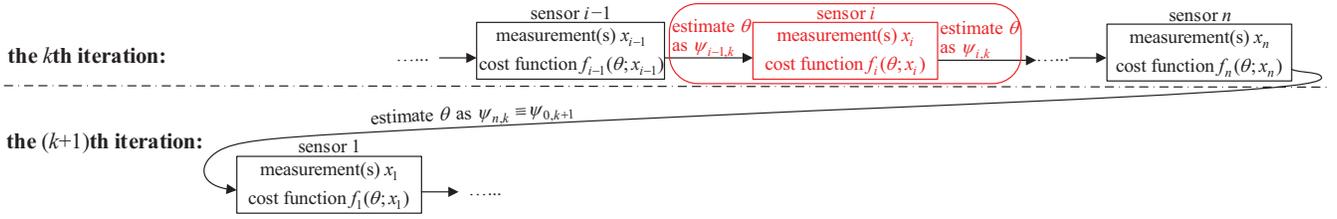}
 \caption{An illustration of the Hamilton-cycle-based distributed in-network parameter estimation by Rabbat and Nowak \cite{rabbat2005quantized}.} \label{fig-hami}
\end{figure*}

Random key graphs \cite{yagan,ZhaoYaganGligor} originally resulted from the modeling of secure sensor networks \cite{adrian,virgil}, and have also been used in other applications including social networks \cite{TCNS-heterogeneous}, the study of epidemics \cite{ball2014}, cryptanalysis \cite{herdingRKG},
 recommender systems \cite{r4}, and circuit design \cite{2013arXiv1301.0466R}. The usual definition of a random key graph with $n$ nodes is as follows  \cite{yagan,ZhaoYaganGligor}: each node \textit{independently} picks a set of $K_n$ different cryptographic keys \emph{uniformly at random} from the same pool of $P_n$ distinct keys, and an \emph{undirected} edge is put between
   any two nodes which share at least one key. In this paper, we consider a more general model than the usual notion above. Specifically, we generalize the definition by requiring two nodes having an edge in between to share at least $q$ key(s) rather than just one key, where $q$ is a positive number. We call this general model as a $q$-composite random key graph and use $G_q(n,K_n,P_n)$ for the notation. Clearly, our model in the special case of $q=1$ reduces to the above traditional notion of random key graph \cite{yagan,ZhaoYaganGligor}. To motivate our study of $G_q(n,K_n,P_n)$, we discuss its applications to secure sensor networks and  social networks.


   \textbf{Applying random key graphs to   secure sensor networks.}
     We explain that ($q$-composite) random key graphs can be used to model secure sensor networks. For wireless sensor networks deployed in hostile environments,  cryptographic protection is needed to ensure secure communications. Random key predistribution \cite{virgil} has been introduced as a suitable security scheme.
 The first random key predistribution scheme, proposed by Eschenauer and Gligor \cite{virgil},   works as follows. For a network of $n$ sensors,
before deployment,
 each sensor is assigned a set of $K_n$ distinct cryptographic keys selected uniformly at random from the same key pool containing $P_n$ different keys. After deployment, two sensors establish secure communication if
and only if they have at least one common key. Chan
\emph{et al.} \cite{adrian} extend the Eschenauer--Gligor (EG) scheme to the so-called $q$-composite key predistribution scheme, by requiring two sensors to share $q$ key(s) rather than just one key for secure communication. Clearly, a secure sensor network employing the $q$-composite scheme induces a topology modeled by a $q$-composite  random key graph, while the induced topology under the EG scheme is represented by a traditional random key graph (i.e., a $q$-composite  random key graph in the case of $q=1$).





 \textbf{Applying random key graphs to social networks.} In addition to secure sensor networks, random key graphs can also used to model social networks  \cite{TCNS-heterogeneous}. To see this, we observe that the concept of ``cryptographic key'' in constructing a $q$-composite random key graph $G_q(n,K_n,P_n)$ can be generalized to any object or interest (e.g.,  watching a video, listening to a song, or reading a novel). Then $G_q(n,K_n,P_n)$ with nodes representing individuals naturally models an interest-based social network, where a link between two people is represented by their selection of at least $q$ common interests, after each of them chooses $K_n$ interests from the same  pool of $P_n$ interests.

To consider more control-related applications, we will mainly focus on using $q$-composite random key graphs for secure sensor networks instead of social networks. Our studied  properties of $q$-composite random key graphs include $k$-connectivity, $k$-robustness, Hamilton cycle containment, and perfect matching containment. We explain their definitions below and will detail their applications to networked control later in Section \ref{secappli}.

 First, $k$-connectivity means that each pair of nodes can find at least $k$
internally node-disjoint path(s) in between \cite{erdoskcon,JansonLuczakRucinski7}. An equivalent definition of
$k$-connectivity is that after the removal of
at most $(k-1)$ nodes, the remaining graph is still connected \cite{erdoskcon,JansonLuczakRucinski7}.  Second, $k$-robustness introduced by Zhang \textit{et al.} \cite{7061412} quantifies the effectiveness of local-information-based consensus algorithms
in the presence of malicious nodes. More formally, a graph with a node set $\mathcal {V}$ is $k$-robust
 if at least one of (a) and (b) below is true for every pair of non-empty, disjoint subsets $A$ and $B$ of $\mathcal {V}$: (a) there exists
 no less than one node $v_a \in A$ such that
  $v_a$ has at least $k$ neighbors inside $\mathcal {V}\setminus
  A$; and (b) there exists no less than one node $v_b \in B$ such that
  $v_b$ has at least $k$ neighbors inside $\mathcal {V}\setminus B$. Third, a perfect matching in a graph with an even number of nodes means a
matching covering all nodes, where a matching in a graph is a set of edges without common nodes  \cite{Perfectmatchings}.
Finally, a Hamilton cycle in a graph is a closed loop that visits each node once~\cite{NikoletseasHM}.

The above four properties are all monotone increasing.
In this paper, we study these properties and show their sharp transitions in a $q$-composite random key graph $G_q(n,K_n,P_n)$. Specifically, we make the following contributions:
\\ $\bullet$ We obtain exact probabilities of
 $G_q(n,K_n,P_n)$ being $k$-connected, having at least one Hamilton cycle, and having at least one perfect matching, respectively. We also derive a zero--one law for $k$-robustness in $G_q(n,K_n,P_n)$.
\\ $\bullet$ Our studies of the above four properties are motivated by a detailed discussion of their applications to \textit{networked control} (see Section \ref{secappli}). Our results show that $G_q(n,K_n,P_n)$ exhibits a sharp transition for each property: as $K_n$ increases, the asymptotic probability that $G_q(n,K_n,P_n)$ has the property sharply increases from $0$ to $1$. These results provide fundamental guidelines to design secure sensor networks for different control-related applications: \textit{distributed in-network parameter estimation}, \textit{fault-tolerant consensus}, and \textit{resilient topology control} (see Section \ref{secappli}).
\\ $\bullet$ To further quantify the sharpness of the transition, we derive the transition width of $G_q(n,K_n,P_n)$ for different properties above, where the transition width measures how should $K_n$ grow to increase the probability of $G_q(n,K_n,P_n)$ having certain property from $\epsilon$ to $1-\epsilon$ for $\epsilon < \frac{1}{2}$. We demonstrate different transitional behavior of the transition width for $q\geq 2$ and $q = 1$ when $G_q(n,K_n,P_n)$ is applied to model secure sensor networks: the transition width can be very small (even $0$ or $1$) for $q = 1$, while no such phenomenon exists for $q\geq 2$. This result shows a fundamental difference between the $q$-composite scheme with $q\geq 2$ and the EG scheme (i.e., the $q$-composite scheme with $q = 1$), and can be used to design secure sensor networks; e.g., the $q$-composite scheme with $q\geq 2$ is preferred over the EG scheme if it is  desired to have stronger \textit{resilience} of $k$-connectivity against key revocation.



The rest of the paper is organized as follows. {\color{black}We discuss the applications of our study to networked control in Section~\ref{secappli}.} Afterwards, Section \ref{sec:main:res:transitional} presents the transitional behavior in the probability of $G_q(n,K_n,P_n)$
  having each property, and Section \ref{sec:sharp:transition:1-composite} investigates the transition width of $G_q(n,K_n,P_n)$ for different properties. We compare this paper with related work in Section \ref{related}. Section \ref{sec-mainproofs} provides technical details.

{\color{black}\section{Applying the Studied Properties of Random Key Graphs to Networked Control} \label{secappli}

Below we discuss the applications of the studied properties in random key graphs to networked control.

\subsection{{Hamilton Cycle for Distributed  Parameter Estimation}}


Hamilton cycle has been used to facilitate  {distributed in-network parameter estimation} in a seminal work of Rabbat and Nowak \cite{rabbat2005quantized}, as detailed below.

In many sensor network applications, sensors often measure quantities
  such as temperature, pressure, water salinity, vibration amplitude \cite{5605238}. The eventual goal is to estimate   environmental parameters from the ``raw'' measurements. To achieve this goal, distributed in-network processing is preferred over a centralized approach (where a fusion center collects data from sensors), since the former makes more efficient use of sensors' limited communication and energy resources.

An algorithm for distributed in-network parameter estimation is proposed by Rabbat and Nowak \cite{rabbat2005quantized}. The algorithm is based on a Hamilton cycle and its basic idea is as follows. An  estimate of certain environmental parameter is passed
from node to node on the Hamilton cycle. Specifically, along the way each node updates the parameter based on its environmental measurements,
and then passes the updated estimate to the next node. It may require several iterations through the Hamilton cycle to obtain the final solution.

The Hamilton-cycle-based algorithm of \cite{rabbat2005quantized} can be formally described   as follows. Without loss of generality, assume
that sensors are numbered by $1,2,\ldots,n$ so that the network has a Hamilton cycle given by $1 \squig\squig \hspace{2pt}2 \squig\squig \hspace{2pt}\ldots \squig\squig\hspace{2pt} n \squig\squig\hspace{2pt} 1$, where ``$\squig\squig\hspace{1pt}$'' represents a  link. As computing the final estimate may require several iterations through the Hamilton cycle, we look at one iteration (say iteration $k$) for illustration. In the $k$th iteration, sensor $i$ receives an estimate $\psi_{i-1,k}$ of $\theta$ from sensor $i-1$, and makes an adjustment to $\theta$ based on its measurement(s) $x_i$ and its local cost function $f_i(\theta, x_i)$. After the adjustment, the estimation of $\theta$ by sensor $i$ is $\psi_{i,k}$, as illustrated inside the  rounded rectangle of Figure \ref{fig-hami}. From the end of the $k$th iteration to the beginning of the $(k+1)$th iteration, sensor $n$ sends its estimate $\psi_{n,k} \equiv \psi_{0,k+1}$ of $\theta$ to sensor $1$, and sensor~$1$ adjusts the estimation of $\theta$ based on its measurement(s) $x_1$ and its local cost function $f_1(\theta, x_1)$. This begins the $(k+1)$th iteration, as shown in Figure~\ref{fig-hami}. In terms of the local adjustments, \mbox{Rabbat and Nowak~\cite{rabbat2005quantized}} consider a gradient \mbox{descent-like} rule and demonstrate its fast convergence.

If the goal of the distributed data processing is to compute the average of sensors' measurements, each $f_i$ can take the quadratic cost function, and the desired average can be obtained after only one iteration. However, more general optimization problems require several ``rounds'' through the network to obtain
a solution \cite{rabbat2005quantized}. Hence, the algorithm depends on finding a cycle that touches each sensor once, and such a cycle is precisely a Hamilton cycle. As   explained, this cycle is $1 \squig\squig \hspace{2pt}2 \squig\squig \hspace{2pt}\ldots \squig\squig\hspace{2pt} n \squig\squig\hspace{2pt} 1$, where each number indexes a sensor.

As explained in Section \ref{sec-Introduction}, the $q$-composite random key graph represents the topology of a secure sensor network under the renowned $q$-composite key predistribution scheme~\cite{adrian}. Then our zero--one law and exact probability results on Hamilton cycle containment in $q$-composite random key graphs provide a precise guideline for setting  parameters of the sensor network to ensure the existence of a Hamilton cycle, which
enables distributed in-network parameter estimation.

\subsection{{$k$-Connectivity for  Resilient Topology Control and Fault-Tolerant Consensus}}

 We explain the applications of
 $k$-connectivity below.
First, $k$-connectivity enables {resilient topology control} against node or link failure, since $k$-connectivity means that connectivity is preserved even after at most $(k-1)$ nodes or links fail. In the application of $q$-composite random key graphs to secure sensor networks in hostile environments, $k$-connectivity is particularly useful for resilient topology control since sensors or links can be compromised by an adversary \cite{adrian,zhao2015resilience}. Second,
$k$-connectivity is useful to achieve {fault-tolerant consensus} in networks, as discussed below.
Sundaram and Hadjicostis \cite{5605238}, and Pasqualetti \textit{et al.} \cite{5779706} show that being $(2h+1)$-connected for a network is the necessary and sufficient condition to ensure that consensus can be reached even if there exist $h$ malicious nodes crafting messages to disrupt the protocol.

\subsection{{$k$-Robustness for Fault-Tolerant Consensus}}

As explained in the previous subsetion, if the network is sufficiently connected, resilient consensus can be achieved. For this, several algorithms have been proposed in the literature~\cite{5605238,5779706}.
However, these algorithms typically assume that nodes know the global network  topology,
which limits application scenarios~\cite{6481629}.
To account for the lack of global topology knowledge in the general case (for example, each node knows only its own neighborhood), Zhang and
Sundaram~\cite{7061412} propose the notion of {\em graph robustness} defined as follows.
A graph with a node set $\mathcal {V}$ is said to be $k$-robust
 if at least one of (a) and (b) below holds for every pair of non-empty, disjoint subsets $A$ and $B$ of $\mathcal {V}$: (a) there exists
at least a node $v_a \in A$ such that
  $v_a$ has no less than $k$ neighbors outside $A$ (i.e., inside $\mathcal {V}\setminus
  A$); and (b) there exists at least a node $v_b \in B$ such that
  $v_b$ has no less than $k$ neighbors outside $B$ (i.e., inside $\mathcal {V}\setminus
  B$).


Zhang \textit{et al.} \cite{7061412} show that $k$-robustness implies $k$-connectivity, while $k$-connectivity may not imply $k$-robustness. Based on \cite{6481629,7061412}, we will explain that $k$-robustness quantifies the effectiveness of local-information-based  {fault-tolerant consensus} algorithms
in the presence of adversarial nodes.

To discuss consensus, we suppose that all nodes are synchronous and the time is divided into different  slots. Each node updates its value as time goes by. Let $x_i[t]$ denote the value of node $v_i$ at time slot $t$ for $t=0,1,\ldots$. For  simplicity, we first consider the case where all nodes are benign. Then consensus can be defined by
 $\lim_{t \to \infty} |x_i[t] - x_j[t]| = 0 $ for each pair of nodes $v_i$ and $v_j$. Each node updates its value in each time slot based on
the following process. With $V_i$ denoting the neighborhood set of each node $v_i$, then $v_i$ updates its value $x_i [t]$ to $x_i[t+1]$ from time slot $t$ to $t+1$ by incorporating every neighbor $v_j$'s value $x_j [t]$ that $v_j$ sends to $v_i$; i.e.,  there is a function $f_i(\cdot)$ such that
$x_i[t+1] = f_i\big( \big\{ x_j[t] \ \big| \ v_j \in V_i \bcup \{ v_i \} \big\}\big) .$
In linear consensus \cite{6481629,7061412}, each $f_i(\cdot)$ is a linear function that assigns appropriate weights to its inputs to compute a weighted summation.

Now we consider the presence of adversarial nodes; i.e., there exist nodes who maliciously deviate from
the nominal consensus protocol. Recall that a benign node $v_i$ sends
$x_i [t]$ to all of its neighbors and applies $f_i(\cdot)$ at every time
slot $t$. In contrast, a malicious node does not follow this protocol; in particular, a malicious node may try various ways (e.g., crafting bad values) to disrupt the consensus evolution. In the presence of malicious nodes, consensus means $\lim_{t \to \infty} |x_i[t] - x_j[t]| = 0 $ for each pair of \emph{benign} nodes $v_i$ and $v_j$.


Assuming each node does not know the global network topology and only knows the number of malicious
nodes in its neighborhood, Zhang and Sundaram \cite{6481629} demonstrate the usefulness of robustness in studying consensus. Specifically, under the adversary model that each benign node has at most $h$ malicious nodes as neighbors, if the graph is $(2h+1)$-robust,  consensus can be achieved according to an algorithm where each node updates its value at each time slot using the values received from its neighbors (see \cite{6481629} for the algorithm details).

 Given the above, in secure sensor network applications of $q$-composite random key graphs, our $k$-robustness result provides guidelines of setting parameters for fault-tolerant consensus.

%
%

\subsection{Perfect Matching for Resilient Data Backup}

 Recently, Tian \textit{et al.} \cite{tian2015network} have used perfect matching to design resilient data backup in sensor networks.
The motivation is that on the one hand, sensors deployed in harsh environments are prone to failure, while on the other hand, data generated by sensors may need to be kept for an extended period of time. The work \cite{tian2015network} proposes to back up each \textit{regular sensor}' data in a randomly selected set of \textit{robust sensors}. The goal of the data-backup scheme is to ensure that even under the failure of regular sensors and a large portion of robust sensors, accessing the remaining small fraction of robust sensors can recover all the data. Then \cite{tian2015network} reduces the above requirement to the existence of a perfect matching in some random graph model. Afterwards, the condition for perfect matching containment is used to derive the   number of robust sensors required by a regular sensor.

We have discussed the applications of  $k$-connectivity, $k$-robustness, Hamilton cycle containment and perfect matching containment. Next, we  present results on transitional behavior of these properties in $q$-composite random key graphs.

   \section{Transitional Behavior of\\[-2pt] $q$-Composite Random Key Graphs} \label{sec:main:res:transitional}

Clearly, each of $k$-connectivity, $k$-robustness, Hamilton cycle containment and perfect matching containment is a monotone increasing graph property.
For each $n$, given $P_n$, the probability that a $q$-composite random key graph $G_q(n,K_n,P_n)$ has a monotone increasing   property increases as $K_n$ increases \cite{zz}. {\color{black} The reason is that stochastically speaking, increasing $K_n$ means adding more edges to   $G_q(n,K_n,P_n)$ as the probability of an edge existence between two nodes increases. With $\mathcal{I}$ denoting  one of $k$-connectivity, $k$-robustness, Hamilton cycle containment, or perfect matching containment, if $K_n =0$, then $G_q(n,K_n,P_n)$ is an empty graph and thus has property $\mathcal{I}$ with probability $0$; if $K_n = P_n$, then $G_q(n,K_n,P_n)$ is an complete graph and thus has property $\mathcal{I}$ with probability $1$ (for all $n$ sufficiently large); and if $K_n$ increases from $0$ to $P_n$, the probability of $G_q(n,K_n,P_n)$ having property $\mathcal{I}$ increases from $0$ to $1$, so there is a transition. In what follows, our Theorem \ref{thm-kcon} on $k$-connectivity, Theorem \ref{thm-krob} on $k$-robustness, Theorem \ref{thm-hc} on Hamilton cycle containment, and   Theorem \ref{thm-pm} on perfect matching containment, show that $q$-composite random key graphs exhibit sharp transitions for  these properties.

   \subsection{Results of $q$-composite random key graphs} \label{sec-results-qRKG}

     We present the main results in Theorems \ref{thm-kcon}--\ref{thm-pm} below. The comparison between them and related results in the literature is given in Section \ref{related}. In this paper, all asymptotics and limits are taken with $n \to \infty$. We use the standard asymptotic notation
 $o(\cdot), O(\cdot), \omega(\cdot),
 \Omega(\cdot), \Theta(\cdot)$; see  \cite[Page 2-Footnote 1]{ZhaoYaganGligor}.
  Also, $\mathbb{P}[\cdot]$
denotes an event probability.

{\color{black} Theorem \ref{thm-kcon} below gives the asymptotically exact probability for $k$-connectivity in $G_q(n,K_n,P_n)$.}

   \begin{thm}[{$k$-Connectivity in $q$-composite random key graphs with \textit{\textbf{improvements over the conference paper \cite{ANALCO}}}}] \label{thm-kcon}
For a $q$-composite random key graph $G_q(n,K_n,P_n)$, if there is a sequence $\alpha_n$ with $\lim_{n \to \infty}{\alpha_n} \in [-\infty, \infty]$
such that
\begin{align}
\frac{1}{q!} \cdot \frac{{K_n}^{2q}}{{P_n}^{q}} & = \frac{\ln  n + {(k-1)} \ln \ln n + {\alpha_n}}{n}, \label{thm-kcon:pe}
\end{align}
then we have
 \begin{align}
& \hspace{-55pt}\lim_{n \to \infty}  \mathbb{P}[\hspace{2pt}G_q(n,K_n,P_n)\text{ is
  $k$-connected.}\hspace{2pt}] \nonumber \\ & \hspace{-55pt}=  e^{-
\frac{e^{- \iffalse \lim\limits_ \fi \lim_{n \to \infty}{\alpha_n}}}{(k-1)!}} \label{expr-kcon-all}
 \end{align}
 \begin{subnumcases}{=} 0,&\text{\hspace{-4pt}if  }$ \lim_{n \to \infty}{\alpha_n} =- \infty$, \label{expr-kcon-0} \\[-2.0pt]
1,&\text{\hspace{-4pt}if  }$ \lim_{n \to \infty}{\alpha_n} = \infty$, \label{expr-kcon-1} \\[-2.0pt] e^{-
\frac{e^{-\alpha^{*}}}{(k-1)!}}  ,&\text{\hspace{-4pt}if  }$ \lim_{n \to \infty}{\alpha_n} = \alpha^{*}\in (-\infty, \infty)$,\label{expr-kcon-exact}  \end{subnumcases}
under
\begin{align}
P_n  &= \begin{cases} \, \Omega(n), &\text{for } q=1,\\
\, \omega\big(n^{2-\frac{1}{q}}(\ln n)^{2+\frac{1}{q}}\big), &\text{for } q \geq 2.
\end{cases}  \label{scalingP-stronger}
\end{align}
\end{thm}

Theorem \ref{thm-kcon} shows that $q$-composite random key graphs exhibit sharp transitions for $k$-connectivity. In particular, it suffices to have an unbounded deviation of $\alpha_n$ in (\ref{thm-kcon:pe}) to ensure that $G_q(n,K_n,P_n)$ is $k$-connected with probability $0$ or $1$, where $\alpha_n$ measures the deviation of $\frac{1}{q!} \cdot \frac{{K_n}^{2q}}{{P_n}^{q}} $ from the critical scaling $\frac{\ln  n + {(k-1)} \ln \ln n}{n}$ as given by (\ref{thm-kcon:pe}).   From~\cite{ANALCO}, the term $\frac{1}{q!} \cdot \frac{{K_n}^{2q}}{{P_n}^{q}}$ in (\ref{thm-kcon:pe}) is an asymptotic value of the edge probability.


 Theorem \ref{thm-krob} below gives a zero--one law for $k$-robustness in a $q$-composite random key graph $G_q(n,K_n,P_n)$. Since the interpretations of Theorems \ref{thm-krob}--\ref{thm-pm} will be similar to that of Theorem~\ref{thm-kcon} above, we omit the details to save space.

   \begin{thm}[{$k$-Robustness in $q$-composite random key graphs}] \label{thm-krob}
For a $q$-composite random key graph $G_q(n,K_n,P_n)$, if there is a sequence $\beta_n$ such that
\begin{align}
\frac{1}{q!} \cdot \frac{{K_n}^{2q}}{{P_n}^{q}} & = \frac{\ln  n + {(k-1)} \ln \ln n + {\beta_n}}{n}, \label{thm-krob:pe}
\end{align}
then it holds that
 \begin{align}
& \lim_{n \to \infty}  \mathbb{P}[\hspace{2pt}G_q(n,K_n,P_n)\text{ is
  $k$-robust.}\hspace{2pt}] \nonumber
 \end{align}
 \begin{subnumcases}{=} 0,&\text{\hspace{-4pt}if  }$ \lim_{n \to \infty}{\beta_n} =- \infty$, \label{expr-krob-0} \\[-2.0pt]
1,&\text{\hspace{-4pt}if  }$ \lim_{n \to \infty}{\beta_n} = \infty$,\label{expr-krob-1}  \end{subnumcases}
under
\begin{align}
P_n  &= \begin{cases} \, \omega\big(n (\ln n)^{5}\big), &\text{for } q=1,\\
\, \omega\big(n^{2-\frac{1}{q}}(\ln n)^{2+\frac{1}{q}}\big). &\text{for } q \geq 2.
\end{cases}  \label{scalingP}
\end{align}
\end{thm}

{\color{black} Theorem \ref{thm-hc} (resp., Theorem \ref{thm-pm}) below gives the asymptotically exact probability for Hamilton cycle containment (resp., perfect matching containment) in $G_q(n,K_n,P_n)$.}

   \begin{thm}[{Hamilton cycle containment in $q$-composite random key graphs}] \label{thm-hc}
For a $q$-composite random key graph $G_q(n,K_n,P_n)$, if there is a sequence $\gamma_n$ with $\lim_{n \to \infty}{\gamma_n} \in [-\infty, \infty]$
such that
\begin{align}
\frac{1}{q!} \cdot \frac{{K_n}^{2q}}{{P_n}^{q}} & = \frac{\ln  n +  \ln \ln n +
 {\gamma_n}}{n}, \label{thm-hc:pe}
\end{align}
then it holds under (\ref{scalingP}) that
 \begin{align}
& \hspace{-1pt}\lim_{n \to \infty}  \mathbb{P}[\hspace{2pt}G_q(n,K_n,P_n)\text{ contains a Hamilton cycle.}\hspace{2pt}] \nonumber \\ &  \hspace{-1pt} = e^{- e^{- \iffalse \lim\limits_ \fi \lim_{n \to \infty}{\gamma_n}}}\label{expr-hc-all}
 \end{align}
 \begin{subnumcases}{=} 0,&\text{\hspace{-4pt}if  }$ \lim_{n \to \infty}{\gamma_n} =- \infty$, \label{expr-hc-0} \\[-2.0pt]
1,&\text{\hspace{-4pt}if  }$ \lim_{n \to \infty}{\gamma_n} = \infty$, \label{expr-hc-1} \\[-2.0pt]
e^{-e^{- \gamma^{*}}},&\text{\hspace{-4pt}if  }$ \lim_{n \to \infty}{\gamma_n} = \gamma^{*}\in (-\infty, \infty)$.\label{expr-hc-exact}  \end{subnumcases}
\end{thm}


   \begin{thm}[{Perfect matching containment in $q$-composite random key graphs}] \label{thm-pm}
For a $q$-composite random key graph
$G_q(n,K_n,P_n)$ with even $n$, if there is a sequence $\xi_n$ with $\lim_{n \to \infty}{\xi_n} \in [-\infty, \infty]$
such that
\begin{align}
\frac{1}{q!} \cdot \frac{{K_n}^{2q}}{{P_n}^{q}} & = \frac{\ln  n   +
 {\xi_n}}{n}, \label{thm-pm:pe}
\end{align}
then it holds under (\ref{scalingP}) that
 \begin{align}
& \hspace{6pt} \lim_{n \to \infty}  \mathbb{P}[\hspace{2pt}G_q(n,K_n,P_n)\text{ contains a perfect matching.}\hspace{2pt}] \nonumber \\ &  \hspace{6pt} = e^{- e^{- \iffalse \lim\limits_ \fi \lim_{n \to \infty}{\xi_n}}} \label{expr-pm-all}
 \end{align}
 \begin{subnumcases}{=} 0,&\text{\hspace{-4pt}if  }$ \lim_{n \to \infty}{\xi_n} =- \infty$, \label{expr-pm-0} \\[-2.0pt]
1,&\text{\hspace{-4pt}if  }$ \lim_{n \to \infty}{\xi_n} = \infty$, \label{expr-pm-1} \\[-2.0pt]
e^{-e^{- \xi^{*}}},&\text{\hspace{-4pt}if  }$ \lim_{n \to \infty}{\xi_n} = \xi^{*}\in (-\infty, \infty)$.\label{expr-pm-exact}  \end{subnumcases}
\end{thm}

%
%

%

We establish Theorems \ref{thm-kcon}--\ref{thm-pm} in Section \ref{sec-mainproofs}. From Theorems \ref{thm-kcon}--\ref{thm-pm}, $k$-robustness (resp., Hamilton cycle containment, and perfect matching containment) in $G_q(n,K_n,P_n)$ has similar asymptotic behavior as $k$-connectivity (resp., $2$-connectivity and $1$-connectivity (i.e., connectivity)).

\subsection{Design guidelines for secure sensor networks} \label{sec-properties-design-guidelines}

 Based on Theorems \ref{thm-kcon}--\ref{thm-pm},   we provide design guidelines
of a secure sensor network employing the $q$-composite key predistribution scheme \cite{adrian} and modeled by a $q$-composite random key graph $G_q(n,K_n,P_n)$. We identify the {critical} value of each parameter given other parameters such that the network has the desired property with probability at least $p$. Taking $k$-connectivity as an example, we set $e^{-
\frac{e^{- \iffalse \lim\limits_ \fi \lim_{n \to \infty}{\alpha_n}}}{(k-1)!}}$ in (\ref{expr-kcon-all}) to be at least $p$ to get $\lim_{n \to \infty}{\alpha_n} \geq - \ln [(k-1)!\ln \frac{1}{p}]$. To use asymptotic results for large network design, we just consider $\alpha_n\geq - \ln [(k-1)!\ln \frac{1}{p}]$, and obtain from (\ref{thm-kcon:pe}) that
 \begin{align}
\textstyle{\frac{1}{q!} \cdot \frac{{K_n}^{2q}}{{P_n}^{q}} \geq \frac{\ln  n + {(k-1)} \ln \ln n - \ln [(k-1)!\ln \frac{1}{p}]}{n}.}\label{pe-critical-condition}
\end{align}
However, (\ref{pe-critical-condition}) may not hold since all parameters are integers. Since $\frac{1}{q!} \cdot \frac{{K_n}^{2q}}{{P_n}^{q}}$ on the left-hand side of (\ref{pe-critical-condition}) increases as the \textit{key ring size} $K_n$ increases or as the \textit{key pool size} $P_n$ decreases, and the term on the right-hand side of (\ref{pe-critical-condition}) decreases as the
number $n$ of nodes increases for large $n$, we define the
critical key ring size (resp., the
critical key pool size, the
critical number of nodes) as the minimal $K_n$ (resp., the maximal $P_n$, the minimal $n$) such that (\ref{pe-critical-condition}) holds. Hence, for $k$-connectivity, the
critical key ring size equals
$ \Big\lceil\sqrt{P_n} \cdot \sqrt[2q]{\frac{q!\cdot\{\ln  n + {(k-1)} \ln \ln n - \ln [(k-1)!\ln \frac{1}{p}]\}}{n}} \hspace{1pt}\Big\rceil $, the
critical key pool size equals $ \Big\lfloor {K_n}^2 \cdot\sqrt[q]{\frac{n}{q!\cdot\{\ln  n + {(k-1)} \ln \ln n - \ln [(k-1)!\ln \frac{1}{p}]\}}}\Big\rfloor $, while the
critical number of nodes can be solved numerically.

We provide some concrete numbers for better understanding of the above guidelines. Typically, we choose $q$ to be no greater than $3$ since larger $q$ means more difficulty for two sensors to satisfy the requirement of key sharing for establishing a secure link. Below we discuss the choices of $P_n$ and $K_n$ for different $q$ and $n$. Again, we focus on $k$-connectivity since the discussions of other properties are similar. Roughly speaking, for small $q$ (e.g., $q=1,2,3$) and $n$ being thousands, we can choose $K_n$ to be dozens and $P_n$ to be tens of thousands to have a $k$-connected network for small $k$ (e.g., $k=1,2,3$) with a relatively high probability $p$ (e.g., $p=0.95$). We can also let $K_n$ to be hundreds and $P_n$ to be hundreds of thousands. As concrete examples, for $q=2$ and $n=1000$, we can choose $K_n=88$ and $P_n=50000$ to have the network $2$-connected with probability $0.99$ or $3$-connected with probability $0.95$, and choose $K_n=92$ and $P_n=50000$ to have the network $3$-connected with probability $0.99$. For $q=3$, $K_n=300$ and $P_n=250000$, we can choose $n=1700$ to have the network $2$-connected with probability $0.99$ or $3$-connected with probability $0.95$, and choose $n=1500$ to have the network $1$-connected with probability $0.99$ or $2$-connected with probability $0.95$.

Comparing (\ref{thm-krob:pe}) (\ref{thm-hc:pe}) (\ref{thm-pm:pe}) with (\ref{thm-kcon:pe}), we know that the critical parameters for $k$-robustness (resp., Hamilton cycle containment and perfect matching containment) are the same as those for  $k$-connectivity (resp., $2$-connectivity and $1$-connectivity). Hence, we can easily use the   guidelines for $k$-connectivity to obtain the corresponding guidelines for $k$-robustness (resp., Hamilton cycle containment and perfect matching containment).

\begin{figure*}[!t]
\addtolength{\subfigcapskip}{-4pt}\centering     
\hspace{0pt}\subfigure[]{\label{transition_q2_varyK}\includegraphics[height=0.17\textwidth]{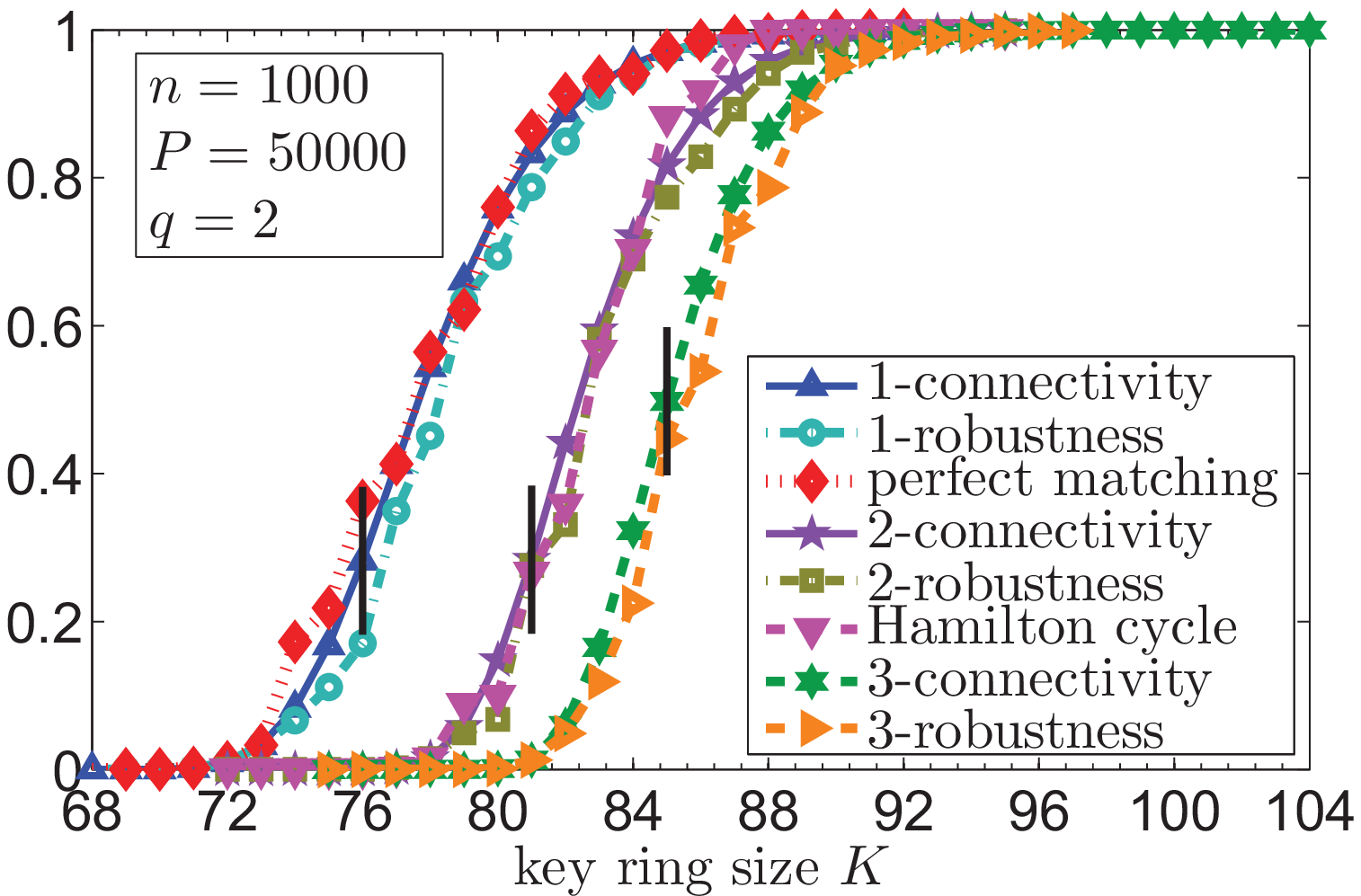}}
\hspace{0pt}~~~~~~\subfigure[]{\label{transition_q2_varyP}\includegraphics[height=0.17\textwidth]{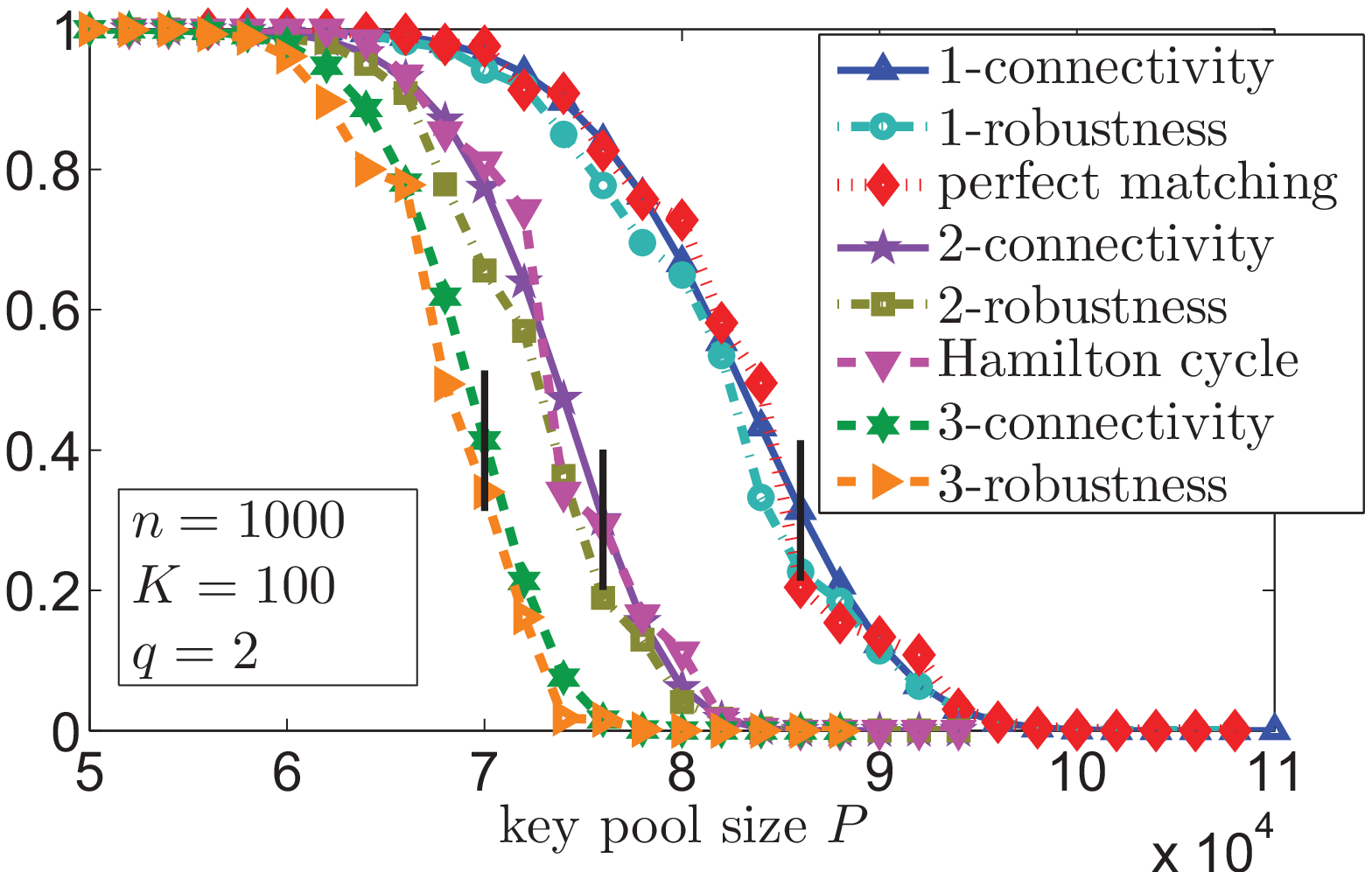}}
\hspace{0pt}~~~~~~\subfigure[]{\label{transition_q2_varyn}\includegraphics[height=0.17\textwidth]{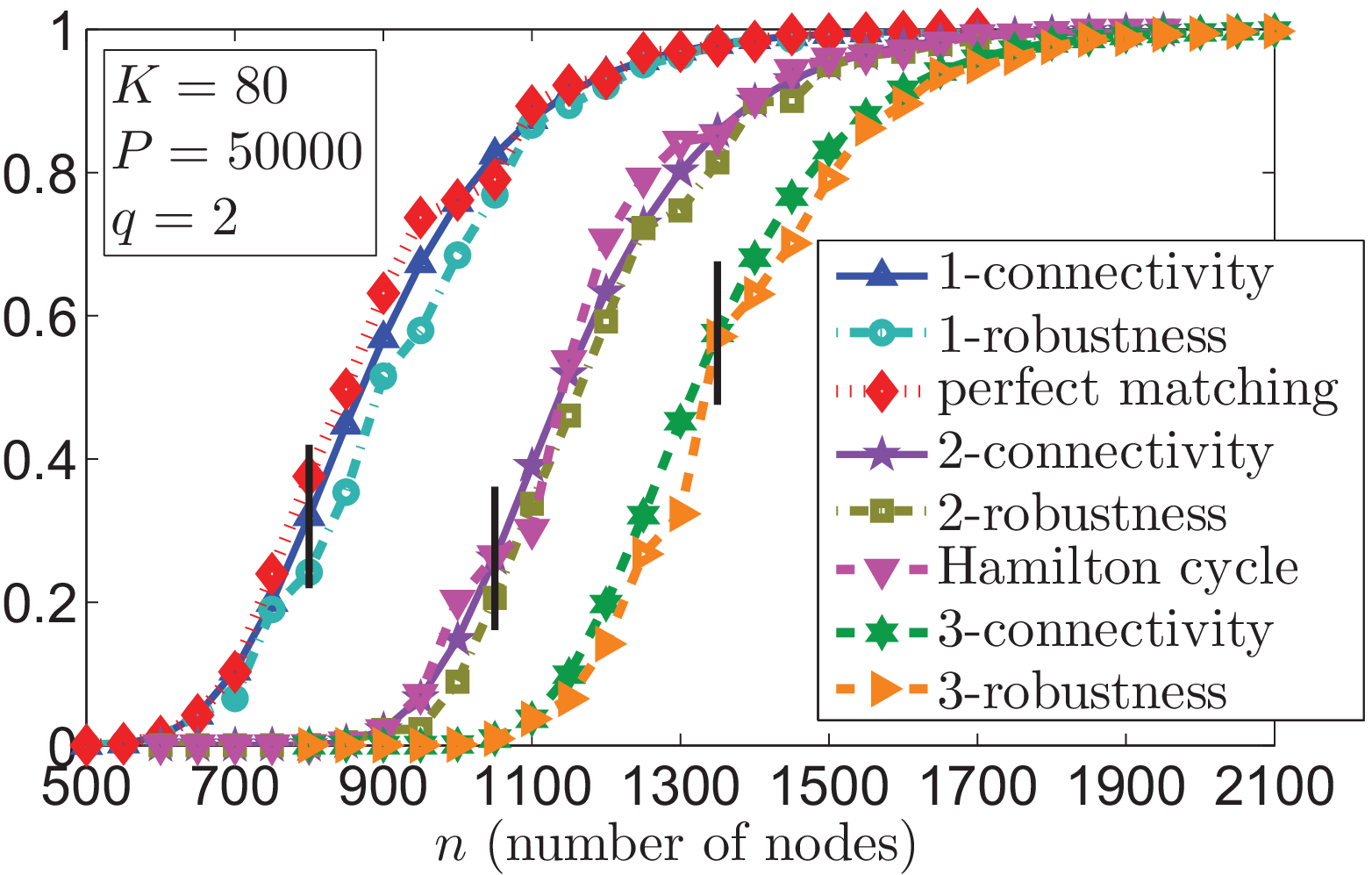}}
  \caption{For $G_q(n,K_n,P_n)$ under $q=2$, we plot its probabilities in terms of $k$-connectivity, $k$-robustness, Hamilton cycle containment and perfect matching containment. In each subfigure, each vertical line presents the
\emph{critical} parameter computed based on Section \ref{sec-properties-design-guidelines}.} \label{transition_q2}
\end{figure*}

\begin{figure*}[!t]
\addtolength{\subfigcapskip}{-4pt}\centering     
\hspace{0pt}\subfigure[]{\label{transition_q3_varyK}\includegraphics[height=0.17\textwidth]{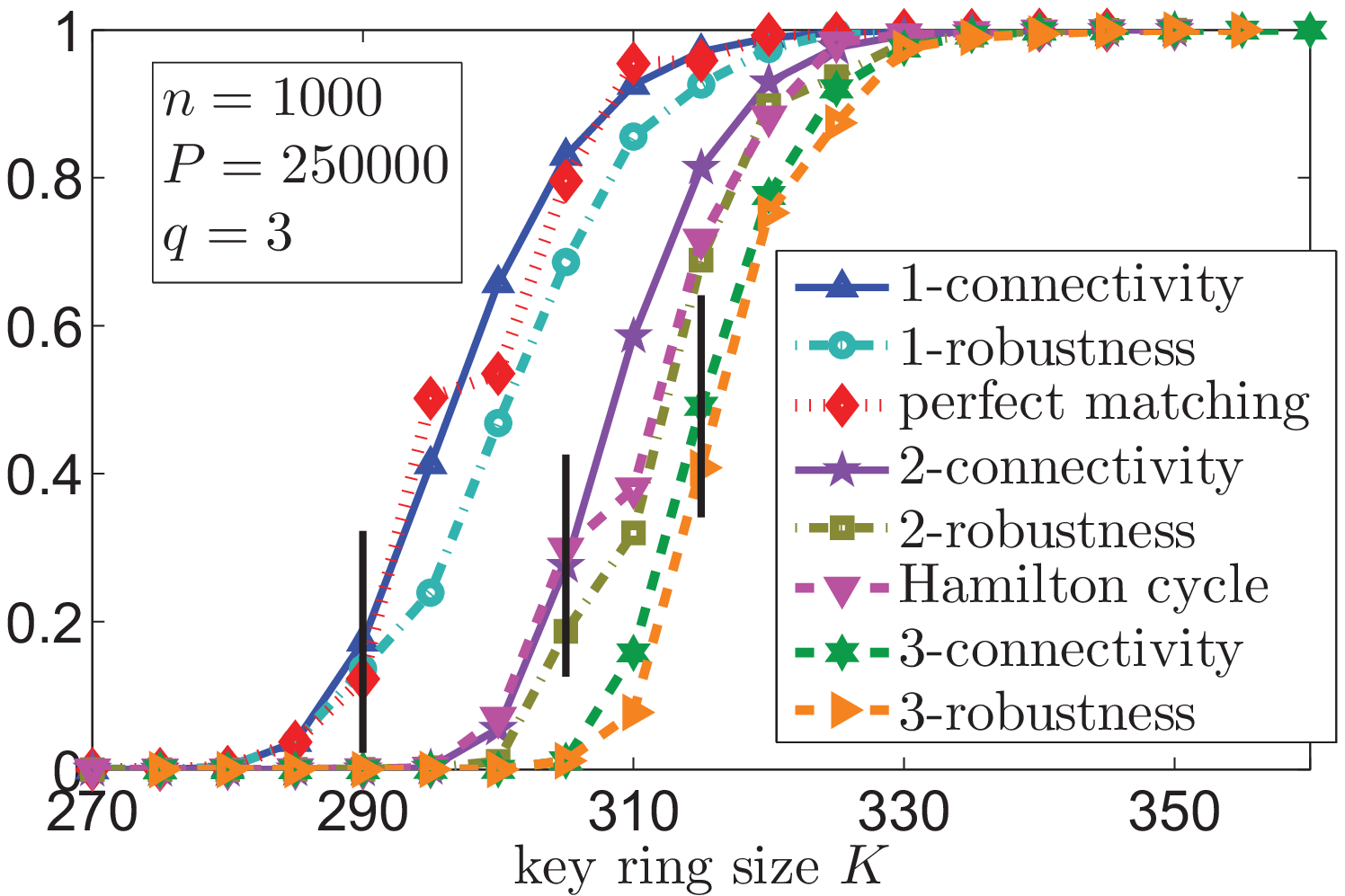}}
\hspace{0pt}~~~~~~\subfigure[]{\label{transition_q3_varyP}\includegraphics[height=0.17\textwidth]{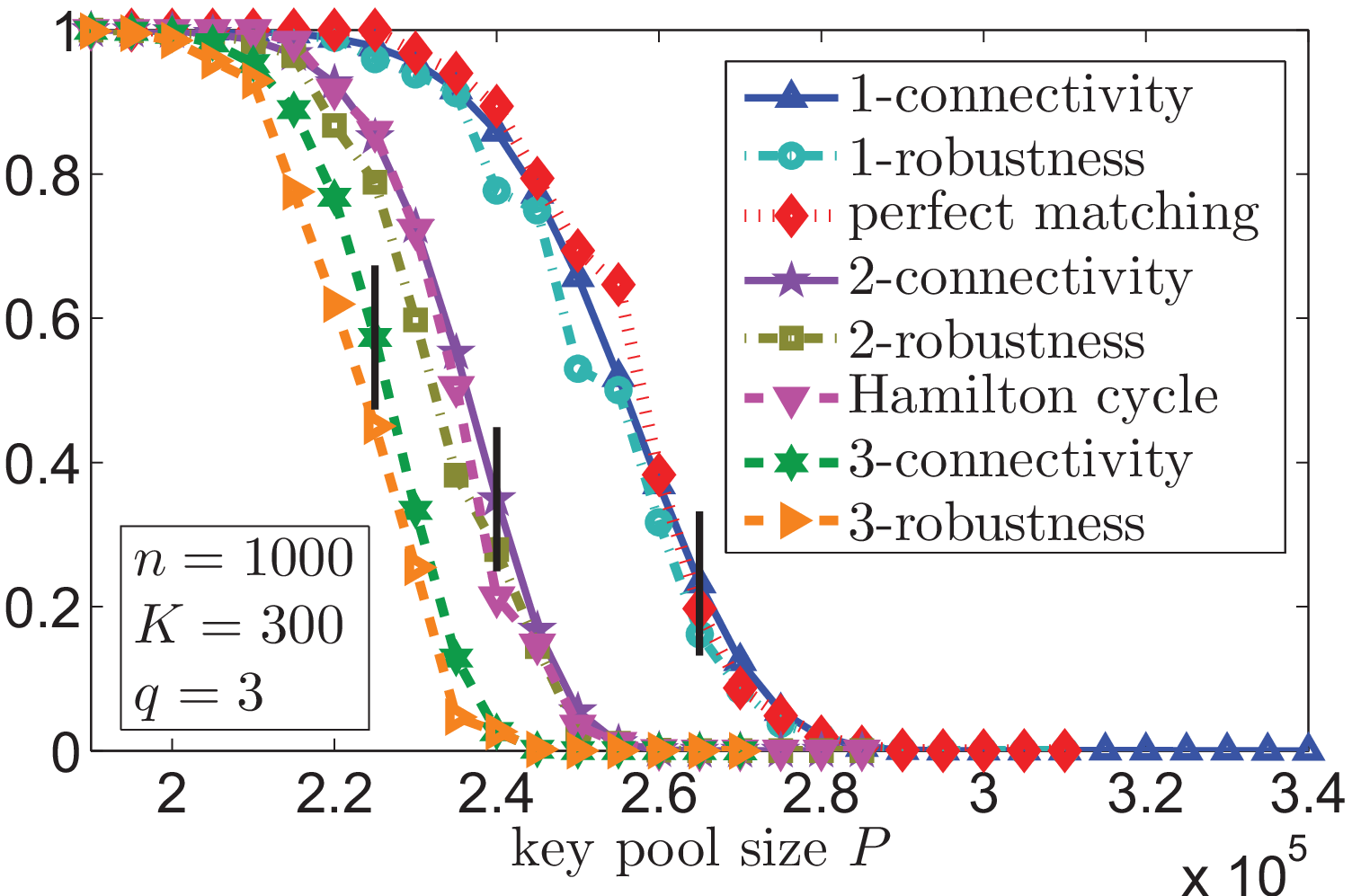}}
\hspace{0pt}~~~~~~\subfigure[]{\label{transition_q3_varyn}\includegraphics[height=0.17\textwidth]{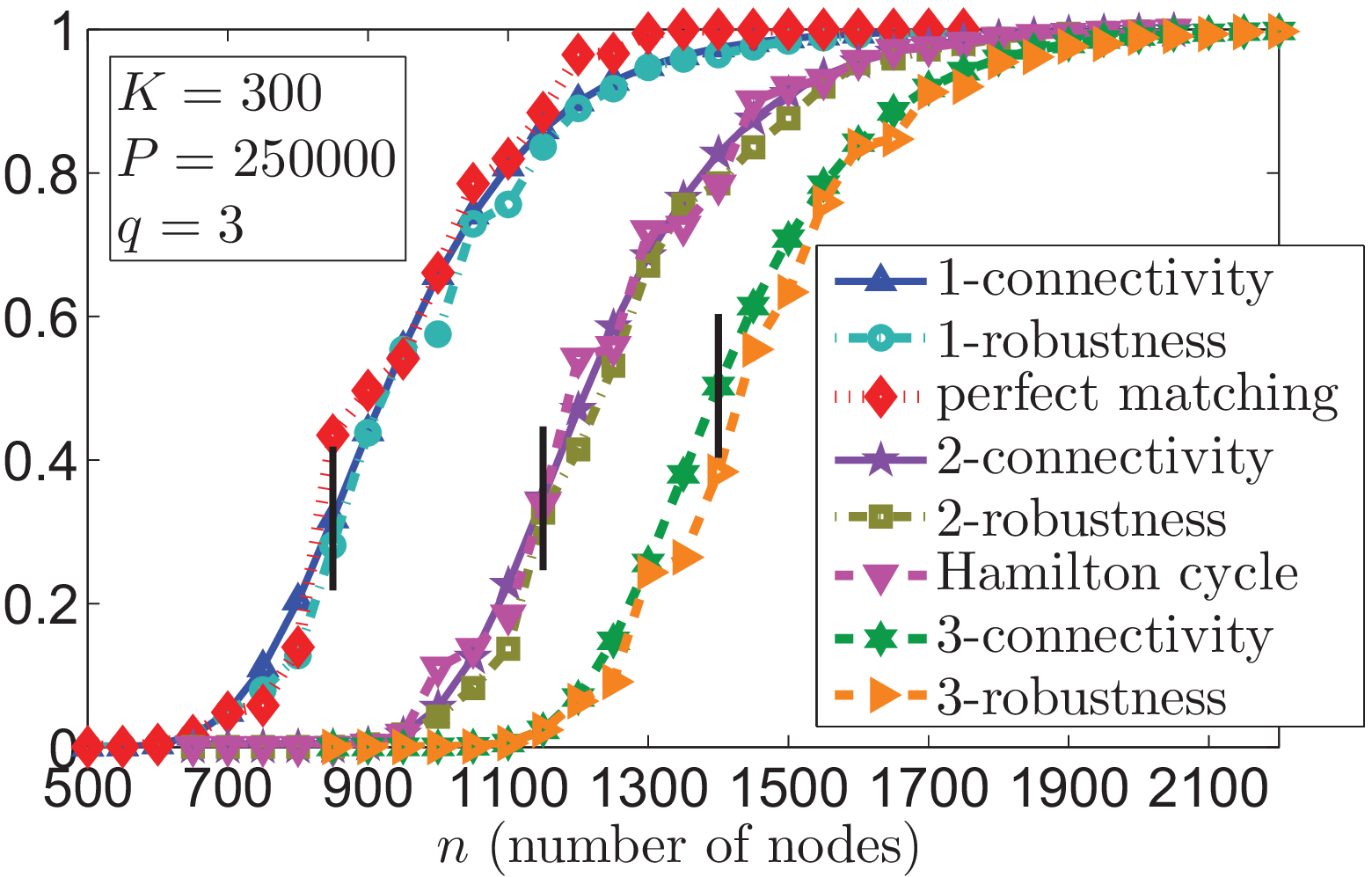}}
 \caption{For $G_q(n,K_n,P_n)$ under $q=3$, we plot its probabilities in terms of $k$-connectivity, $k$-robustness, Hamilton cycle containment and perfect matching containment. In each subfigure, each vertical line presents the
\emph{critical} parameter computed based on Section \ref{sec-properties-design-guidelines}.} \label{transition_q3}
\end{figure*}

\subsection{Experimental results}  \label{experiments-transition}

We present experiments below to confirm our theoretical results in Theorems \ref{thm-kcon}--\ref{thm-pm}. For $G_q(n,K_n,P_n)$, we plot its probabilities in terms of $k$-connectivity, $k$-robustness, Hamilton cycle containment and perfect matching containment in Figure  \ref{transition_q2} for $q=2$, and in Figure \ref{transition_q3}  for $q=3$, \\
$\bullet$ when the key ring size $K$ varies in Figures \ref{transition_q2_varyK} and \ref{transition_q3_varyK},
 \\
$\bullet$ when the key pool size $P$ varies in Figures \ref{transition_q2_varyP} and \ref{transition_q3_varyP},
 \\
$\bullet$ when the number $n$ of nodes varies in Figures \ref{transition_q2_varyn} and \ref{transition_q3_varyn}.
\\
For each data point, we
generate $500$ independent samples of $G_q(n,K_n,P_n)$, record the count that the obtained graph has the studied property, and then divide the count by $500$ to obtain the corresponding empirical probability. In each  figure, we   see the transitional behavior. Also, we observe that the probability
 \\
$\bullet$   increases with $K$ (resp., $n$) while fixing other parameters,
  \\
$\bullet$  decreases with $P$ while fixing other parameters.
\\
Moreover, in each figure, each vertical line presents the
{critical} parameter computed based on Section \ref{sec-properties-design-guidelines} above with probability $p$ being $0.5$: the critical key ring size in Figures \ref{transition_q2_varyK} and \ref{transition_q3_varyK}, the critical key pool size in Figures \ref{transition_q2_varyP} and \ref{transition_q3_varyP}, and the critical number nodes in Figures \ref{transition_q2_varyn} and \ref{transition_q3_varyn}. Summarizing the above, the experiments have confirmed our Theorems \ref{thm-kcon}--\ref{thm-pm}.


\section{Using the Transition Width to Quantify the Transitional Behavior in Section \ref{sec:main:res:transitional}} \label{sec:sharp:transition:1-composite}

In the previous section, we have presented the sharp transitions in $q$-composite random key graphs in terms of the studied graph properties. To further quantify the sharpness, we aim to understand how should $K_n$ grow to raise the probability of $G_q(n,K_n,P_n)$ having property $\mathcal{I}$ from $\epsilon$ to $1-\epsilon$ for a positive constant $\epsilon < \frac{1}{2}$, where $\mathcal{I}$ denotes one of $k$-connectivity,  Hamilton cycle containment, or perfect matching containment. To this end,
   noting that there may not exist  $K_n$ such that $G_q(n,K_n,P_n)$ has property $\mathcal{I}$  with probability \emph{exactly} $\epsilon$ or $1-\epsilon$ since $K_n$ is an integer,
 we quantify $K_n$ that renders $G_q(n,K_n,P_n)$ having property $\mathcal{I}$ with probability \emph{at least} $\epsilon$ or $1-\epsilon$.} To this end, we formally define for $\epsilon < \frac{1}{2}$ that
\begin{align}
K_{q,n}^{-}(\mathcal{I}, \epsilon) := \min \bigg\{K_n \bigg| \begin{array}{l}G_q(n,K_n,P_n)\text{ has property }\mathcal{I}\\ \text{  with probability {at least} }\epsilon .\end{array}\bigg\} \label{KnI-minus}
\end{align}
and
\begin{align}
\hspace{-10pt}K_{q,n}^{+}(\mathcal{I}, \epsilon) := \min \bigg\{K_n \bigg| \begin{array}{l}G_q(n,K_n,P_n)\text{ has property }\mathcal{I}\\ \text{  with probability {at least} }1-\epsilon .\end{array}\bigg\}.\label{KnI-plus}
\end{align}
Then the transition width $d_{q,n}(\mathcal{I}, \epsilon)$ of graph $G_q(n,K_n,P_n)$ for
property $\mathcal{I}$ and $\epsilon < \frac{1}{2}$ is defined by
\begin{align}
d_{q,n}(\mathcal{I}, \epsilon) = K_{q,n}^{+}(\mathcal{I}, \epsilon) - K_{q,n}^{-}(\mathcal{I}, \epsilon). \label{phase-transition-width}
\end{align}

   \subsection{Transition widths for $k$-connectivity, perfect matching containment and Hamilton cycle containment} \label{sec-width-guideline}

{\color{black}
Theorem \ref{thm-PHW} later in this section presents the result  of the transition width $d_{q,n}(\mathcal{I}, \epsilon)$ for a $q$-composite random key graph $G_q(n,K_n,P_n)$. For $G_q(n,K_n,P_n)$ modeling a secure sensor network employing the $q$-composite key predistribution scheme \cite{adrian} in practice (so that $P_n = \Omega(n)$ from \cite[Equation (2)]{zhao2015resilience}), we have:
\begin{itemize}
\item[\ding{172}] for $q=1$, the transition width $d_{q,n}(\mathcal{I}, \epsilon)$ can be very small (even $0$ or $1$ as detailed below) for $\mathcal{I}$ being $k$-connectivity,
\item[\ding{173}] for $q\geq 2$, $d_{q,n}(\mathcal{I}, \epsilon)$ scales with $n$ and can be written as $\omega(1)$ (i.e., it converges to $\infty$ as $n \to \infty$).
\end{itemize}
Roughly speaking, the transitional behavior for $q=1$ can be much sharper than that for $q\geq 2$. We now discuss the implication to secure sensor network applications of $G_q(n,K_n,P_n)$.

Recall from Section \ref{sec-Introduction} that the $q$-composite key predistribution scheme \cite{adrian} in the special case of $q=1$ becomes the Eschenauer--Gligor (EG) key predistribution scheme \cite{virgil}. Then the result above shows
a fundamental difference between the EG scheme and the $q$-composite scheme (with $q\geq 2$) in terms of the transition width. We can interpret the difference as a result that the EG scheme is more fragile than the $q$-composite scheme in terms of preserving $k$-connectivity under key revocation, where key revocation means removing keys that have been compromised \cite{ChenGligorPerrigMuralidharanTDSC2005}. In addition to cryptographic exposure, another significant reason for keys being compromised is a sensor-capture attack resulting in that all secret keys of a captured
node are
discovered by the adversary. Sensors
deployed in hostile environments are particularly prone to capture because their protection is limited by low-cost considerations (in fact their operation
is often unattended)
\cite{ChenGligorPerrigMuralidharanTDSC2005,virgil,adrian}. For a $k$-connected secure sensor network under the EG scheme, since $d_{q,n}(\text{$k$-connectivity},  \epsilon)$ may be $0$ or $1$  from result \ding{172} above, then even revoking a single key may induce losing $k$-connectivity (this is confirmed by experiments of Figure \ref{sharp_q1} explained in Section \ref{experiments-width}). In contrast, such extreme phenomenon does not happen for a secure sensor network under the $q$-composite scheme with $q\geq 2$. This fundamental difference between the EG scheme and the $q$-composite scheme with $q\geq 2$ can be useful for the design of secure sensor networks; e.g., $q\geq 2$ is preferred over $q = 1$ if one desires stronger resilience of $k$-connectivity against key revocation.




\begin{thm} \label{thm-PHW}
For a $q$-composite random key graph $G_q(n,K_n,P_n)$, we have:\\
\textbullet~For $q=1$, we have results (i.1) (i.2a) and (i.2b) below:\\
\textbf{(i.1)} $d_{1,n}(\text{$k$-connectivity},  \epsilon) = $
\begin{subnumcases}{\hspace{-25pt}}
\text{$0$ or $1$ for each $n$ sufficiently large}, \nonumber \\ ~~~~~~~~ \text{if $P_n = \Omega(n)$ and $P_n = o(n\ln n)$}, \label{transition-width-kcon-1} \\ \Theta(1) , \text{ if $P_n = \Theta(n\ln n)$}, \label{transition-width-kcon-2} \\ \omega(1) , \text{ if $P_n = \omega(n\ln n)$}; \label{transition-width-kcon-3}
\end{subnumcases}
  \textbf{(i.2a)} $d_{1,n}(\text{Hamilton cycle containment}, \epsilon)$ and \\ $d_{1,n}(\text{perfect matching containment}, \epsilon)$ can both be written as $\omega(1)$, if $P_n  = \omega\big(n (\ln n)^5\big)$. \textbf{(i.2b)} Moreover, if we  improve Theorem \ref{thm-hc} (resp., Theorem \ref{thm-pm}) by weakening the condition of $P_n$ for $q=1$ from $P_n  = \omega\big(n (\ln n)^5\big)$ to $P_n = \Omega(n)$, then $d_{1,n}(\text{Hamilton cycle containment}, \epsilon)$ (resp., $d_{1,n}(\text{perfect matching containment}, \epsilon)$) also satisfies (\ref{transition-width-kcon-1}) (\ref{transition-width-kcon-2})  (\ref{transition-width-kcon-3}) above.\\
\textbullet~For $q\geq 2$, we have the following results (ii.a) and (ii.b):
\\  \textbf{(ii.a)} $d_{q,n}(\text{$k$-connectivity},  \epsilon)$, $d_{q,n}(\text{Hamilton cycle containment}, \epsilon)$, \mbox{and $d_{q,n}(\text{perfect matching containment}, \epsilon)$} can all be written as $\omega(1)$, if $P_n = \omega\big(n^{2-\frac{1}{q}}(\ln n)^{2+\frac{1}{q}}\big)$. \textbf{(ii.b)} Furthermore, if we  improve Theorem \ref{thm-hc} (resp., Theorem \ref{thm-pm}) by weakening the condition of $P_n$ for $q\geq 2$ from $P_n = \omega\big(n^{2-\frac{1}{q}}(\ln n)^{2+\frac{1}{q}}\big)$ to $P_n = \Omega(n)$, then $d_{q,n}(\text{Hamilton cycle containment}, \epsilon)$ (resp., $d_{q,n}(\text{perfect matching containment}, \epsilon)$) is still $\omega(1)$ if $P_n = \Omega(n)$.
\end{thm}



We establish Theorem \ref{thm-PHW} in the Appendix. 

\begin{figure}[!t]
\addtolength{\subfigcapskip}{-4pt}\centering     
\hspace{0pt}\subfigure[]{\label{sharp_q1_Kvary_n1000}\includegraphics[height=0.155\textwidth]{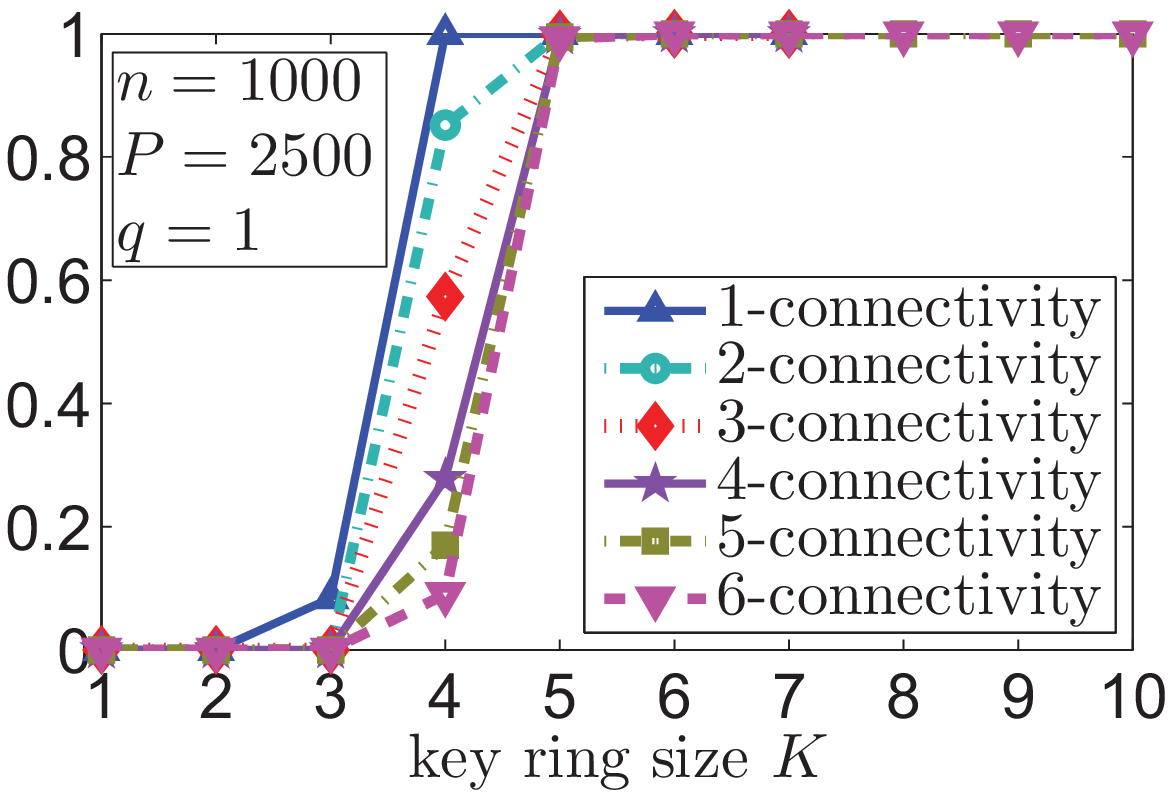}}
\hspace{0pt}\subfigure[]{\label{sharp_q1_Kvary_n10000}\includegraphics[height=0.155\textwidth]{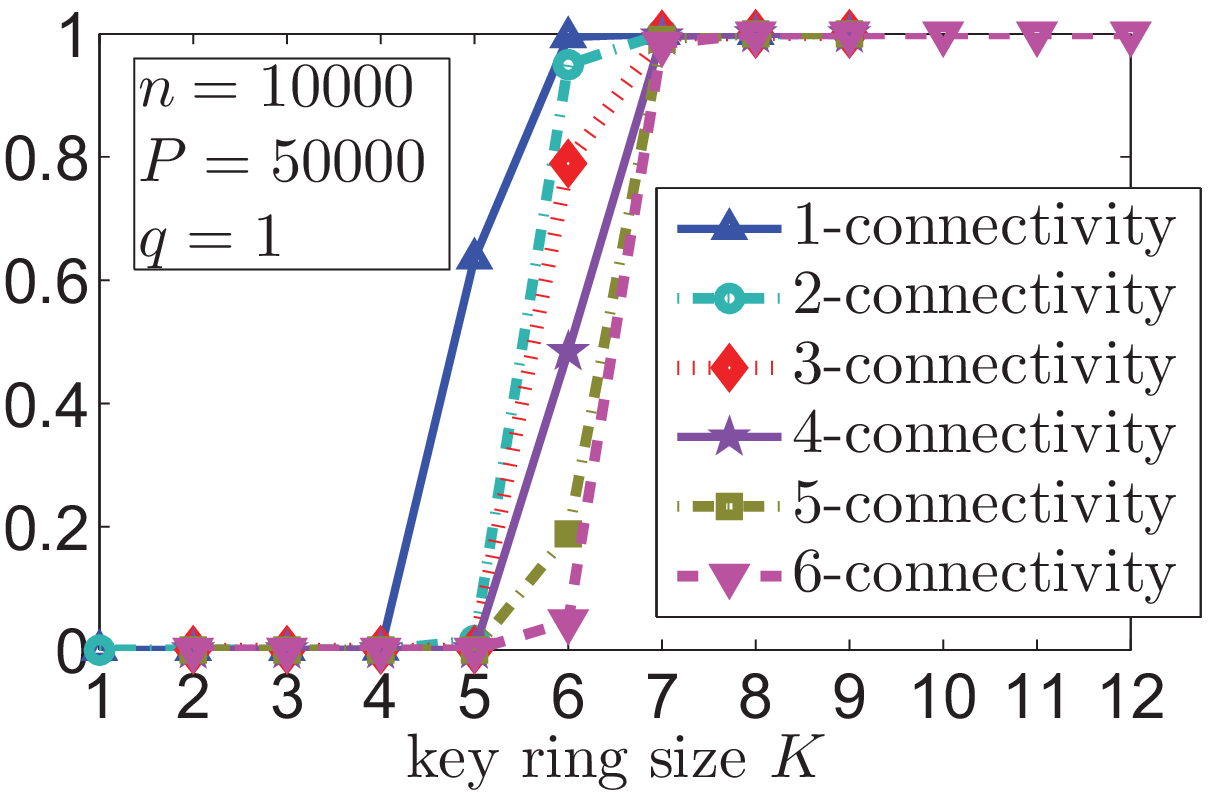}}
 \caption{For $G_1(n,K_n,P_n)$ (i.e., $G_q(n,K_n,P_n)$ under $q=1$), we plot its $k$-connectivity probabilities for $k=1,2,3,4,5,6$. The subfigure (a) (resp., (b)) considers $n=1000$ (resp., $n=10000$). In each curve here, $d_{1,n}(\text{$k$-connectivity},  \epsilon)$ is just $0$ or $1$.} \label{sharp_q1}
\end{figure}

\begin{figure}[!t]
\addtolength{\subfigcapskip}{-4pt}\centering     
\hspace{0pt}\subfigure[]{\label{sharp_no_q1_Kvary_n1000}\includegraphics[height=0.152\textwidth]{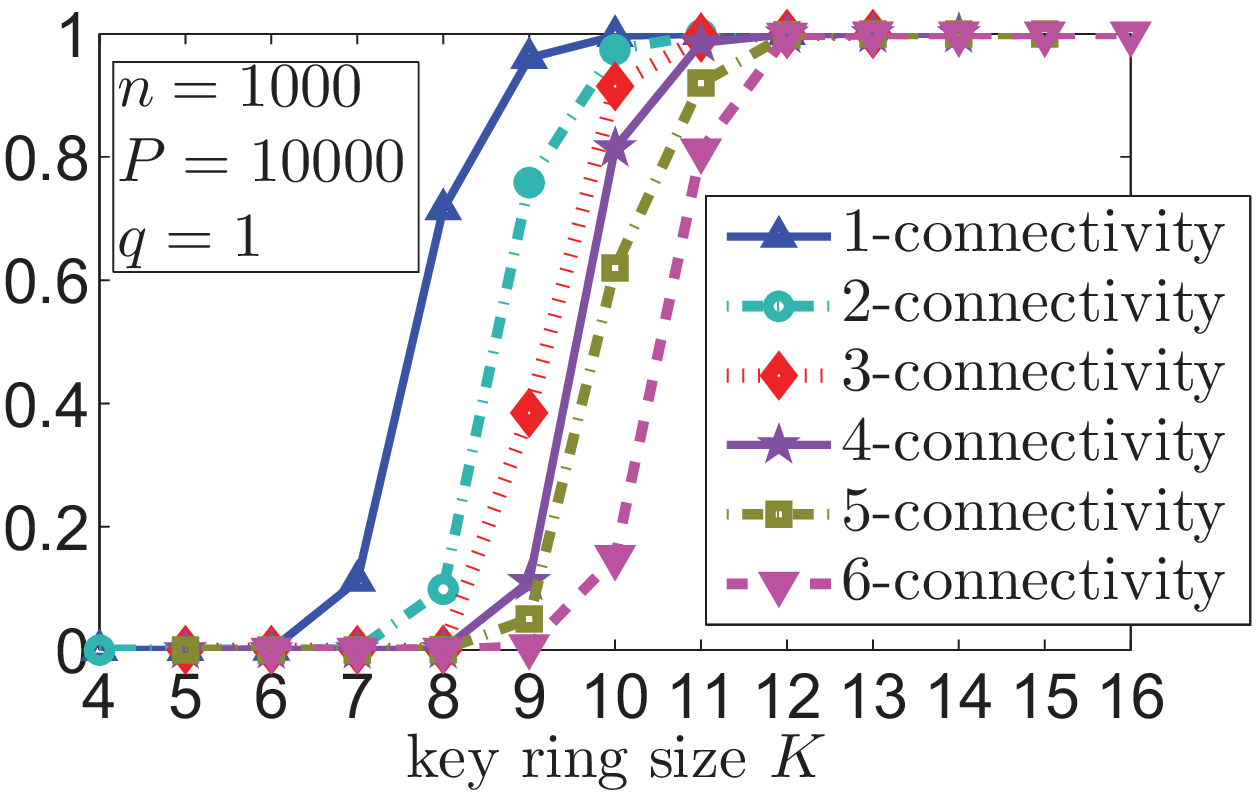}}
\hspace{0pt}\subfigure[]{\label{sharp_no_q1_Kvary_n10000}\includegraphics[height=0.152\textwidth]{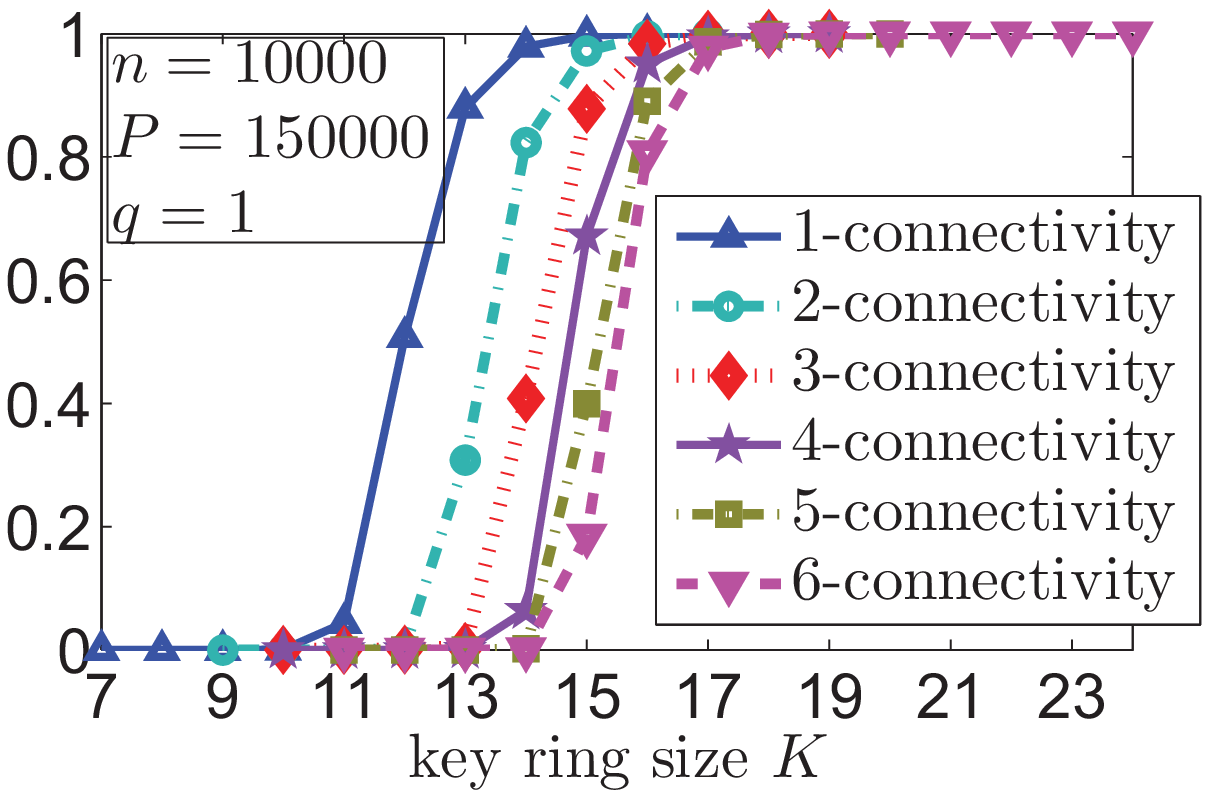}}
 \caption{For $G_1(n,K_n,P_n)$, we plot its $k$-connectivity probabilities for $k=1,2,3,4,5,6$. The subfigure (a) (resp., (b)) considers $n=1000$ (resp., $n=10000$). Comparing Figures \ref{sharp_q1_Kvary_n1000} and \ref{sharp_no_q1_Kvary_n1000} (or comparing Figures \ref{sharp_q1_Kvary_n10000} and \ref{sharp_no_q1_Kvary_n10000}), we see that when $P_n$ increases, $d_{1,n}(\text{$k$-connectivity},  \epsilon)$ can increase from being just $0$ or $1$ to being greater than $1$.} \label{sharp_no_q1}
\end{figure}

 \subsection{Experimental results} \label{experiments-width}

 We present experiments to confirm different behavior of $d_{q,n}(\text{$k$-connectivity},  \epsilon)$ for $q = 1$ and $q\geq 2$, as explained above. Figures \ref{sharp_q1} and \ref{sharp_no_q1} here consider $q=1$, while the case of $q\geq 2$ has already been addressed by Figures  \ref{transition_q2} and \ref{transition_q3}.

In Figures \ref{sharp_q1} and \ref{sharp_no_q1} for $q=1$, we plot the  probabilities of $G_1(n,K_n,P_n)$ being $k$-connected for $k=1,2,3,4,5,6$. For each data point, we
generate $500$ independent samples of $G_1(n,K_n,P_n)$, record the count that the obtained graph is $k$-connected, and then divide the count by $500$ to obtain the corresponding empirical probability.
 Comparing Figures \ref{sharp_q1_Kvary_n1000} and \ref{sharp_no_q1_Kvary_n1000} for $n=1000$ (or comparing Figures \ref{sharp_q1_Kvary_n10000} and \ref{sharp_no_q1_Kvary_n10000} for $n=10000$), we see that when $P_n$ increases, $d_{1,n}(\text{$k$-connectivity},  \epsilon)$ can increase from being just $0$ or $1$ to being greater than $1$.

Compared with Figures \ref{sharp_q1_Kvary_n1000} and \ref{sharp_q1_Kvary_n10000} for $q=1$, Figures \ref{transition_q2_varyK} and \ref{transition_q3_varyK} for $q \geq 2$ present much larger $d_{q,n}(\text{$k$-connectivity},  \epsilon)$.

To summarize, the experiments are useful to illustrate  different behavior of $d_{q,n}(\text{$k$-connectivity},  \epsilon)$ for $q = 1$ and $q\geq 2$.

 \section{Comparing This Paper with Related Work} \label{related}

We first elaborate the improvements of this paper over our recent work \cite{ZhaoTAC}:
\begin{itemize}
\item[i)] This paper considers $G_q(n,K_n,P_n)$ for $\underline{\textit{general $q$}}$ (which makes the analysis challenging), while \cite{ZhaoTAC} addresses $G_1(n,K_n,P_n)$ (i.e., $G_q(n,K_n,P_n)$ in the case of $\underline{q=1}$).
\item[ii)] This paper studies  $\underline{\textit{four}}$ properties: $k$-connectivity, $k$-robustness, Hamilton cycle containment, and perfect matching containment for general $q$ and discusses their applications to networked control, while \cite{ZhaoTAC} tackles only the first $\underline{\textit{two}}$ properties for $q=1$.
\item[iii)] This paper further examines the transition widths of various properties, and demonstrates   different behavior for $q = 1$ and $q\geq 2$ (which provides a useful guideline for resilient design of secure sensor networks as discussed in Section \ref{sec-width-guideline}), while \cite{ZhaoTAC} does not study the transition widths.
\end{itemize}

Now we discuss the improvements of this paper over other related work in terms of different graph properties respectively.

  \textbf{($k$-)Connectivity.}
 For connectivity (i.e., $k$-connectivity in the case of $k=1$) in  $G_1(n,K_n,P_n)$ (i.e., $G_q(n,K_n,P_n)$ in the case of $q=1$), Blackburn and Gerke \cite{r1}, and Ya\u{g}an and Makowski
\cite{yagan} obtain different granularities of zero--one laws;
 Rybarczyk
\cite{ryb3} establishes the asymptotically exact probability result; and earlier studies by Di Pitero \textit{et al.}
\cite{DiPietroTissec}
 report results weaker than the above work \cite{r1,yagan,ryb3}. For $k$-connectivity in  $G_1(n,K_n,P_n)$,  Rybarczyk \cite{zz}
implicitly shows a zero--one law, and we \cite{ZhaoTAC} derive the asymptotically exact probability. For $G_q(n,K_n,P_n)$ with constant $q$, Bloznelis and {\L}uczak \cite{Perfectmatchings} (resp., Bloznelis and Rybarczyk
\cite{Bloznelis201494}) have recently derived the asymptotically exact
probability for $k$-connectivity (resp., connectivity), but both results after a rewriting \vspace{2pt} address only the narrow range of $P_n $ satisfying both $o\big(n^{\frac{1}{q}} (\ln n)^{\frac{2}{5}-\frac{1}{q}}\big)$ and $\Omega\big(n^{\frac{1}{q}} (\ln n)^{-\frac{1}{q}}\big)$. Then their range $P_n = o\big(n^{\frac{1}{q}} (\ln n)^{\frac{2}{5}-\frac{1}{q}}\big) = o(n)$ is  impractical in secure sensor networks modeled by $G_q(n,K_n,P_n)$ where $P_n = \Omega(n)$ holds from \cite[Equation (2)]{zhao2015resilience}. In contrast, our Theorem \ref{thm-kcon} investigates a more practical range of $P_n$ given by (\ref{scalingP-stronger}): $P_n = \Omega(n)$ for $q=1$, and $P_n = \omega\big(n^{2-\frac{1}{q}}(\ln n)^{2+\frac{1}{q}}\big)$ for $q\geq 2$.

  \textbf{($k$-)Robustness.}  Zhang and Sundaram \cite{7061412} present
a zero--one law for $k$-robustness in an Erd\H{o}s--R\'{e}nyi graph \cite{erdoskcon}, where each node pair has an edge independently with the same probability. For  $G_1(n,K_n,P_n)$ (i.e., $G_q(n,K_n,P_n)$ in the case of $q=1$), we \cite{ZhaoTAC} analyze its $k$-robustness, while this paper considers $G_q(n,K_n,P_n)$ for general $q$.

  \textbf{Hamilton cycle containment.}
 In terms of Hamilton cycle containment in $G_1(n,K_n,P_n)$ (i.e., $G_q(n,K_n,P_n)$ in the case of $q=1$), Nikoletseas \emph{et al. } \cite{NikoletseasHM} prove that
 $G_q(n,K_n,P_n)$ under $K_n \geq 2$ has a
    Hamilton cycle with a probability converging to $1$ as $n\to\infty$,
     if it holds for some constant $\delta>0$ that $n \geq (1+\delta) \binom{P_n}{K_n} \ln \binom{P_n}{K_n} $, which implies a condition  of $P_n$ not applicable
 to practical secure sensor networks modeled by $G_q(n,K_n,P_n)$; specifically, the condition implied by \cite{NikoletseasHM} is that $P_n$ is much smaller than $n$ ($P_n = O(\sqrt{n}\hspace{1.5pt})$ given $K_n \geq 2$,
    $P_n = O(\sqrt[3]{n}\hspace{1.5pt})$ if $K_n \geq 3$, $P_n = O(\sqrt[4]{n}\hspace{1.5pt})$ if $K_n \geq 4$, etc.). From \cite[Equation (2)]{zhao2015resilience},  $P_n = \Omega(n)$ holds in practical sensor network applications.  Different from the result of \cite{NikoletseasHM}, our Theorem \ref{thm-hc} \textbf{(i)} applies to general $q$ rather than only the special case of $q=1$, \textbf{(ii)} presents the asymptotically exact probability which is stronger than the zero--one law (and thus further stronger than the one-law of \cite{NikoletseasHM}), and \textbf{(iii)} considers a more practical range of $P_n$ given by (\ref{scalingP}): $P_n = \omega\big(n (\ln n)^{5}\big)$ for $q=1$, and $P_n = \omega\big(n^{2-\frac{1}{q}}(\ln n)^{2+\frac{1}{q}}\big)$ for $q\geq 2$.


  \textbf{Perfect matching containment.} For perfect matching containment in $G_1(n,K_n,P_n)$ (i.e., $G_q(n,K_n,P_n)$ under $q=1$), Blackburn \textit{et al.} \cite{herdingRKG} present a zero--one law, but their scaling is in the form of $c \frac{\ln n}{n}$ for $c<1$ or $c>1$, while our much stronger scaling is $\frac{\ln n + \xi_n}{n}$ for $\xi_n \to -\infty$ or $\infty$ as $n \to \infty$ since the case of $c=1$ not covered by \cite{herdingRKG} is addressed by us. Moreover, our result is for general $q$ while \cite{herdingRKG} is for $q=1$ only.
For perfect matching containment in $G_q(n,K_n,P_n)$, Bloznelis and {\L}uczak \cite{Perfectmatchings} give the asymptotically exact probability result,
but they tackle only the narrow range of $P_n $ satisfying both $o\big(n^{\frac{1}{q}} (\ln n)^{\frac{2}{5}-\frac{1}{q}}\big)$ and $\Omega\big(n^{\frac{1}{q}} (\ln n)^{-\frac{1}{q}}\big)$. Hence, their range is also impractical in secure sensor networks modeled by $G_q(n,K_n,P_n)$ where $P_n = \Omega(n)$ holds from \cite[Equation (2)]{zhao2015resilience}. In contrast, our Theorem \ref{thm-pm} investigates a more practical range (\ref{scalingP}) where $P_n = \Omega(n)$ is implied. 

 \section{Establishing Theorems \ref{thm-kcon}--\ref{thm-pm}} \label{sec-mainproofs}

 To establish Theorems \ref{thm-kcon}--\ref{thm-pm},   we first explain the basic ideas in Section \ref{section-basicprf} and then provide additional proof details.

 {\color{black}
 \subsection{Basic ideas for proving Theorems \ref{thm-kcon}--\ref{thm-pm}} \label{section-basicprf}

}



The basic ideas to show Theorems \ref{thm-kcon}--\ref{thm-pm}  are as follows. We decompose the theorem results into lower and upper bounds, where the lower bounds are proved by associating our studied $q$-composite random key graph with an Erd\H{o}s--R\'enyi graph, while the upper bounds are obtained by associating the studied graph property in each theorem with minimum node degree.

\subsubsection{{Decomposing the results into lower and upper bounds}} \label{sec-Decomposing}

We discuss the decomposition for Theorems \ref{thm-kcon}--\ref{thm-pm}, respectively.

\begin{itemize}
\item[\ding{202}] For {Theorem \ref{thm-kcon}}, we prove (\ref{expr-kcon-all}) by  showing that the probability $\mathbb{P}[\hspace{2pt}G_q(n,K_n,P_n)\text{ is
  $k$-connected.}\hspace{2pt}]$ has a lower bound $e^{-
\frac{e^{- \iffalse \lim\limits_ \fi \lim_{n \to \infty}{\alpha_n}}}{(k-1)!}} \times [1 -o(1)]$ and an upper bound $e^{-
\frac{e^{- \iffalse \lim\limits_ \fi \lim_{n \to \infty}{\alpha_n}}}{(k-1)!}} \times  [1 + o(1)]$ {(afterwards, the obtained (\ref{expr-kcon-all})   implies (\ref{expr-kcon-0}) (\ref{expr-kcon-1})   (\ref{expr-kcon-exact})).}
\item[\ding{203}] For {Theorem \ref{thm-krob}}, we prove (\ref{expr-krob-1}) (resp., (\ref{expr-krob-0})) by  showing that the probability $\mathbb{P}[\hspace{2pt}G_q(n,K_n,P_n)\text{ is
  $k$-robust.}\hspace{2pt}]$ has a lower bound $1-o(1)$ (resp., an upper bound $o(1)$) for  $\lim_{n \to \infty} \beta_n = \infty$ (resp., $\lim_{n \to \infty} \beta_n = -\infty$). Given the above, (\ref{expr-krob-1}) and (\ref{expr-krob-0}) immediately follow.
\item[\ding{204}] For {Theorem \ref{thm-hc}}, we prove (\ref{expr-hc-all}) by  showing that the probability $\mathbb{P}[\hspace{2pt}G_q(n,K_n,P_n)\text{ has a Hamilton cycle.}\hspace{2pt}]$ has a lower bound $e^{- e^{- \iffalse \lim\limits_ \fi \lim_{n \to \infty}{\gamma_n}}} \times [1-o(1)]$ and an upper bound $e^{- e^{- \iffalse \lim\limits_ \fi \lim_{n \to \infty}{\gamma_n}}} \times  [1+o(1)]$ (afterwards, the obtained (\ref{expr-hc-all}) implies (\ref{expr-hc-0}) (\ref{expr-hc-1})   (\ref{expr-hc-exact})).
\item[\ding{205}] For {Theorem \ref{thm-pm}}, we prove (\ref{expr-pm-all}) by  showing that the probability $\mathbb{P}[\hspace{2pt}G_q(n,K_n,P_n)\text{ has a perfect matching.}\hspace{2pt}]$ has a lower bound $ e^{- e^{- \iffalse \lim\limits_ \fi \lim_{n \to \infty}{\xi_n}}} \times [1-o(1)]$ and an upper bound $ e^{- e^{- \iffalse \lim\limits_ \fi \lim_{n \to \infty}{\xi_n}}} \times [1\hspace{-2pt}+\hspace{-2pt}o(1)]$ (afterwards, the obtained (\ref{expr-pm-all}) implies (\ref{expr-pm-0}) (\ref{expr-pm-1})   (\ref{expr-pm-exact})).
\end{itemize}

\subsubsection{{Proving the lower bounds by showing that a $q$-composite random key graph contains an Erd\H{o}s--R\'enyi graph}} \label{sec-prf-lower-bounds}

To prove the above lower bounds of Section \ref{sec-Decomposing} for our studied $q$-composite random key graph, we will show that the studied graph contains an Erd\H{o}s--R\'enyi graph as its spanning subgraph with probability $1-o(1)$, and   show that the lower bounds also hold for the Erd\H{o}s--R\'enyi graph. More specifically, the Erd\H{o}s--R\'enyi graph under the corresponding conditions is $k$-connected with probability $e^{-
\frac{e^{- \iffalse \lim\limits_ \fi \lim_{n \to \infty}{\alpha_n}}}{(k-1)!}} \times [1-o(1)]$, is $k$-robust with probability $1-o(1)$, has a Hamilton cycle with probability $e^{- e^{- \iffalse \lim\limits_ \fi \lim_{n \to \infty}{\gamma_n}}} \times [1-o(1)]$, and has a perfect matching with probability $ e^{- e^{- \iffalse \lim\limits_ \fi \lim_{n \to \infty}{\xi_n}}} \times [1-o(1)]$ ({note that the conditions for the Erd\H{o}s--R\'enyi graph are different for different properties since they are derived from (\ref{thm-kcon:pe}) (\ref{thm-krob:pe}) (\ref{thm-hc:pe}) (\ref{thm-pm:pe}) respectively}).

We provide more details for the above idea in Section \ref{sec-prf-lower-bounds-details}.

\subsubsection{{Proving the upper bounds by considering minimum node degree}} \label{sec-prf-upper-bounds}
To prove the upper bounds of  Section \ref{sec-Decomposing} for the studied graph properties, we leverage the necessary conditions on the minimum (node) degree enforced by the studied properties, and explain that the upper bounds also hold for the requirements  of the minimum degree. Specifically, we use the following results:
\begin{itemize}
\item[\ding{172}] A necessary condition for a graph to be $k$-connected is that the minimum degree is at least $k$ \cite{erdoskcon}.
\item[\ding{173}] A necessary condition for a graph to be $k$-robust is $k$-connectivity, which further requires that the minimum degree is at least $k$ \cite{7061412}.
\item[\ding{174}] A necessary condition for a graph to contain a Hamilton cycle is that the minimum degree is at least $2$ \cite{erdosHC}.
\item[\ding{175}] A necessary condition for a graph to contain a perfect matching is that the minimum degree is at least $1$ \cite{erdosPF}.
\end{itemize}
 We provide more details in Appendix \ref{sec-prf-upper-bounds-details}.

In addition to the proof ideas above, we also find it useful to confine the deviations $\alpha_n$, $\beta_n$, $\gamma_n$, $\xi_n$ in Theorems \ref{thm-kcon}--\ref{thm-pm}.

\subsubsection{\mbox{Confining $\alpha_n$, $\beta_n$, $\gamma_n$, $\xi_n$ in Theorems \ref{thm-kcon}--\ref{thm-pm}}} \label{sec-Confining}

We will show that to prove Theorems \ref{thm-kcon}--\ref{thm-pm}, the deviations $\alpha_n$, $\beta_n$, $\gamma_n$, and $\xi_n$ in the theorem statements can all be confined as $\pm  o(\ln n)$. More specifically, if Theorem \ref{thm-kcon} (resp., \ref{thm-krob}, \ref{thm-hc}, \ref{thm-pm}) holds under the extra condition $|\alpha_n |= o(\ln n)$ (resp., $|\beta_n |= o(\ln n)$, $|\gamma_n |= o(\ln n)$, $|\xi_n |= o(\ln n)$), then the result also holds regardless of the extra condition. These extra conditions will be useful for the aforementioned steps in Sections \ref{sec-prf-lower-bounds} and \ref{sec-prf-upper-bounds}. We present more details in Appendix \ref{app-confining}.

\subsection{More details for proving the lower bounds of   Section \ref{sec-Decomposing}} \label{sec-prf-lower-bounds-details}

The idea has been explained in Section \ref{sec-prf-lower-bounds}.
 Lemma \ref{cp_rig_er} relates an Erd\H{o}s--R\'enyi graph $G_{ER}(n,s_n)$ with a $q$-composite random key graph $G_q(n,K_n,P_n)$, where $G_{ER}(n,s_n)$  is defined on $n$ nodes such that each node pair has an edge \textit{independently} with   probability $s_n$.

\begin{lem}  \label{cp_rig_er}

 If $ \frac{{K_n}^2}{P_n}  =
 o\left( \frac{1}{\ln n} \right)$, $ \frac{K_n}{P_n} = o\left( \frac{1}{n\ln n} \right)$\vspace{1pt} and $K_n = \omega\big((\ln n)^3\big)$, then there exists a sequence $s_n$ satisfying
\begin{align}
\textstyle{s_n =
\frac{1}{q!} \cdot \frac{{K_n}^{2q}}{{P_n}^{q}} \cdot \left[1-
 o\left(\frac{1}{ \ln n}\right)\right] } \label{ERgraph-sn-defn}
\end{align}
 such that
a $q$-composite random key graph $G_q(n,K_n,P_n)$ is a spanning supergraph of an Erd\H{o}s--R\'enyi graph $G_{ER}(n,s_n)$ with probability $1-o(1)$.
 \end{lem}

\begin{rem}
From \cite{zz}, (\ref{ERgraph-sn-defn}) further implies that with  for any monotone increasing graph property $\mathcal{I}$,
\begin{align}
\hspace{-12pt}  \mathbb{P}[\hspace{2pt}G_q(n,K_n,P_n)
  \textrm{ has }\mathcal{I}.
\hspace{2pt}]
&  \geq
 \mathbb{P}[\hspace{2pt}G_{ER}(n,s_n)
 \textrm{ has }\mathcal{I}.\hspace{2pt} ] - o(1).  \label{cp_res_rig_er}
 \end{align}
 \end{rem}
Lemma \ref{cp_rig_er} is proved in Appendix \ref{app-prf-cp_rig_er}.

We evaluate $s_n$ given by (\ref{ERgraph-sn-defn}) under different theorems. First, as explained in Section \ref{sec-Confining}, to prove  Theorem \ref{thm-kcon} (resp., \ref{thm-krob}, \ref{thm-hc}, \ref{thm-pm}), we can introduce the extra condition $|\alpha_n |= o(\ln n)$ (resp., $|\beta_n |= o(\ln n)$, $|\gamma_n |= o(\ln n)$, $|\xi_n |= o(\ln n)$). Then we obtain:
\begin{itemize}
\item[i)] Under the condition (\ref{thm-kcon:pe}) of Theorem \ref{thm-kcon} with the extra condition $| \alpha_n | = o(\ln n)$, $s_n$ given by (\ref{ERgraph-sn-defn}) satisfies
\begin{align}
\textstyle{s_n   = \frac{\ln  n + {(k-1)} \ln \ln n + {\alpha_n}-o(1)}{n}.} \label{ER-sn-kcon}
\end{align}
\item[ii)] Under the condition (\ref{thm-krob:pe}) of Theorem \ref{thm-krob} with the extra condition $| \beta_n | = o(\ln n)$, $s_n$ given by (\ref{ERgraph-sn-defn}) satisfies
\begin{align}
\textstyle{s_n   = \frac{\ln  n + {(k-1)} \ln \ln n + {\beta_n}-o(1)}{n}.} \label{ER-sn-krob}
\end{align}
\item[iii)] Under the condition (\ref{thm-hc:pe}) of Theorem \ref{thm-hc} with the extra condition $| \gamma_n | = o(\ln n)$, $s_n$ given by (\ref{ERgraph-sn-defn}) satisfies
\begin{align}
\textstyle{s_n   = \frac{\ln  n + \ln \ln n + {\gamma_n}-o(1)}{n}. }\label{ER-sn-hc}
\end{align}
\item[iv)] Under the condition (\ref{thm-pm:pe}) of Theorem \ref{thm-pm} with the extra condition $| \xi_n | = o(\ln n)$, $s_n$ given by (\ref{ERgraph-sn-defn}) satisfies
\begin{align}
\textstyle{s_n   = \frac{\ln  n  + {\xi_n}-o(1)}{n}.} \label{ER-sn-pm}
\end{align}
\end{itemize}
Furthermore, we can show that all conditions of Lemma \ref{cp_rig_er} hold (For Theorem \ref{thm-kcon}, we replace (\ref{scalingP-stronger}) by (\ref{scalingP}) and address the additional part using \cite[Theorem 1]{ZhaoYaganGligor}). Then we apply Lemma \ref{cp_rig_er} to obtain (\ref{cp_res_rig_er}), which we now use to establish the lower bounds given in   Section \ref{sec-Decomposing}.

\textbf{Lower bound of $k$-connectivity.}
For $s_n$ satisfying (\ref{ER-sn-kcon}), we obtain from  \cite[Theorem 1]{erdoskcon} that probability of $G_{ER}(n,s_n)$ being $k$-connected
 can be written as $e^{-
\frac{e^{- \iffalse \lim\limits_ \fi \lim_{n \to \infty}{\alpha_n}}}{(k-1)!}} \times [1\pm o(1)]$. This result and (\ref{cp_res_rig_er}) (with $\mathcal{I}$ therein set as $k$-connectivity) induce that $G_q(n,K_n,P_n)$ under the conditions of Theorem \ref{thm-kcon} with $| \alpha_n | = o(\ln n)$ is $k$-connected with probability at least $e^{-
\frac{e^{- \iffalse \lim\limits_ \fi \lim_{n \to \infty}{\alpha_n}}}{(k-1)!}} \times [1- o(1)]$. This proves the lower bound in
Bullet \ding{202} of Section \ref{sec-Decomposing}.

\textbf{Lower bound of $k$-robustness.}
For $s_n$ satisfying (\ref{ER-sn-krob}), we obtain from \cite[Lemma 3]{ZhaoTAC} that probability of $G_{ER}(n,s_n)$ being $k$-robust converges to $1$ as $n\to \infty$ and hence can be written as $1-o(1)$. This result and (\ref{cp_res_rig_er}) (with $\mathcal{I}$ therein set as $k$-robustness) induce that $G_q(n,K_n,P_n)$ under the conditions of Theorem \ref{thm-krob} with $| \beta_n | = o(\ln n)$ is $k$-robust with probability at least $1-o(1)$. This proves the lower bound in
Bullet \ding{203} of Section \ref{sec-Decomposing}.

\textbf{Lower bound of Hamilton cycle containment.}
For $s_n$ satisfying (\ref{ER-sn-hc}), we obtain from \cite[Theorem 1]{erdosHC} that probability of $G_{ER}(n,s_n)$ having a Hamilton cycle
 can be written as $e^{e^{- \iffalse \lim\limits_ \fi \lim_{n \to \infty}{\gamma_n}}} \times [1\pm o(1)]$. This result and (\ref{cp_res_rig_er}) (with $\mathcal{I}$ therein set as Hamilton cycle containment) induce that $G_q(n,K_n,P_n)$ under the conditions of Theorem \ref{thm-hc} with $| \gamma_n | = o(\ln n)$ \vspace{1pt} has a Hamilton cycle with probability at least $e^{e^{- \iffalse \lim\limits_ \fi \lim_{n \to \infty}{\gamma_n}}} \times [1- o(1)]$. This proves the lower bound in
Bullet \ding{204} of Section \ref{sec-Decomposing}.

\textbf{Lower bound of perfect matching containment.}
For $s_n$ satisfying (\ref{ER-sn-pm}), we obtain from \cite[Theorem 1]{erdosPF} that probability of $G_{ER}(n,s_n)$ having a perfect matching
 can be written as $e^{e^{- \iffalse \lim\limits_ \fi \lim_{n \to \infty}{\xi_n}}} \times [1\pm o(1)]$. This result and (\ref{cp_res_rig_er}) (with $\mathcal{I}$ therein set as perfect matching containment) induce that $G_q(n,K_n,P_n)$ under the conditions of Theorem \ref{thm-pm} with $| \xi_n | = o(\ln n)$ \vspace{1pt} has a perfect matching with probability at least $e^{e^{- \iffalse \lim\limits_ \fi \lim_{n \to \infty}{\xi_n}}} \times [1- o(1)]$. This proves the lower bound in
Bullet~\ding{205} of Section \ref{sec-Decomposing}. \pfe

\appendix

 \subsection{Proof of Theorem \ref{thm-PHW}}

We recall from (\ref{KnI-minus}) (resp., (\ref{KnI-plus})) that $K_{q,n}^{-}(\mathcal{I}, \epsilon)$ (resp., $K_{q,n}^{+}(\mathcal{I}, \epsilon)$) denotes   the minimal $K_n$ such that $G_q(n,K_n,P_n)$ has property $\mathcal{I}$  with probability at least $\epsilon$ (resp., $1-\epsilon$). We prove Theorem \ref{thm-PHW} by analyzing $d_{q,n}(\mathcal{I}, \epsilon)$ defined as $K_{q,n}^{+}(\mathcal{I}, \epsilon) - K_{q,n}^{-}(\mathcal{I}, \epsilon)$ from (\ref{phase-transition-width}). To do so, we will bound $K_{q,n}^{-}(\mathcal{I}, \epsilon)$ and $K_{q,n}^{+}(\mathcal{I}, \epsilon)$ using Theorems \ref{thm-kcon}, \ref{thm-hc} and \ref{thm-pm}, where $\mathcal{I}$ is  $k$-connectivity, Hamilton cycle containment, or perfect matching containment.

We define $\kappa(\mathcal{I})$ by
\begin{align}
 \kappa(\mathcal{I})  &= \begin{cases} k,~\hspace{2.5pt}\text{if $\mathcal{I}$ is $k$-connectivity}, \\
2,~\hspace{2.5pt}\text{if $\mathcal{I}$ is Hamilton cycle containment}, \\
1,~\hspace{2.5pt}\text{if $\mathcal{I}$ is perfect matching containment},\end{cases} \label{eq-FnIkappa}
 \end{align}
 and define $F_n(\mathcal{I})$ by
  \begin{align}
 & F_n(\mathcal{I})   =  \ln n + [\kappa(\mathcal{I})-1] \ln \ln n .
 \label{eq-FnI}
 \end{align}

Recalling $K_{q,n}^{-}(\mathcal{I}, \epsilon)$  in (\ref{KnI-minus}), we  define $\xi_{q,n}^{-}(\mathcal{I}, \epsilon)$ to ensure
\begin{align}
\textstyle{\frac{1}{q!} \cdot  \frac{{[K_{q,n}^{-}(\mathcal{I}, \epsilon)]}^{2q}}{{P_n}^{q}}   = \frac{F_n(\mathcal{I})   +
 {\xi_{q,n}^{-}(\mathcal{I}, \epsilon)}}{n},}\label{KnI-minus-alpha}
\end{align}
and now use \begin{align}
 \mathbb{P}[\hspace{1pt}G_q(n, K_{q,n}^{-}(\mathcal{I}, \epsilon), P_{n})\text{ has property $\mathcal{I}$.}\hspace{1pt}]  &  \geq \epsilon \label{geqepsilon}
 \end{align} to
 prove for any positive constant $\delta_1 < \epsilon $ that
 \begin{align}
 K_{q,n}^{-}(\mathcal{I}, \epsilon) & \geq \sqrt{P_n} \times \sqrt[2q]{q![F_n(\mathcal{I}) -\ln (-[\kappa(\mathcal{I})-1]!\ln \delta_1)]/n} \nonumber \\  &  \quad\text{ for all $n$ sufficiently large,}  \label{PMalpha-lowerboundsb}
\end{align}
where (\ref{geqepsilon}) holds from the definition of $K_{q,n}^{-}(\mathcal{I}, \epsilon)$  in (\ref{KnI-minus}). Given (\ref{KnI-minus-alpha}), we will obtain (\ref{PMalpha-lowerboundsb}) once proving (\ref{PMalpha-lowerbound}) below:
\begin{align}
\xi_{q,n}^{-}(\mathcal{I}, \epsilon) & \geq  -\ln (-[\kappa(\mathcal{I})-1]!\ln \delta_1)  \text{ for all $n$ sufficiently large.} \label{PMalpha-lowerbound}
\end{align}
By contradiction, if (\ref{PMalpha-lowerbound}) is not true,
 there exists a subsequence $N_i |_{i=1,2,\ldots}$ of $\mathbb{N}$ ($\mathbb{N}$ denotes the set of
 all positive integers) such that  
  $\xi_{N_i}^{-}(\mathcal{I}, \epsilon)  <  -\ln (-[\kappa(\mathcal{I})-1]!\ln \delta_1)$ for $i=1,2,\ldots$. By \cite[Lemma 1]{Xuan}, there exists a subsequence $M_j |_{j=1,2,\ldots}$ of $N_i |_{i=1,2,\ldots}$ such that $\lim_{j \to \infty} \xi_{M_j}^{-}(\mathcal{I}, \epsilon) \in [-\infty, \infty]$. From $\xi_{N_i}  < -\ln (-[\kappa(\mathcal{I})-1]!\ln \delta_1)$ for $i=1,2,\ldots$, we have
  \begin{align}
\lim_{j \to \infty} \xi_{M_j}^{-}(\mathcal{I}, \epsilon)\in [-\infty,  -\ln (-[\kappa(\mathcal{I})-1]!\ln \delta_1)] \label{xiMjeqnexpr}.
 \end{align}
Given (\ref{eq-FnIkappa}) (\ref{eq-FnI}) and (\ref{KnI-minus-alpha}), we use Theorems \ref{thm-kcon}--\ref{thm-pm} and the subsequence principle to obtain
  \begin{align}
 & \lim_{j \to \infty}   \mathbb{P}[\hspace{1pt}G_q(M_j, K_{q,n}^{-}(\mathcal{I}, \epsilon), P_n)\text{ has property $\mathcal{I}$.}\hspace{1pt}] \nonumber \\  &  \quad = e^{- \big[e^{- \iffalse \lim\limits_ \fi \lim_{j \to \infty}{\xi_{M_j}^{-}(\mathcal{I}, \epsilon)}}\big]/[\kappa(\mathcal{I})-1]!}\nonumber \\  &  \quad\leq e^{-[e^{\ln (-[\kappa(\mathcal{I})-1]!\ln \delta_1)}]/[\kappa(\mathcal{I})-1]!}  = \delta_1 \label{PMalpha-lowerboundsbaska},
 \end{align}
where the inequality uses (\ref{xiMjeqnexpr}). Since (\ref{PMalpha-lowerboundsbaska}) contradicts (\ref{geqepsilon}) given $\delta_1 < \epsilon $, we have proved (\ref{PMalpha-lowerbound}). Then (\ref{KnI-minus-alpha}) and (\ref{PMalpha-lowerbound})   imply (\ref{PMalpha-lowerboundsb}).


Similar to the  analysis of using (\ref{geqepsilon}) to prove (\ref{PMalpha-lowerboundsb}), we   use
  \begin{align}
 \mathbb{P}[\hspace{1pt}G_q(n, K_{q,n}^{-}(\mathcal{I}) - 1, \epsilon), P_{n})\text{ has property $\mathcal{I}$.}\hspace{1pt}]  &  < \epsilon \label{geqepsilon-a}
 \end{align}
 to prove
for any positive constant $\delta_2 > \epsilon$ that
\begin{align}
K_{q,n}^{-}(\mathcal{I}, \epsilon)
&  \hspace{-1pt} \leq \hspace{-1pt} \sqrt{P_n} \hspace{-2pt} \times \hspace{-2pt} \sqrt[2q]{q![F_n(\mathcal{I})\hspace{-2pt} -\hspace{-2pt}\ln (-[\kappa(\mathcal{I})-1]!\ln \delta_2)]/n}\hspace{-1pt}+\hspace{-2pt}1 \nonumber \\  &  \quad \text{ for all $n$ sufficiently large,}  \label{PMalpha-lowerboundsb-a}
\end{align}
use
   \begin{align}
 \mathbb{P}[\hspace{1pt}G_q(n, K_{q,n}^{+}(\mathcal{I}, \epsilon), P_{n})\text{ has property $\mathcal{I}$.}\hspace{1pt}]  &  \geq 1-\epsilon \label{geqepsilon-b}
 \end{align}
 to prove
 for any positive constant $\delta_3 < 1 - \epsilon$ that
\begin{align}
K_{q,n}^{+}(\mathcal{I}, \epsilon)
&   \geq \sqrt{P_n} \times \sqrt[2q]{q![F_n(\mathcal{I}) -\ln (-[\kappa(\mathcal{I})-1]!\ln \delta_3)]/n}\nonumber \\  &  \quad\text{ for all $n$ sufficiently large,}  \label{PMalpha-lowerboundsb-b}
\end{align}
and use%
   \begin{align}
 \mathbb{P}[\hspace{1pt}G_q(n, K_{q,n}^{+}(\mathcal{I}) - 1, \epsilon), P_{n})\text{ has property $\mathcal{I}$.}\hspace{1pt}]  &  < 1-\epsilon \label{geqepsilon-c}
 \end{align}
  to prove
for any positive constant $\delta_4 > 1 - \epsilon$ that
\begin{align}
K_{q,n}^{+}(\mathcal{I}, \epsilon)
&  \hspace{-1pt} \leq\hspace{-1pt} \sqrt{P_n} \hspace{-2pt}\times \hspace{-2pt}\sqrt[2q]{q![F_n(\mathcal{I})\hspace{-2pt} -\hspace{-2pt}\ln (-[\kappa(\mathcal{I})-1]!\ln \delta_4)]/n}\hspace{-1pt}+\hspace{-2pt}1\nonumber \\  &  \quad\text{ for all $n$ sufficiently large,}  \label{PMalpha-lowerboundsb-c}
\end{align}

With the transition width $ d_{q,n}(\mathcal{I},  \epsilon)$ defined in (\ref{phase-transition-width}), we obtain from (\ref{PMalpha-lowerboundsb})  (\ref{PMalpha-lowerboundsb-a}) (\ref{PMalpha-lowerboundsb-b}) and (\ref{PMalpha-lowerboundsb-c}) that
\begin{align}
 & \hspace{-10pt}  d_{q,n}(\mathcal{I},  \epsilon)  + 1   \nonumber \\
& \hspace{-10pt}   \geq \hspace{-2pt}  \sqrt{P_n}  \sqrt[2q]{q!/n}  \times \left[\hspace{-4pt} \begin{array}{l} \sqrt[2q]{F_n(\mathcal{I}) -\ln (-[\kappa(\mathcal{I})-1]!\ln \delta_3)} \\[1pt] -   \sqrt[2q]{F_n(\mathcal{I}) -\ln (-[\kappa(\mathcal{I})-1]!\ln \delta_2)} \end{array} \hspace{-4pt} \right] \label{dnlower}
\end{align}
%
and
\begin{align}
& \hspace{-10pt}  d_{q,n}(\mathcal{I},  \epsilon)   -  1   \nonumber \\
& \hspace{-10pt}   \leq  \hspace{-2pt}  \sqrt{P_n}  \sqrt[2q]{q!/n}  \hspace{-2pt} \times \hspace{-2pt}  \left[\hspace{-4pt} \begin{array}{l} \sqrt[2q]{F_n(\mathcal{I}) -\ln (-[\kappa(\mathcal{I})-1]!\ln \delta_4)} \\[1pt] -   \sqrt[2q]{F_n(\mathcal{I}) -\ln (-[\kappa(\mathcal{I})-1]!\ln \delta_1)} \end{array} \hspace{-4pt} \right] .  \label{dnupper}
\end{align}
 It is straightforward to show that the  {right-hand side (RHS)} of (\ref{dnlower}) and the RHS of (\ref{dnupper}) can both be written as
\begin{align}
 \textstyle{{P_n}^{\frac{1}{2}} n^{-\frac{1}{2q}}  (\ln n)^{\frac{1-2q}{2q}} \times  \frac{(c_1-c_2) \cdot \sqrt[2q]{q!}}{2q} [1\pm o(1)]} ;\label{dnboundfinal}
\end{align}
in other words, from (\ref{dnlower}) and (\ref{dnupper}), we can write
\begin{align}
&\text{RHS of (\ref{dnboundfinal})} - 1 \text{ with $c_1$ and $c_2$ in case \ding{172} above} \leq d_{q,n}(\mathcal{I},  \epsilon) \nonumber \\ & \leq \text{RHS of (\ref{dnboundfinal})} + 1 \text{ with $c_1$ and $c_2$ in case \ding{173} above}.  \label{dnboundfinalsub}
\end{align}
After analyzing RHS of (\ref{dnboundfinal}) for different $P_n$, we finally use (\ref{dnboundfinalsub}) to establish Theorem \ref{thm-PHW}. \pfe

\subsection{More details for proving the upper bounds of   Sections \ref{sec-Decomposing} and \ref{sec-prf-upper-bounds}} \label{sec-prf-upper-bounds-details}

The idea has been explained in Section \ref{sec-prf-upper-bounds}.

{\color{black} Lemma \ref{lem-mnd} below gives the asymptotically exact probability for the property of minimum   degree being at least $k$ in a $q$-composite random key graph $G_q(n,K_n,P_n)$.}

\begin{lem}[\textbf{Minimum   degree in $q$-composite random key graphs}] \label{lem-mnd}
For a $q$-composite random key graph $G_q(n,K_n,P_n)$, if there is a sequence $\phi_n$ with $\lim_{n \to \infty}{\phi_n} \in [-\infty, \infty]$
such that
\begin{align}
\textstyle{\frac{1}{q!} \cdot \frac{{K_n}^{2q}}{{P_n}^{q}}   = \frac{\ln  n + {(k-1)} \ln \ln n + {\phi_n}}{n},} \label{lem-mnd:pe}
\end{align}
then it holds under (\ref{scalingP}) that
 \begin{align}
& \lim_{n \to \infty}  \mathbb{P}[\hspace{2pt}G_q(n,K_n,P_n)\text{ has a minimum   degree at least $k$.}\hspace{2pt}] \nonumber \\ &=  e^{-
\frac{e^{- \iffalse \lim\limits_ \fi \lim_{n \to \infty}{\phi_n}}}{(k-1)!}} \label{expr-mnd-all}
 \end{align}
 \begin{subnumcases}{\hspace{-35pt}=} 0,&\text{\hspace{-4pt}if  }$ \lim_{n \to \infty}{\phi_n} =- \infty$, \label{expr-mnd-0} \\
1,&\text{\hspace{-4pt}if  }$ \lim_{n \to \infty}{\phi_n} = \infty$, \label{expr-mnd-1} \\ e^{-
\frac{e^{-\phi^{*}}}{(k-1)!}}  ,&\text{\hspace{-4pt}if  }$ \lim_{n \to \infty}{\phi_n} = \phi^{*}\in (-\infty, \infty)$.\label{expr-mnd-exact}  \end{subnumcases}
\end{lem}

We defer the proof of Lemma \ref{lem-mnd} to Appendix \ref{app-mnd-thm}. With $\kappa(\mathcal{I})$ defined by
\begin{align}
 \kappa(\mathcal{I})  &= \begin{cases} k,~\hspace{2.5pt}\text{if $\mathcal{I}$ is $k$-connectivity or $k$-robustness}, \\
2,~\hspace{2.5pt}\text{if $\mathcal{I}$ is Hamilton cycle containment}, \\
1,~\hspace{2.5pt}\text{if $\mathcal{I}$ is perfect matching containment},\end{cases} \label{eq-FnIkappa}
 \end{align}
 from results \ding{172}--\ding{175} in Section \ref{sec-prf-upper-bounds}, we have
 \begin{align}
& \mathbb{P}[\hspace{2pt}G_q(n,K_n,P_n)\text{ has $\mathcal{I}$.}\hspace{2pt}] \nonumber \\ & \leq   \mathbb{P}[\hspace{2pt}G_q(n,K_n,P_n)\text{ has a minimum   degree at least $\kappa(\mathcal{I})$.}\hspace{2pt}]
   \label{all-vs-mnd}
 \end{align}
 More specifically, we can write (\ref{all-vs-mnd}) as the following (\ref{kcon-vs-mnd})--(\ref{pm-vs-mnd}): \begin{align}
& \mathbb{P}[\hspace{2pt}G_q(n,K_n,P_n)\text{ is
  $k$-connected.}\hspace{2pt}] \nonumber \\ & \leq   \mathbb{P}[\hspace{2pt}G_q(n,K_n,P_n)\text{ has a minimum   degree at least $k$.}\hspace{2pt}], \label{kcon-vs-mnd} \\ & \mathbb{P}[\hspace{2pt}G_q(n,K_n,P_n)\text{ is
  $k$-robust.}\hspace{2pt}] \nonumber \\ & \leq    \mathbb{P}[\hspace{2pt}G_q(n,K_n,P_n)\text{ has a minimum   degree at least $k$.}\hspace{2pt}], \label{krob-vs-mnd} \\ &  \mathbb{P}[\hspace{2pt}G_q(n,K_n,P_n)\text{ has a Hamilton cycle.}\hspace{2pt}] \nonumber \\ & \leq   \mathbb{P}[\hspace{2pt}G_q(n,K_n,P_n)\text{ has a minimum   degree at least $2$.}\hspace{2pt}], \label{hc-vs-mnd} \\ &\mathbb{P}[\hspace{2pt}G_q(n,K_n,P_n)\text{ has a perfect matching.}\hspace{2pt}]  \nonumber \\ & \leq   \mathbb{P}[\hspace{2pt}G_q(n,K_n,P_n)\text{ has a minimum   degree at least $1$.}\hspace{2pt}], \label{pm-vs-mnd}
 \end{align}
  Then clearly (\ref{kcon-vs-mnd})--(\ref{pm-vs-mnd}) and Lemma \ref{lem-mnd} together prove the upper bounds in
Bullet \ding{202}--\ding{205} of Section \ref{sec-Decomposing}. More precisely, we have:
\begin{itemize}
\item (\ref{kcon-vs-mnd}) along with (\ref{expr-mnd-all}) of Lemma \ref{lem-mnd} proves the upper bound in
Bullet \ding{202} of Section \ref{sec-Decomposing}.
\item (\ref{krob-vs-mnd}) along with (\ref{expr-mnd-0}) of Lemma \ref{lem-mnd} proves the upper bound in
Bullet \ding{203} of Section \ref{sec-Decomposing}.
\item (\ref{hc-vs-mnd}) along with (\ref{expr-mnd-all}) of Lemma \ref{lem-mnd} proves the upper bound in
Bullet \ding{204} of Section \ref{sec-Decomposing}.
\item (\ref{pm-vs-mnd}) along with (\ref{expr-mnd-all}) of Lemma \ref{lem-mnd} proves the upper bound in
Bullet \ding{205} of Section \ref{sec-Decomposing}.\pfe
\end{itemize}

\subsection{Confining $|\alpha_n |$, $|\beta_n |$, $|\gamma_n |$, $|\xi_n |$ as $o(\ln n)$ in Theorems \ref{thm-kcon}, \ref{thm-krob}, \ref{thm-hc}, \ref{thm-pm}} \label{app-confining}

We will show that to prove Theorems \ref{thm-kcon}--\ref{thm-pm}, the deviations $\alpha_n$, $\beta_n$, $\gamma_n$, and $\xi_n$ in the theorem statements can all be confined as $\pm  o(\ln n)$. More specifically, if Theorem \ref{thm-kcon} (resp., \ref{thm-krob}, \ref{thm-hc}, \ref{thm-pm}) holds under the extra condition $|\alpha_n |= o(\ln n)$ (resp., $|\beta_n |= o(\ln n)$, $|\gamma_n |= o(\ln n)$, $|\xi_n |= o(\ln n)$), then the result also holds regardless of the extra condition.

\textbf{Notation for coupling between random graphs:}

We will couple different random graphs together. The idea is  converting a problem of one random graph to the corresponding problem in another random graph, in order to solve the original problem.
Formally, a coupling
\cite{zz,2013arXiv1301.0466R,Krzywdzi} of two random graphs
$G_1$ and $G_2$ means a probability space on which random graphs
$G_1'$ and $G_2'$ are defined such that $G_1'$ and $G_2'$ have the
same distributions as $G_1$ and $G_2$, respectively. If $G_1'$ is a spanning subgraph
(resp., supergraph)\footnote{A graph
$G_a$ is a spanning subgraph (resp., spanning supergraph) of a graph
$G_b$ if $G_a$ and $G_b$ have the same node set, and the edge set of
$G_a$ is a subset (resp., superset) of the edge set of $G_b$.} of $G_2'$, we say that under the coupling, $G_1$ is a spanning subgraph
(resp., supergraph) of $G_2$, which yields that for any monotone increasing property $\mathcal {I}$, the probability of $G_1$ having $\mathcal {I}$ is at most (resp., at least) the probability of $G_2$ having $\mathcal {I}$.

 Following Rybarczyk's notation \cite{zz}, we
write
\begin{align}
G_1 \succeq & G_2 \quad (\textrm{resp.}, G_1 \succeq_{1-o(1)} G_2)
\label{g1g2coupling}
\end{align}
if there exists a coupling under which $G_2$ is a spanning subgraph
of $G_1$ with probability $1$ (resp., $1-o(1)$).

Note that $k$-connectivity, $k$-robustness, Hamilton cycle containment, or perfect matching containment are all monotone increasing\footnote{
A graph
property is called monotone increasing if it holds under the
addition of edges \cite{JansonLuczakRucinski7}.}.
For any monotone increasing property $\mathcal {I}$, the probability that a spanning subgraph
(resp., supergraph) of graph $G$ has $\mathcal {I}$ is at most (resp., at least) the probability of $G$ having $\mathcal {I}$. Therefore, to show \begin{align}
& \textrm{Theorem \ref{thm-kcon} under }|\alpha_n | = o(\ln n)  \\ & \nonumber \Longrightarrow   \textrm{Theorem \ref{thm-kcon} regardless of }|\alpha_n | = o(\ln n),  \nonumber\\ & \textrm{Theorem \ref{thm-krob} under }|\beta_n | = o(\ln n)  \\ & \nonumber \Longrightarrow   \textrm{Theorem \ref{thm-krob} regardless of }|\beta_n | = o(\ln n),  \nonumber\\& \textrm{Theorem \ref{thm-hc} under }|\gamma_n | = o(\ln n)  \\ & \nonumber \Longrightarrow   \textrm{Theorem \ref{thm-hc} regardless of }|\gamma_n | = o(\ln n),  \nonumber\\& \textrm{Theorem \ref{thm-pm} under }|\xi_n | = o(\ln n)  \\ & \nonumber \Longrightarrow   \textrm{Theorem \ref{thm-pm} regardless of }|\xi_n | = o(\ln n).
\end{align} it suffices to prove the following lemma.

\begin{lem} \label{graph_Gs_cpl}

\textbf{(a)} For graph $G_q(n,K_n,P_n)$ under \begin{align}
P_n  &= \begin{cases} \, \Omega(n), &\text{for } q=1,\\
\, \omega\big(n^{2-\frac{1}{q}}(\ln n)^{2+\frac{1}{q}}\big), &\text{for } q \geq 2.
\end{cases}  \label{scalingP-strongercopyv}
\end{align} and
\begin{align}
\frac{1}{q!} \cdot \frac{{K_n}^{2q}}{{P_n}^{q}}  &  = \frac{\ln  n + {(k-1)} \ln \ln n + {\nu_n}}{n} \label{al1-parta}
\end{align}
with $\lim_{n \to \infty}\nu_n = -\infty$, there exists graph $G_q(n,\widetilde{K_n},\widetilde{P_n})$ under
\begin{align}
\widetilde{P_n}   &= \begin{cases} \, \Omega(n), &\text{for } q=1,\\
\, \omega\big(n^{2-\frac{1}{q}}(\ln n)^{2+\frac{1}{q}}\big), &\text{for } q \geq 2.
\end{cases}  \label{scalingP-strongercopyv-widetilde}
\end{align}
and
\begin{align}
\frac{1}{q!} \cdot \frac{{\widetilde{K_n}}^{2q}}{{\widetilde{P_n}}^{q}}  &  = \frac{\ln  n + {(k-1)} \ln \ln n + {\widetilde{\nu_n}}}{n} \label{al0-parta}
\end{align}
with $\lim_{n \to \infty}\widetilde{\nu_n} = -\infty$ and $\widetilde{\nu_n} = -o(\ln n)$,
such that there exists a graph coupling under which
$G_q(n,K_n,P_n)$ is a spanning subgraph of $G_q(n,\widetilde{K_n},\widetilde{P_n})$.

\textbf{(b)} For graph $G_q(n,K_n,P_n)$ under (\ref{scalingP-strongercopyv-widetilde}) and
\begin{align}
\frac{1}{q!} \cdot \frac{{K_n}^{2q}}{{P_n}^{q}}  &  = \frac{\ln  n + {(k-1)} \ln \ln n + {\nu_n}}{n} \label{al1}
\end{align}
with $\lim_{n \to \infty}\nu_n = \infty$, there exists graph $G_q(n,\widehat{K_n},\widehat{P_n})$ under \begin{align}
\widehat{P_n}   &= \begin{cases} \, \Omega(n), &\text{for } q=1,\\
\, \omega\big(n^{2-\frac{1}{q}}(\ln n)^{2+\frac{1}{q}}\big), &\text{for } q \geq 2.
\end{cases}  \label{scalingP-strongercopyv-widehat}
\end{align} and
\begin{align}
\frac{1}{q!} \cdot \frac{{\widehat{K_n}}^{2q}}{{\widehat{P_n}}^{q}}  &  = \frac{\ln  n + {(k-1)} \ln \ln n + {\widehat{\nu_n}}}{n} \label{al0}
\end{align}
with $\lim_{n \to \infty}\widehat{\nu_n} = \infty$ and $\widehat{\nu_n} = o(\ln n)$,
such that there exists a graph coupling under which
$G_q(n,K_n,P_n)$ is a spanning supergraph of $G_q(n,\widehat{K_n},\widehat{P_n})$.

\end{lem}

\subsection{Proof of Lemma \ref{graph_Gs_cpl}} \label{sec_graph_Gs_cpl}

\paragraph{Proving property (a).}

 We
 define $\widetilde{\nu_n}^*$ by
 \begin{align}
\widetilde{\nu_n}^* &  = \max\{\nu_n, -\ln \ln n\}, \label{al2-parta}
\end{align}
and
define $\widetilde{K_n}^*$ such that
\begin{align}
\frac{1}{q!} \cdot \frac{({\widetilde{K_n}^{*}})^{2q}}{{{P_n}}^{q}}  &  = \frac{\ln  n + {(k-1)} \ln \ln n + \widetilde{\nu_n}^*}{n}. \label{al3-parta}
\end{align}
We set
\begin{align}
\widetilde{K_n} & : = \big\lfloor \widetilde{K_n}^* \big\rfloor,  \label{al4-parta}
\end{align}
and
\begin{align}
\widetilde{P_n} & : = P_n.  \label{al5-parta}
\end{align}

From (\ref{al1}) (\ref{al2-parta}) and (\ref{al3-parta}), it holds that
\begin{align}
K_n \leq \widetilde{K_n}^*.  \label{Kn1-parta}
\end{align}
Then by (\ref{al4-parta}) (\ref{Kn1-parta}) and the fact that $K_n$ and $\widetilde{K_n}$ are both integers, it follows that
\begin{align}
K_n \leq \widetilde{K_n}.  \label{al6-parta}
\end{align}
 From (\ref{al5-parta}) and (\ref{al6-parta}), by  \cite[Lemma 3]{Rybarczyk}, there exists a graph coupling under which $G_q(n,K_n,P_n)$ is a spanning subgraph of $G_q(n,\widetilde{K_n},\widetilde{P_n})$. Therefore, the proof of property (a) is completed once we show
$\widetilde{\nu_n}$ defined in $(\ref{al0-parta})$ satisfies
\begin{align}
  \lim_{n \to \infty}\widetilde{\nu_n} & = - \infty, \label{al8-parta} \\
 \widetilde{\nu_n} & = - o(\ln n).  \label{al7-parta}
\end{align}

We first prove (\ref{al8-parta}). From (\ref{al0-parta}) (\ref{al3-parta}) and (\ref{al4-parta}), it holds that
\begin{align}
\widetilde{\nu_n} \leq \widetilde{\nu_n}^*, \label{haa-parta}
\end{align}
which together with (\ref{al2-parta}) and $\lim_{n \to \infty}\nu_n = -\infty$ yields (\ref{al8-parta}).

Now we establish (\ref{al7-parta}). From (\ref{al4-parta}), we have $\widetilde{K_n} > \widetilde{K_n}^* - 1$. Then from (\ref{al0-parta}) and (\ref{al5-parta}), it holds that
\begin{align}
 \widetilde{\nu_n} & = n \cdot \frac{1}{q!} \cdot \frac{{\widetilde{K_n}}^{2q}}{{{P_n}}^{q}}  - [\ln  n + {(k-1)} \ln \ln n] \nonumber \\
 & >  n \cdot \frac{1}{q!} \cdot \frac{{(\widetilde{K_n}^* - 1)}^{2q}}{{{P_n}}^{q}}  - [\ln  n + {(k-1)} \ln \ln n]  . \label{aph1-parta}
\end{align}
By $\lim_{n \to \infty}\nu_n =- \infty$, it holds that $\nu_n \leq 0$ for all $n$ sufficiently large. Then from (\ref{al2-parta}), it follows that
\begin{align}
 \widetilde{\nu_n}^* = - o(\ln n),  \label{widetilde-al2-parta}
\end{align}
which along with Lemma \ref{lem:logn2}, equation (\ref{al3-parta}) and condition $P_n = \Omega(n)$ induces
\begin{align}
 \widetilde{K_n}^* & = \Omega\big( (\ln n)^{\frac{1}{2q}} \big). \label{aph5-parta}
\end{align}
Hence, we have $\lim_{n \to \infty} \widetilde{K_n}^* = \infty$ and it further holds for all $n$ sufficient large that
\begin{align}
{(\widetilde{K_n}^* - 1)}^{2q} > ({\widetilde{K_n}^{*}})^{2q} - 3s  ({\widetilde{K_n}^{*}})^{2q-1}. \label{aph2-parta}
\end{align}
Applying (\ref{aph2-parta}) to (\ref{aph1-parta}) and then using (\ref{al3-parta}), Lemma \ref{lem:logn2} and $P_n = \Omega(n)$, it follows that
\begin{align}
 \widetilde{\nu_n} & > n \cdot \frac{1}{q!} \cdot \frac{ ({\widetilde{K_n}^{*}})^{2q} - 3s  ({\widetilde{K_n}^{*}})^{2q-1}}{{{P_n}}^{q}}  - [\ln  n + {(k-1)} \ln \ln n]  \nonumber \\
 &    = \widetilde{\nu_n}^* - \frac{3s}{q!} \cdot n \cdot \Theta\big({P_n}^{-\frac{1}{2}} n^{-\frac{2q-1}{2q}} (\ln n)^{\frac{2q-1}{2q}} \big)   \nonumber \\
 &    = \widetilde{\nu_n}^* - O\big(n^{-\frac{1}{2}+\frac{1}{2q}} (\ln n)^{1-\frac{1}{2q}}\big). \label{widetilde-al-parta}
\end{align}

We only need to consider $q \geq 2$ here since the case of $q = 1$ is already proved by us as Lemma 5 of \cite{ZhaoTAC}. Using $q \geq 2$ in (\ref{widetilde-al-parta}), it holds that $ \widetilde{\nu_n} > \widetilde{\nu_n}^* + o(1)$, which along with (\ref{haa-parta}) and (\ref{widetilde-al2-parta}) yields (\ref{al7-parta}).

\paragraph{Proving property (b).}

 We
 define $\widehat{\nu_n}^*$ by
 \begin{align}
\widehat{\nu_n}^* &  = \min\{\nu_n, \ln \ln n\}, \label{al2}
\end{align}
and
define $\widehat{K_n}^*$ such that
\begin{align}
\frac{1}{q!} \cdot \frac{({\widehat{K_n}^{*}})^{2q}}{{{P_n}}^{q}}  &  = \frac{\ln  n + {(k-1)} \ln \ln n + \widehat{\nu_n}^*}{n}. \label{al3}
\end{align}
We set
\begin{align}
\widehat{K_n} & : = \big\lceil \widehat{K_n}^* \big\rceil,  \label{al4}
\end{align}
and
\begin{align}
\widehat{P_n} & : = P_n.  \label{al5}
\end{align}

From (\ref{al1}) (\ref{al2}) and (\ref{al3}), it holds that
\begin{align}
K_n \geq \widehat{K_n}^*.  \label{Kn1}
\end{align}
Then by (\ref{al4}) (\ref{Kn1}) and the fact that $K_n$ and $\widehat{K_n}$ are both integers, it follows that
\begin{align}
K_n \geq \widehat{K_n}.  \label{al6}
\end{align}
 From (\ref{al5}) and (\ref{al6}), by  \cite[Lemma 3]{Rybarczyk}, there exists a graph coupling under which $G_q(n,K_n,P_n)$ is a spanning supergraph of $G_q(n,\widehat{K_n},\widehat{P_n})$. Therefore, the proof of property (b) is completed once we show
$\widehat{\nu_n}$ defined in $(\ref{al0})$ satisfies
\begin{align}
  \lim_{n \to \infty}\widehat{\nu_n} & = \infty, \label{al8} \\
 \widehat{\nu_n} & = o(\ln n).  \label{al7}
\end{align}

We first prove (\ref{al8}). From (\ref{al0}) (\ref{al3}) and (\ref{al4}), it holds that
\begin{align}
\widehat{\nu_n} \geq \widehat{\nu_n}^*, \label{haa}
\end{align}
which together with (\ref{al2}) and $\lim_{n \to \infty}\nu_n = \infty$ yields (\ref{al8}).

Now we establish (\ref{al7}). From (\ref{al4}), we have $\widehat{K_n} < \widehat{K_n}^* + 1$. Then from (\ref{al0}) and (\ref{al5}), it holds that
\begin{align}
 \widehat{\nu_n} & = n \cdot \frac{1}{q!} \cdot \frac{{\widehat{K_n}}^{2q}}{{{P_n}}^{q}}  - [\ln  n + {(k-1)} \ln \ln n] \nonumber \\
 & <  n \cdot \frac{1}{q!} \cdot \frac{{(\widehat{K_n}^* + 1)}^{2q}}{{{P_n}}^{q}}  - [\ln  n + {(k-1)} \ln \ln n]  . \label{aph1}
\end{align}
By $\lim_{n \to \infty}\nu_n = \infty$, it holds that $\nu_n \geq 0$ for all $n$ sufficiently large. Then from (\ref{al2}), it follows that
\begin{align}
 \widehat{\nu_n}^* = o(\ln n),  \label{widehat-al2}
\end{align}
which along with Lemma \ref{lem:logn2}, equation (\ref{al3}) and condition $P_n = \Omega(n)$ induces
\begin{align}
 \widehat{K_n}^* & = \Omega\big( (\ln n)^{\frac{1}{2q}} \big). \label{aph5}
\end{align}
Hence, we have $\lim_{n \to \infty} \widehat{K_n}^* = \infty$ and it further holds for all $n$ sufficient large that
\begin{align}
{(\widehat{K_n}^* + 1)}^{2q}< ({\widehat{K_n}^{*}})^{2q} + 3s  ({\widehat{K_n}^{*}})^{2q-1}. \label{aph2}
\end{align}
Applying (\ref{aph2}) to (\ref{aph1}) and then using (\ref{al3}), Lemma \ref{lem:logn2} and $P_n = \Omega(n)$, it follows that
\begin{align}
 \widehat{\nu_n} & <  n \cdot \frac{1}{q!} \cdot \frac{ ({\widehat{K_n}^{*}})^{2q} + 3s  ({\widehat{K_n}^{*}})^{2q-1}}{{{P_n}}^{q}}  - [\ln  n + {(k-1)} \ln \ln n]  \nonumber \\
 &    = \widehat{\nu_n}^* + \frac{3s}{q!} \cdot n \cdot \Theta\big({P_n}^{-\frac{1}{2}} n^{-\frac{2q-1}{2q}} (\ln n)^{\frac{2q-1}{2q}} \big)   \nonumber \\
 &    = \widehat{\nu_n}^* + O\big(n^{-\frac{1}{2}+\frac{1}{2q}} (\ln n)^{1-\frac{1}{2q}}\big). \label{widehat-al}
\end{align}

We only need to consider $q \geq 2$ here since the case of $q = 1$ is already proved by us as Lemma 5 of \cite{ZhaoTAC}. Using $q \geq 2$ in (\ref{widehat-al}), it holds that $ \widehat{\nu_n} < \widehat{\nu_n}^* + o(1)$, which along with (\ref{haa}) and (\ref{widehat-al2}) yields (\ref{al7}).

\begin{lem}  \label{lem:logn2}
If
$\frac{1}{q!} \cdot \frac{{K_n}^{2q}}{{P_n}^{q}}    = \frac{\ln  n \pm o( \ln n) }{n}$ and $P_n = \Omega(n^c)$ for constant $c$, then $K_n = \Omega\big( n^{\frac{c}{2} - \frac{1}{2q}} (\ln n)^{\frac{1}{2q}} \big)$.
\end{lem}
\noindent \textbf{Proof of Lemma \ref{lem:logn2}:}

From condition
\begin{align}
\frac{1}{q!} \cdot \frac{{K_n}^{2q}}{{P_n}^{q}}   = \frac{\ln  n \pm o( \ln n) }{n} \sim \frac{\ln  n}{n}, \label{eqslnn-Kpsn}
\end{align}
it holds that
\begin{align}
\frac{{K_n}^2}{P_n}\  & = \Theta\big( n^{-\frac{1}{q}} (\ln n)^{\frac{1}{q}} \big) , \label{tp-KnPn}
\end{align}
which along with condition $P_n = \Omega(n^c)$ yields
$K_n   = \sqrt{P_n \cdot \Theta\big( n^{-\frac{1}{q}} (\ln n)^{\frac{1}{q}} \big)}
= \Omega\Big( n^{\frac{c}{2} - \frac{1}{2q}} (\ln n)^{\frac{1}{2q}} \Big) . $ \pfe

\subsection{Proof of Lemma \ref{cp_rig_er}} \label{app-prf-cp_rig_er}

\textbf{Lemma \ref{cp_rig_er} (Restated).}
{\em  If $ \frac{{K_n}^2}{P_n}  =
 o\left( \frac{1}{\ln n} \right)$, $ \frac{K_n}{P_n} = o\left( \frac{1}{n\ln n} \right)$\vspace{1pt} and $K_n = \omega\big((\ln n)^3\big)$, then there exists a sequence $s_n$ satisfying
\begin{align}
\textstyle{s_n =
\frac{1}{q!} \cdot \frac{{K_n}^{2q}}{{P_n}^{q}} \cdot \left[1-
 o\left(\frac{1}{ \ln n}\right)\right] } \label{ERgraph-sn-defn-restated}
\end{align}
 such that
a $q$-composite random key graph $G_q(n,K_n,P_n)$ is a spanning supergraph of an Erd\H{o}s--R\'enyi graph $G_{ER}(n,s_n)$ with probability $1-o(1)$.}

We have explained the notation for coupling between random graphs in Appendix \ref{app-confining}.
Then the conclusion in Lemma \ref{cp_rig_er} means
$$G_q(n,K_n,P_n)   \succeq_{1-o(1)}G_{ER}(n,s_n) .$$

\textbf{Proof of Lemma \ref{cp_rig_er}:}

 To prove Lemma \ref{cp_rig_er}, we
introduce an auxiliary graph called a \emph{binomial
$q$-intersection
 graph} $H_q(n,x_n,P_n)$ \cite{Rybarczyk,bloznelis2013}, which
 can be defined on $n$ nodes by the following process.
  There exists a key pool of size $P_n$. Each key in the
pool is added to each node \emph{independently} with probability
$x_n$. After each node obtains a set of keys, two nodes establish an edge in between if and only if they share at least $q$ keys. Clearly, the only difference between a binomial $q$-intersection
 graph $H_q(n,x_n,P_n)$ and a $q$-composite random key graph $G_q(n,K_n,P_n)$ for the $q$-composite key predistribution scheme is that in the former,
  the number of keys assigned to each
 node obeys a binomial distribution with $P_n$ as
the number of trials, and with $x_n$ as the success probability in
each trial, while in the latter graph, such number equals $K_n$ with
probability $1$.

In Appendix \ref{app-prf-lem-cpgraph-rigrig} below, we prove Lemma \ref{cp_rig_er} by using
Lemmas \ref{brig_urig} and \ref{er_brig} below.




\begin{lem} \label{brig_urig}
If $K_n = \omega(\ln n)$ and $ \frac{{K_n}^2}{P_n}  =
 o\left( 1\right)$, with
$x_n$ set by
\begin{align}
 x_n   = \textstyle{\frac{K_n}{P_n}
 \Big(1 - \sqrt{\frac{3\ln
n}{K_n }}\hspace{2pt}\Big)}, \label{pnKn}
 \end{align}
then 
 it holds that
\begin{align}
  G_q(n,K_n,P_n) & \succeq_{1-o(1)}H_q(n,x_n,P_n). \label{eq_brig_urig}
\end{align}

\end{lem}

We establish Lemma \ref{brig_urig} in
Appendix \ref{app-prf-brig_urig}.

\begin{lem} \label{er_brig}
If
 \begin{align}
\textstyle{ x_n  P_n } & = \omega\big((\ln n)^3\big), \label{er_brig-eq1} \\  \textstyle{ {x_n} } & = \textstyle{o\left( \frac{1}{n\ln n} \right)}, \label{er_brig-eq2} \\ \textstyle{{x_n}^2 P_n} &   =\textstyle{o\left( \frac{1}{\ln n} \right)}, \text{ and} \label{er_brig-eq3} \\ \textstyle{{x_n}^2 P_n} & = \textstyle{ \omega\big(\frac{(\ln n)^6}{n^2}\big)}, \label{er_brig-eq4}
 \end{align}
 then
there exits some $s_n$ satisfying
\begin{align}
s_n & = \textstyle{\frac{(P_n{x_n}^2)^q}{q!}} \cdot \big[1- o\left( \frac{1}{\ln n} \right)\big]
\label{pnpb01}
\end{align}
such that Erd\H{o}s--R\'{e}nyi graph $G_{ER}(n,s_n)$
\cite{citeulike:4012374} obeys
\begin{align}
 H_q(n,x_n,P_n)& \succeq_{1-o(1)} G_{ER}(n,s_n) . \label{GerGb}
\end{align}

\end{lem}

We prove Lemma \ref{er_brig} in
Appendix \ref{app-prf-er_brig}.

\subsubsection{\textbf{Proof of Lemma \ref{cp_rig_er} Using
Lemmas \ref{brig_urig} and \ref{er_brig}}} \label{app-prf-lem-cpgraph-rigrig}~

We complete the proof of Lemma \ref{cp_rig_er} by using
Lemmas \ref{brig_urig} and \ref{er_brig}.
 We first explain that given the conditions of Lemma \ref{cp_rig_er}:
 \begin{align}
 \textstyle{\frac{{K_n}^2}{P_n} } & = \textstyle{ o\left( \frac{1}{\ln n} \right)}, \label{pnKn2-toneq1} \\  \textstyle{\frac{K_n}{P_n}} & = \textstyle{o\left( \frac{1}{n\ln n} \right)}, \label{pnKn2-toneq2} \\ \textstyle{ K_n } & = \omega\big((\ln n)^3\big), \label{pnKn2-toneq3} \\ \textstyle{\frac{{K_n}^2}{P_n}} & = \textstyle{\omega\big(\frac{(\ln n)^6}{n^2}\big)}, \label{pnKn2-toneq4}
 \end{align}
 all
conditions in Lemmas \ref{brig_urig} and \ref{er_brig} are true;
i.e.,
 \begin{align}
\textstyle{ K_n } & = \omega(\ln n), \label{pnKn2-toneqa1} \\  \textstyle{ \frac{{K_n}^2}{P_n} } & = \textstyle{o\left( 1 \right)}, \label{pnKn2-toneqa2} \\ \textstyle{{x_n}} &   =\textstyle{o\left( \frac{1}{n\ln n} \right)}, \label{pnKn2-toneqa3} \\ \textstyle{{x_n}^2 P_n} & = \textstyle{ o\left( \frac{1}{\ln n} \right)}, \text{ and} \label{pnKn2-toneqa4}  \\ \textstyle{ {x_n}^2 P_n} & = \textstyle{\omega\big(\frac{(\ln n)^6}{n^2}\big)}, \label{pnKn2-toneqa5}
 \end{align}
all hold, where $x_n$ is defined in (\ref{pnKn}).

Clearly, (\ref{pnKn2-toneq3}) implies (\ref{pnKn2-toneqa1}). Also, (\ref{pnKn2-toneqa2}) implies (\ref{pnKn2-toneq1}).
Using (\ref{pnKn2-toneq3}) in (\ref{pnKn}), \f
\begin{align}
 x_n & = \textstyle{\frac{K_n}{P_n}  \cdot \Big[1 - o\Big(\sqrt{\frac{3\ln
n}{(\ln n)^3}}\Big)\hspace{1pt}\Big] } \label{pnKn2-tonabc} \\  & = \textstyle{\frac{K_n}{P_n}  \cdot \big[1 - \textstyle{ o\left( \frac{1}{\ln n} \right)}\big]} \label{pnKn2}.
 \end{align}
Then we obtain the following. First, (\ref{pnKn2}) and (\ref{pnKn2-toneq2}) together yield (\ref{pnKn2-toneqa3}).
Second, (\ref{pnKn2}) and (\ref{pnKn2-toneq1}) induce
(\ref{pnKn2-toneqa4}). Third,
(\ref{pnKn2}) and (\ref{pnKn2-toneq4}) lead
to (\ref{pnKn2-toneqa5}). Therefore, all conditions in Lemmas \ref{brig_urig}
and \ref{er_brig} hold.

We use $s_n$ defined in (\ref{pnpb01}). By \cite[Fact
3]{2013arXiv1301.0466R} on the transitivity of graph coupling, we
use (\ref{eq_brig_urig}) in Lemma \ref{brig_urig} and (\ref{GerGb})
in Lemma \ref{er_brig} to obtain
\begin{align}
G_q(n,K_n,P_n)   \succeq_{1-o(1)}G_{ER}(n,s_n) . \label{GerGurig}
\end{align}
From (\ref{pnKn2-tonabc}) and (\ref{pnpb01}), we derive
\begin{align}s_n  =
\textstyle{\frac{1}{q!}  \cdot \frac{{K_n}^{2q}}{{P_n} ^{q}}} \cdot
 \big[1 - \textstyle{ o\left( \frac{1}{\ln n} \right)}\big]^{2q} =
\textstyle{\frac{1}{q!}  \cdot \frac{{K_n}^{2q}}{{P_n} ^{q}}} \cdot
\big[1- o\left( \frac{1}{\ln n} \right)\big], \label{GerGurigsba}
\end{align}
where the last step uses the fact that for a sequence $a_n = o\big(\frac{1}{\ln n}\big)$, we have $(1-a_n)^{2q} = 1-o\big(\frac{1}{\ln n}\big)$. To see this, given $a_n = o\big(\frac{1}{\ln n}\big)$ and thus $0\leq a_n <1$ for all $n$ sufficiently large, we use \cite[Fact 2]{ZhaoYaganGligor} to obtain $1- a_n \cdot 2q \leq (1-a_n)^{2q} \leq 1-a_n \cdot 2q + \frac{1}{2} \cdot {a_n}^2 \cdot ({2q })^2$.

To summarize, the proof of Lemma \ref{cp_rig_er} is
completed. \pfe

\subsubsection{\textbf{Proof of Lemma \ref{brig_urig}}}\label{app-prf-brig_urig}~

By \cite[Lemma 4]{Rybarczyk}, if $x_n P_n = \omega\left( \ln n
\right)$, and for all $n$ sufficiently large,
\begin{align}
K_{n}  & \geq x_n P_n + \sqrt{3(x_n P_n + \ln n) \ln n}  ,
\label{Knbig}
\end{align}
then
\begin{align}
 G_q(n,K_n,P_n) & \succeq_{1-o(1)}H_q(n,x_n,P_n) .
\end{align}

Therefore, the proof of Lemma \ref{brig_urig} is completed once we
show $x_n P_n = \omega\left( \ln n \right)$ and (\ref{Knbig}) with
$x_n$ defined in (\ref{pnKn}). From conditions\vspace{1pt} $K_n =
\omega\left( \ln n \right)$ and $x_n = \frac{K_n}{P_n}
 \left(1 - \sqrt{\frac{3\ln
n}{K_n }}\hspace{2pt}\right)$, we first obtain $x_n P_n  =
\omega\left( \ln n \right)$ and then
 for all $n$ sufficiently large,
\begin{align}
&  K_n - \left[ x_n P_n + \sqrt{3(x_n P_n + \ln n) \ln n}
\hspace{1.5pt}\right] \nonumber \\ & = K_n \sqrt{\frac{3\ln n}{K_n
}} - \sqrt{3\left[ K_n \left(1 - \sqrt{\frac{3\ln n}{K_n
}}\hspace{2pt}\right) + \ln n\right] \ln n} \nonumber
\\  & = \sqrt{3K_n\ln n}  -
\sqrt{3\left[K_n  \hspace{-1pt}+ \hspace{-1pt} \sqrt{\ln n} \left(
\sqrt{\ln n} \hspace{-1pt}- \hspace{-1pt}
\sqrt{3K_n}\hspace{2pt}\right) \right ] \hspace{-1pt} \ln n}
\nonumber \\  & \geq \sqrt{3K_n\ln n} - \sqrt{3K_n\ln n} \nonumber
\\  & =  0,
\end{align}
where we use $K_n \geq \ln n$ for all $n$ sufficiently large (this
holds from condition $K_n = \omega\left( \ln n \right)$). Then it is
clear that Lemma \ref{brig_urig} is proved. \pfe

\subsubsection{\textbf{Proving Lemma \ref{er_brig}}} \label{app-prf-er_brig}~

We number the keys in the key pool of size $P_n$ by $1,2,\ldots,P_n$.
In binomial $q$-intersection graph $H_q(n,P_n,x_n)$, let $\mathcal
{U}_i$ be the set of sensors assigned with key $\kappa_i$
($i=1,2,\ldots,P_n$). Then $U_i$ denoting the cardinality of $\mathcal
{U}_i$ (i.e., $U_i:=|\mathcal {U}_i|$) obeys a binomial distribution
$\textrm{Bin}(n, x_n)$, with $n$ as the number of trials, and $x_n$
as the success probability in each trial. Clearly, we can generate
the random set $\mathcal {U}_i$ in the following equivalent manner:
First draw the cardinality $U_i$ from the distribution
$\textrm{Bin}(n, x_n)$, and then choose $U_i$ distinct nodes
uniformly at random from the set $\mathcal{V}_n$ of all $n$ nodes ($\mathcal{V}_n=\{v_1,v_2,\ldots,v_n\}$).

Given  $\mathcal {U}_i$ defined above, we generate a graph $H(\mathcal{U}_i)$ on node set $\mathcal{V}_n$ as follows. We construct
the graph $H(\mathcal{U}_i)$ by establishing edges between any and only pair
of nodes in $\mathcal {U}_i$; i.e., $H(\mathcal{U}_i)$ has a clique on
$\mathcal {U}_i$ and no edges between nodes outside of this clique.
If a given realization of the random variable $U_i$ satisfies $U_i <
2$, then the corresponding instantiation of $H(\mathcal{U}_i)$ will be an
empty graph.

We now explain the connection between $H(\mathcal{U}_i)$ and the binomial
  $q$-intersection graph $H_q(n,P_n,x_n)$.
We let an operator ${\cal{O}}_q$ take a multigraph
\cite{citeulike:505396} with possibly multiple edges between two
nodes as its argument. The operator returns a simple graph with an
undirected edge between two nodes $i$ and $j$, if and only if the
input multigraph has at least $q$ edges between these nodes. Recall
that two nodes in $H_q(n,P_n,x_n)$
  need to share at least $q$ keys to
have an edge in between. Then, with $H(\mathcal{U}_1), \ldots, H(\mathcal{U}_{P_n})$
generated independently, it is straightforward to see
\begin{equation}
{\cal{O}}_q \left(\bigcup_{i=1}^{P_n}  H(\mathcal{U}_i) \right)
=_{\textrm{st}}  H_q(n,P_n,x_n), \label{eq:osy_new_1}
\end{equation}
with $=_{\textrm{st}}$ denoting statistical equivalence.

We will introduce auxiliary random graphs $L(n,B)$ and $L_q(n,B)$,
both defined on the $n$-size node set $\mathcal{V}_n = \{v_1,v_2,\ldots,v_n\}$, where $B$ is a random integer variable. The motivation for defining $L(n,B)$ and $L_q(n,B)$ is that they serve as an intermediate step to build the connection between the above binomial
  $q$-intersection graph $H_q(n,P_n,x_n)$ and an Erd\H{o}s--R\'enyi graph. More specifically,
\begin{itemize}
\item on the one hand, given $U_i$ defined above, we build the connection between $L(n,\big\lfloor U_i/2\big\rfloor)$ and $H(\mathcal{U}_i)$, in order to find the relationship between $L_q\big(n,\sum_{i=1}^{P_n}\big\lfloor U_i/2\big\rfloor \big)$ and the binomial
  $q$-intersection graph $H_q(n,P_n,x_n)$;
\item on the other hand, when $Z$ is a Poisson random variable, $L(n,Z)$ becomes an
Erd\H{o}s--R\'{e}nyi graph;
\item given the above two points, we further find  the relationship between $L_q\big(n,\sum_{i=1}^{P_n}\big\lfloor U_i/2\big\rfloor \big)$ and $L(n,Z)$ for a Poisson random variable $Z$. Then summarizing all points, we build the connection between the binomial
  $q$-intersection graph $H_q(n,P_n,x_n)$ and an Erd\H{o}s--R\'enyi graph.
\end{itemize}
We now define $L(n,B)$ and $L_q(n,B)$ on the node set $\mathcal{V}_n = \{v_1,v_2,\ldots,v_n\}$ for a random integer variable $B$. For different nodes $v_i$ and $v_j$, we use $\text{edge}(v_i, v_j)$ to denote an undirected edge between nodes $v_i$ and $v_j$ so there is no difference between $\text{edge}(v_i, v_j)$ and $\text{edge}(v_j, v_i)$. For the $n$ nodes in $\mathcal{V}_n = \{v_1,v_2,\ldots,v_n\}$, the number of possible edges is $\binom{n}{2}$ (i.e., the number of ways to select two unordered nodes from $n$ nodes). Among these $\binom{n}{2}$ edges, we select one edge uniformly at random at each time. We repeat the selection $b$ times independently for an integer $b$. Note that at each time, an edge is selected from the $\binom{n}{2}$ edges, so we have that even if an edge has already been selected, it may get selected again next time. In other words, the selections are done \emph{with repetition} since it is possible that an edge gets selected multiple times. After the $b$ times of selection, we obtain $b$ edges where several edges may be the same. These $b$ edges constitute
a multiset $\mathcal{M}(b)$, where a multiset is a generalization of a set such that  unlike a set, a multiset allows multiple elements to take the same value. Given an integer $b$, after obtaining a  multiset $\mathcal{M}(b)$ according to the above procedure, we now construct graphs $L(n,b)$ and $L_q(n,b)$, which are both defined on the node set $\mathcal{V}_n = \{v_1,v_2,\ldots,v_n\}$. An edge is put in graph $L(n,b)$ if and only if it appears at least once in the multiset $\mathcal{M}(b)$, while an edge is put in graph $L_q(n,b)$ if and only if it appears at least $q$ times in the multiset $\mathcal{M}(b)$. Now given graphs $L(n,b)$ and $L_q(n,b)$ for an integer $b$, we define graphs $L(n,B)$ and $L_q(n,B)$ for an integer-valued random variable $B$ as follows: we let $L(n,B)$ be $L(n,b)$ with probability $\mathbb{P}[B=b]$, and let $L_q(n,B)$ be $L_q(n,b)$ with probability $\mathbb{P}[B=b]$.

With $H(\mathcal{U}_i)$ and $L(n,B)$ given above, we show a coupling below
under which random graph $L(n,\big\lfloor U_i/2\big\rfloor)$ is a
subgraph of random graph $H(\mathcal{U}_i)$; i.e.,
\begin{align}
 H(\mathcal{U}_i) & \succeq L(n,\big\lfloor U_i/2\big\rfloor) . \label{cpHG}
\end{align}
By definition, graph $L(n,\big\lfloor U_i/2\big\rfloor)$ has at most
$\big\lfloor U_i/2\big\rfloor$ edges and thus contains non-isolated
nodes with a number (denoted by $\ell$) at most $2\cdot \big\lfloor
U_i/2\big\rfloor \leq U_i$, where a node is non-isolated if it has a link with at least another node, and a node is isolated if it has no link with any other node. Given an instance $\mathcal {L}$ of
random graph $L(n,\big\lfloor U_i/2\big\rfloor)$, we construct set
$\mathcal {U}_i$ as the union of the $\ell$ number non-isolated
nodes in $\mathcal {L}$ and the rest $(U_i -\ell)$ nodes selected
uniformly at random from the rest $(n-\ell)$ isolated nodes in
$\mathcal {L}$. Since graph $H(\mathcal{U}_i)$ contains a clique of $\mathcal
{U}_i$, it is clear that the induced instance of $H(\mathcal{U}_i)$ is a
supergraph of the instance $\mathcal {L}$ of graph $L(n,\big\lfloor
U_i/2\big\rfloor)$. Then the proof of (\ref{cpHG}) is completed.

Now based on $L(n,\big\lfloor U_i/2\big\rfloor)$, we construct a
graph defined on node set $\mathcal{V}_n$. We add an edge between two
nodes in this graph if and only if there exist at least $q$
different number of $i$ such that the two nodes have an edge in each
of these $L(n,\big\lfloor U_i/2\big\rfloor)$. By the independence of
$U_i$ ($i=1,2,\ldots,P_n$) and the definition of $L_q(n,B)$ above,
it is clear that such induced graph is statistically equivalent to
$L_q\big(n,\sum_{i=1}^{P_n}\big\lfloor U_i/2\big\rfloor \big)$.
Namely, we have
\begin{align}
{\cal{O}}_q \left(\bigcup_{i=1}^{P_n}  L(n,\big\lfloor
U_i/2\big\rfloor) \right) =_{\textrm{st}}
L_q\big(n,\sum_{i=1}^{P_n}\big\lfloor U_i/2\big\rfloor \big)
\label{eq:osy_new_2}
\end{align}

In view of (\ref{eq:osy_new_1}), (\ref{cpHG}), and
(\ref{eq:osy_new_2}), we see
\begin{align}
 H_q(n,P_n,x_n) & \succeq L_q(n, Y), \label{HqnY}
\end{align}
where $Y$ is defined via
\begin{align}
 Y: = \sum_{i=1}^{P_n}W_i,\label{newzY}
\end{align}
with
\begin{align}
W_i : = \big\lfloor U_i/2\big\rfloor = \m{\frac{1}{2}}(U_i-
\textrm{I}_{[U_i\textrm{ is odd}]}). \label{eqWi}
\end{align}

We now explore a bound of $Y$ based on (\ref{newzY}) and
(\ref{eqWi}). For a random variable $\mathcal {R}$, we denote its
expected value (i.e., mean) and variance by $\mathbb{E}[\mathcal
{R}]$ and $\textrm{Var}[\mathcal {R}]$, respectively. As noted,
$U_i$ obeys a binomial distribution $\textrm{Bin}(n, x_n)$. Then
\begin{align}
\mathbb{E}[U_i] & = \sum_{a=0,1,\ldots, n} \Bigg[ a \cdot
\binom{n}{a}{x_n}^a(1-x_n)^{n-a}\Bigg]  \nonumber
\\  & = nx_n \sum_{a=0,1,\ldots, n} \Bigg[
\binom{n-1}{a-1}{x_n}^{a-1}(1-x_n)^{n-a}\Bigg]  \nonumber
\\  & = nx_n [x_n+(1-x_n)]^{n-1} \nonumber
\\  & =  nx_n,  \label{evi1}
\end{align}
and
\begin{align}
 &\mathbb{E}\big[ \textrm{I}_{[U_i\textrm{ is odd}]}\big]   \nonumber
\\  & \quad = \mathbb{P}[U_i\textrm{ is odd}]  \nonumber
\\  & \quad = \sum_{a=1,3,\ldots, n -
\textrm{I}_{[U_i\textrm{ is even}]}}
\binom{n}{a}{x_n}^a(1-x_n)^{n-a} \nonumber
\\  & \quad = \frac{1}{2} \sum_{a=0,1,\ldots, n}
\binom{n}{a}{x_n}^a(1-x_n)^{n-a}  \nonumber
\\  &  \quad \quad  - \frac{1}{2} \sum_{a=0,1,\ldots,
n} \binom{n}{a}(-x_n)^{a}(1-x_n)^{n-a}  \nonumber
\\  & \quad = \m{\frac{1}{2}} [x_n + (1-x_n)]^{n} - \m{\frac{1}{2}} [-x_n +
(1-x_n)]^{n}  \nonumber
\\  & \quad = \m{\frac{1}{2}}[1 - (1-2x_n)^{n}] .  \label{evi2}
\end{align}
Applying (\ref{evi1}) and (\ref{evi2}) to (\ref{eqWi}), and using the condition (\ref{er_brig-eq2})
 (i.e., $ {x_n}=o\big(\frac{1}{n\ln n}\big)$), we derive
\begin{align}
& \mathbb{E}[W_i] \nonumber
\\   &  = \m{\frac{1}{2}}\mathbb{E}[U_i] -
\m{\frac{1}{2}}\mathbb{E}\big[ \textrm{I}_{[U_i\textrm{ is
odd}]}\big]  \label{Ewi}
\\  & = \m{\frac{1}{2}}nx_n  - \m{\frac{1}{4}} + \m{\frac{1}{4}}(1-2x_n)^{n}   \nonumber
\\  & = \m{\frac{1}{2}}nx_n  - \m{\frac{1}{4}} + \m{\frac{1}{4}}\big[1-2nx_n
+ 2n(n-1){x_n}^2 \pm O\big(n^3{x_n}^3\big)\big]  \nonumber
\\  & = \m{\frac{1}{2}} n(n-1){x_n}^2 \pm O\big(n^3{x_n}^3\big) \nonumber
\\  & =  \m{\frac{1}{2}} n(n-1){x_n}^2 \cdot [1 \pm o(n{x_n})]
\label{exWi-stronger}
\end{align}


From (\ref{eqWi}), it holds that
\begin{align}
& \textrm{Var}[2W_i]  \nonumber
\\ & = \textrm{Var}\big[U_i- \textrm{I}_{[U_i\textrm{
is odd}]}\big]  \nonumber
\\  &= \textrm{Var}[U_i]+ \textrm{Var}[\textrm{I}_{[U_i\textrm{ is
odd}]}]  - 2 \textrm{Cov}[U_i, \textrm{I}_{[U_i\textrm{ is odd}]}],
\label{var2wi}
\end{align}
where $\textrm{Cov}[U_i, \textrm{I}_{[U_i\textrm{ is odd}]}]$
denoting the covariance between $U_i$ and $\textrm{I}_{[U_i\textrm{
is odd}]}$ is given by
\begin{align}
& \textrm{Cov}[U_i, \textrm{I}_{[U_i\textrm{ is odd}]}]  \nonumber
\\ & = \mathbb{E}\big[(U_i-\mathbb{E}[U_i])\big(\textrm{I}_{[U_i\textrm{ is odd}]}
- \mathbb{E}[\textrm{I}_{[U_i\textrm{ is odd}]}]\big)\big] \nonumber
\\ & = \mathbb{E}[U_i\textrm{I}_{[U_i\textrm{ is odd}]}] - \mathbb{E}[U_i]\mathbb{E}[\textrm{I}_{[U_i\textrm{ is
odd}]}].
   \label{var2wi1}
\end{align}
Clearly, it holds that $U_i\textrm{I}_{[U_i\textrm{ is odd}]} \geq
\textrm{I}_{[U_i\textrm{ is odd}]}$, inducing
\begin{align}
\mathbb{E}[U_i\textrm{I}_{[U_i\textrm{ is odd}]}] & \geq
\mathbb{E}[\textrm{I}_{[U_i\textrm{ is odd}]}].
   \label{var2wi1ViIVi}
\end{align}
From (\ref{evi1}) and (\ref{evi2}), we further obtain
\begin{align}
&\mathbb{E}[U_i]\mathbb{E}[\textrm{I}_{[U_i\textrm{ is odd}]}] -
\mbox{$\frac{3}{2}$} \cdot (\mathbb{E}[U_i] -
\mathbb{E}[\textrm{I}_{[U_i\textrm{ is odd}]}]) \nonumber
\\ & = n x_n \cdot \mbox{$\frac{1}{2}$}[1 - (1-2x_n)^{n}]  - \mbox{$\frac{3}{2}$} \big\{ n x_n - \mbox{$\frac{1}{2}$}[1 -
(1-2x_n)^{n}]\big\}  \nonumber
\\ & = -nx_n + \m{\frac{3}{4}} -
(\m{\frac{1}{2}}nx_n+\m{\frac{3}{4}})(1-2x_n)^{n}  \nonumber
\\ & \leq -nx_n + \m{\frac{3}{4}} -
(\m{\frac{1}{2}}nx_n+\m{\frac{3}{4}})(1-2nx_n+\m{\frac{4}{3}}n^2{x_n}^2)
\nonumber
\\ & = - \m{\frac{2}{3}} n^2{x_n}^2 \leq 0,  \label{var2wi2}
\end{align}
where the step involving the first ``$\leq$'' uses the inequality
$(1-2x_n)^{n} \geq 1-2nx_n+\m{\frac{4}{3}}n^2{x_n}^2$ for all $n$
sufficiently large, which is derived from a Taylor expansion of the
binomial series $(1-2x_n)^{n}$, given the condition (\ref{er_brig-eq2})
 (i.e., $ {x_n}=o\big(\frac{1}{n\ln n}\big)$).

Using (\ref{var2wi1ViIVi}) and (\ref{var2wi2}) in (\ref{var2wi1}),
it follows that
\begin{align}
 \textrm{Cov}[U_i, \textrm{I}_{[U_i\textrm{ is odd}]}]  & \geq \mbox{$\frac{5}{2}$} \mathbb{E}[\textrm{I}_{[U_i\textrm{ is odd}]}] - \mbox{$\frac{3}{2}$} \mathbb{E}[U_i].  \label{var2wi3}
\end{align}
For binomial random variable $U_i$ and Bernoulli random variable
$\textrm{I}_{[U_i\textrm{ is odd}]}$, it is clear that
\begin{align}
\textrm{Var}[U_i] & \leq \mathbb{E}[U_i], \label{varvi}
\end{align}
and
\begin{align}
\textrm{Var}[\textrm{I}_{[U_i\textrm{ is odd}]}] & \leq
\mathbb{E}[\textrm{I}_{[U_i\textrm{ is odd}]}]. \label{varviI}
\end{align}

Applying (\ref{var2wi3}) (\ref{varvi}) and (\ref{varviI}) to
(\ref{var2wi}), we have
\begin{align}
 \textrm{Var}[2W_i] & \leq \mathbb{E}[U_i] +
\mathbb{E}[\textrm{I}_{[U_i\textrm{ is odd}]}] \nonumber
\\ & \quad  - 5\mathbb{E}[\textrm{I}_{[U_i\textrm{ is odd}]}] + 3\mathbb{E}[U_i]  \nonumber
\\ & = 4 (\mathbb{E}[U_i] -
\mathbb{E}[\textrm{I}_{[U_i\textrm{ is odd}]}]),
\end{align}
which along with (\ref{Ewi}) yields $\textrm{Var}[2W_i] \leq 8
\mathbb{E}[W_i]$; i.e.,
\begin{align}
 \textrm{Var}[W_i]  & \leq  2 \mathbb{E}[W_i]. \label{varWi}
\end{align}

Considering the independence of $W_i$ ($i=1,2,\ldots,P_n$), for $Y =
\sum_{i=1}^{P_n}W_i$ given in (\ref{Y}), we use (\ref{varWi}) to
derive
\begin{align}
 \textrm{Var}[Y]  & \leq  2 \mathbb{E}[Y]. \label{varYEY}
\end{align}

From $Y = \sum_{i=1}^{P_n}W_i$, (\ref{exWi-stronger}), and the fact that $\mathbb{E}[W_i]$ for each $i$ is the same, we obtain
\begin{align}
\mathbb{E}[Y]  & =   \m{\frac{1}{2}} n(n-1)P_n {x_n}^2 \cdot [1 \pm o(n{x_n})]
 \label{varYEY2-stronger-sbsbabcd}.
\end{align}
Note that Lemma \ref{er_brig} has conditions (\ref{er_brig-eq2}) and (\ref{er_brig-eq4}) (i.e., ${x_n} = o\left( \frac{1}{n\ln n} \right)$ and ${x_n}^2 P_n = \omega\big(\frac{(\ln n)^6}{n^2}\big)$). Using these in (\ref{varYEY2-stronger-sbsbabcd}), we have
\begin{align}
\mathbb{E}[Y]  = \textstyle{\frac{1}{2}n(n-1)P_n {x_n}^2 \cdot \big[1 \pm o\big(\frac{1}{\ln n}\big)\big]}\label{GerGb-ton-tac4}
\end{align}
and
\begin{align}
\mathbb{E}[Y]  =\textstyle{\omega\big((\ln n)^6\big)}.\label{GerGb-ton-tac2}
\end{align}


Now based on (\ref{varYEY}) and (\ref{GerGb-ton-tac2}), we provide a lower
bound on $Y$ with high probability. By Chebyshev's inequality, \f
for any $\phi > 0$,
\begin{align}
\mathbb{P}\big[\hspace{1pt}|Y-\mathbb{E}[Y]|\geq
\phi\sqrt{\textrm{Var}[Y]}\hspace{1pt}\big] \leq {\phi}^{-2}.
\end{align}
We select
\begin{align}
 \phi & =
\frac{\big\{\mathbb{E}[Y]\big\}^{\frac{5}{6}}}{2\sqrt{\textrm{Var}[Y]}},
\label{phieq}
\end{align}
 which with (\ref{varYEY}) and (\ref{GerGb-ton-tac2}) results in $\phi  =
\omega(1)$ and hence
\begin{align}
\mathbb{P}\big[ \hspace{1pt}Y < \mathbb{E}[Y] -
\phi\sqrt{\textrm{Var}[Y]}\hspace{1pt}\big] = o(1). \label{Y}
\end{align}

Let $Z$ be a Poisson random variable with mean 
\begin{align}
{\lambda_n} & : = \mathbb{E}[Y] -
\big\{\mathbb{E}[Y]\big\}^{\frac{5}{6}}. \label{lam}
\end{align}
With ${\psi_n}$ defined by
\begin{align}
 {\psi_n} & : =
\m{\frac{1}{2}} \big\{\mathbb{E}[Y]\big\}^{\frac{1}{3}}, \label{psi}
\end{align}
 we conclude from (\ref{GerGb-ton-tac2}) (\ref{lam}) and (\ref{psi}) that
${\psi_n} = \omega(1)$ and ${\psi_n} =
o\big(\sqrt{{\lambda_n}}\hspace{2pt}\big)$.

By \cite[Lemma 1.2]{citeulike:505396}, it holds that
\begin{align}
 \mathbb{P}\big[\hspace{1pt}Z \geq {\lambda_n} + {\psi_n}
\sqrt{{\lambda_n}}\hspace{2pt}\big]  \leq e^{{\psi_n} \sqrt{{\lambda_n}}
-({\lambda_n} + {\psi_n} \sqrt{{\lambda_n}})
\ln(1+\frac{{\psi_n}}{\sqrt{{\lambda_n}}})}. \label{poisax}
\end{align}
From ${\psi_n} = o\big(\sqrt{{\lambda_n}}\hspace{2pt}\big)$, then for all
$n$ sufficiently large, we have
$\ln\big(1+\frac{{\psi_n}}{\sqrt{{\lambda_n}}}\big) \geq
\frac{{\psi_n}}{\sqrt{{\lambda_n}}}-\frac{{\psi_n}^2}{2{\lambda_n}}$ (derived from a
Taylor expansion), which is used in (\ref{poisax}) to
yield
\begin{align}
 \mathbb{P}\big[\hspace{1pt}Z \geq {\lambda_n} + {\psi_n}
\sqrt{{\lambda_n}}\hspace{2pt}\big]  & \leq e^{{\psi_n} \sqrt{{\lambda_n}}
-({\lambda_n} + {\psi_n} \sqrt{{\lambda_n}})
\big(\frac{{\psi_n}}{\sqrt{{\lambda_n}}}-\frac{{\psi_n}^2}{2{\lambda_n}}\big)}
\nonumber
\\ & =
e^{\frac{{\psi_n}^2}{2}\big(\frac{{\psi_n}}{\sqrt{{\lambda_n}}}-1\big)}.
\label{pois}
\end{align}
Applying ${\psi_n} = \omega(1)$ and ${\psi_n} =
o\big(\sqrt{{\lambda_n}}\hspace{2pt}\big)$ to (\ref{pois}), we
obtain
\begin{align}
 \mathbb{P}\big[\hspace{1pt}Z \geq {\lambda_n} + {\psi_n}
\sqrt{{\lambda_n}}\hspace{2pt}\big] &  = o(1). \label{Z}
\end{align}

From (\ref{phieq}) (\ref{lam}) and (\ref{psi}), we
establish
\begin{align}
 {\lambda_n} + {\psi_n} \sqrt{{\lambda_n}} & \leq \mathbb{E}[Y] -
\big\{\mathbb{E}[Y]\big\}^{\frac{5}{6}} + \m{\frac{1}{2}}
\big\{\mathbb{E}[Y]\big\}^{\frac{1}{3}} \cdot \sqrt{\mathbb{E}[Y]}
\nonumber
\\   & =  \mathbb{E}[Y] -
\phi\sqrt{\textrm{Var}[Y]}. \label{lambda}
\end{align}

Given (\ref{Y}) (\ref{Z}) and (\ref{lambda}), we obtain\vspace{-2pt}
\begin{align}
 &\mathbb{P}[ Y \geq Z] \nonumber
\\  & \geq \mathbb{P}\Big[
 \big(Y \geq \mathbb{E}[Y] - \phi\sqrt{\textrm{Var}[Y]}\hspace{1pt}
\big) \bcap  \hspace{1pt} ({\lambda_n} + {\psi_n} \sqrt{{\lambda_n}} \geq
Z\hspace{1pt} )\Big] \nonumber
\\ & \geq 1 - \mathbb{P}\big[Y < \mathbb{E}[Y] -
\phi\sqrt{\textrm{Var}[Y]}\hspace{1pt}\big] - \mathbb{P}\big[
{\lambda_n} + {\psi_n} \sqrt{{\lambda_n}} < Z\hspace{1pt}\big] \nonumber
\\ & \to 1,\textrm{ as }n \to \infty, \label{YZ}
\end{align}
where in the second to the last step, we use a union bound.

Given (\ref{YZ}), by the definition of graph $L_q(n,X)$, it is easy
to construct a coupling such that $L_q(n,Z)$ is a subgraph of
$L_q(n,Y)$ with probability $1-o(1)$; namely,
\begin{align}
 L_q(n,Y)& \succeq_{1-o(1)}L_q(n,Z) . \label{HqnZ}
\end{align}

From \cite[Proof of Claim 1]{Fill:2000:RIG:340808.340814}, for
Poisson random variable $Z$ with mean ${\lambda_n}$, in sampling $Z$
edges with repetition from all possible $\binom{n}{2}$ edges of an
$n$-size node set, the numbers of draws for different edges are
independent Poisson random variables with mean
\begin{align}
{\mu_n}:={\lambda_n}\Bigg/\binom{n}{2},\label{defmu}
\end{align}
where ``with repetition'' means that  at each time, an edge is selected from the $\binom{n}{2}$ edges, so we have that even if an edge has already been selected, it may get selected again next time.
 Therefore, $L_q(n,Z)$ with $Z \in \textrm{Poisson}({\lambda_n})$ is an
Erd\H{o}s--R\'{e}nyi graph \cite{citeulike:4012374} in which each
edge independently appears with a probability that a Poisson random
variable with mean ${\mu_n}$ is at least $q$, i.e., a probability of
\begin{align}
\varrho_n &  : =\sum_{x=q}^{\infty} \frac{{{\mu_n}}^x e^{-{\mu_n}}}{x!}.
\label{pndef}
\end{align}

In view that $L_q(n,Z)$ is equivalent to $G_{ER}(n,\varrho_n)$, then
from (\ref{HqnY}) and (\ref{HqnZ}), \f
\begin{align}
 H_q(n,P_n,x_n)& \succeq_{1-o(1)}G_{ER}(n,\varrho_n) , \label{gercp}
\end{align}
which is exactly (\ref{GerGb}) in Lemma
\ref{er_brig}. Therefore, to complete proving
Lemmas
\ref{er_brig}, we now analyze $\varrho_n$ in (\ref{pndef}).


From \cite[Proposition 1]{PES:36149}, $\varrho_n$ in (\ref{pndef})
can be bounded by\vspace{-1pt}
\begin{align}
\frac{{{\mu_n}}^q e^{-{\mu_n}}}{q!}  & < \varrho_n < \frac{{{\mu_n}}^q
e^{-{\mu_n}}}{q!} \cdot\bigg(1-\frac{{\mu_n}}{q+1}\bigg)^{-1}.\vspace{-1pt}
\label{pnbound}
\end{align}

To evaluate $\varrho_n$ based on (\ref{pnbound}), we now assess $\mu_n$ in (\ref{defmu}), and   analyze $\lambda_n$ in (\ref{lam}).
Applying (\ref{GerGb-ton-tac4}) and (\ref{GerGb-ton-tac2}) to (\ref{lam}), and noting that $\big[1 \pm o\big(\frac{1}{\ln n}\big)\big]\cdot \big[1 \pm o\big(\frac{1}{\ln n}\big)\big]$ can also be written as $\big[1 \pm o\big(\frac{1}{\ln n}\big)\big]$, we obtain
\begin{align}
 {\lambda_n} &    = \mathbb{E}[Y] -
\big\{\mathbb{E}[Y]\big\}^{\frac{5}{6}} \nonumber \\    & = \mathbb{E}[Y] \cdot \Big[1-
\big\{\mathbb{E}[Y]\big\}^{-\frac{1}{6}}\Big]\nonumber \\    & =\textstyle{\frac{1}{2}n(n-1)P_n {x_n}^2 \cdot \big[1 \pm o\big(\frac{1}{\ln n}\big)\big]}.\label{GerGb-ton-tac6}
\end{align}
The application of (\ref{GerGb-ton-tac6}) to (\ref{defmu}) gives
\begin{align}
{\mu_n}  =\textstyle{ P_n {x_n}^2 \cdot \big[1 \pm o\big(\frac{1}{\ln n}\big)\big]}.\label{GerGb-ton-tac8}
\end{align}
Note that Lemma \ref{er_brig} has condition (\ref{er_brig-eq3}) (i.e., ${x_n}^2 P_n = o\left( \frac{1}{\ln n} \right)$). Using (\ref{er_brig-eq3}) in (\ref{GerGb-ton-tac8}), we have
\begin{align}
{\mu_n}  =
\textstyle{o\left( \frac{1}{\ln n} \right)}.\label{GerGb-ton-tac10}
\end{align}

For any sequence $a_n$ satisfying ${a_n} = \pm o(1)$, we explain below $(1+a_n)^q = 1 \pm \Theta(a_n)$ since $q$ is a constant. To see this, given $|a_n|<1$ for all $n$ sufficiently large from ${a_n} = \pm o(1)$, we obtain: on the one hand, $(1+a_n)^q \leq (1+|a_n|)^q = 1 + \sum_{i=1}^q \big[ \binom{q}{i} |a_n|^i \big] \leq 1 + |a_n|\sum_{i=1}^q   \binom{q}{i} = 1 + (2^q-1) |a_n| =  1 + \Theta(a_n)$; on the other hand, $(1-a_n)^q \leq (1+|a_n|)^q = 1 + \sum_{i=1}^q \big[ \binom{q}{i} (-|a_n|)^i \big] \geq 1 - |a_n|\sum_{i=1}^q   \binom{q}{i} = 1 - (2^q-1) |a_n| =  1 - \Theta(a_n)$. Summarizing $ 1 - \Theta(a_n) \leq  (1-a_n)^q \leq  1 + \Theta(a_n)$, we obtain
 \begin{align}
(1+a_n)^q = 1 \pm \Theta(a_n)\text{ for }{a_n} = \pm o(1).  \label{pnbounda-ton-tac-tonasab}
\end{align}
From (\ref{GerGb-ton-tac8}) and (\ref{pnbounda-ton-tac-tonasab}), it holds  that
\begin{align}
{\mu_n}^q  =
\textstyle{ (P_n {x_n}^2)^q \cdot \big[1 \pm o\big(\frac{1}{\ln n}\big)\big]}.\label{GerGb-ton-tac8-tif}
\end{align}

For ${\mu_n} = o(1)$, we explain below $e^{-{\mu_n}} = 1 - \Theta({\mu_n})$. To see this, on the one hand, it holds that $e^{-{\mu_n}} \geq 1 - {\mu_n}$. On the other hand, given ${\mu_n} < 0.5$ for all $n$ sufficiently large (which holds from ${\mu_n} = o(1)$), we can easily show $e^{-{\mu_n}} \leq 1 - 0.5 {\mu_n}$ by taking the derivative of $e^{-{\mu_n}} - (1 - 0.5 {\mu_n})$ to investigate its monotonicity. Summarizing $ 1 - {\mu_n} \leq  e^{-{\mu_n}} \leq 1 - 0.5 {\mu_n}$, we obtain
 \begin{align}
e^{-{\mu_n}} = 1 - \Theta({\mu_n}).  \label{pnbounda-ton-tac-tona}
\end{align}

From ${\mu_n} = o(1)$, we have $\big(1-\frac{{\mu_n}}{q+1}\big)^{-1} = 1 + \frac{{\mu_n}}{q+1-{\mu_n}} = 1 + \Theta({\mu_n})$, which along with (\ref{pnbounda-ton-tac-tona}) is used in (\ref{pnbound}) to derive
\begin{align}
\varrho_n & \hspace{-1pt}=\hspace{-1pt} \frac{{{\mu_n}}^q e^{-{\mu_n}}}{q!} \hspace{-1pt} \cdot \hspace{-1pt}\big[1\hspace{-1pt} +\hspace{-1pt} \Theta({\mu_n})\big]\hspace{-1pt} = \hspace{-1pt} \frac{{{\mu_n}}^q}{q!}  \hspace{-1pt}\cdot \hspace{-1pt}\big[1\hspace{-1pt} - \hspace{-1pt}\Theta({\mu_n})\big]\hspace{-1pt}  \cdot\hspace{-1pt}\big[1\hspace{-1pt} + \hspace{-1pt}\Theta({\mu_n})\big].  \label{pnbounda-ton-tac-tonabc}
\end{align}
For any two sequences $c_n$ and $d_n$ satisfying
$c_n = \Theta({\mu_n})$ and $d_n = \Theta({\mu_n})$ with ${\mu_n} = o(1)$, we have $(1-c_n) (1+d_n) = 1 - c_n + d_n - c_n d_n = 1 \pm \Theta({\mu_n})$, which we use in (\ref{pnbounda-ton-tac-tonabc}) to get
\begin{align}
\varrho_n & = \frac{{{\mu_n}}^q}{q!}  \cdot \big[1 \pm  \Theta({\mu_n})\big] .  \label{pnbounda-ton-tac-tonabc5}
\end{align}
Then applying (\ref{GerGb-ton-tac8-tif}) and (\ref{GerGb-ton-tac10}) to (\ref{pnbounda-ton-tac-tonabc5}), and noting that $\big[1 \pm o\big(\frac{1}{\ln n}\big)\big]\cdot \big[1 \pm o\big(\frac{1}{\ln n}\big)\big]$ can also be written as $\big[1 \pm o\big(\frac{1}{\ln n}\big)\big]$, we obtain
\begin{align}
\varrho_n  & = \textstyle{\frac{(P_n{x_n}^2)^q}{q!}  \cdot \big[1 \pm o\big(\frac{1}{\ln n}\big)\big]} .  \label{GerGb-ton-tac11}
\end{align}





From \cite[Fact 3]{zz}, for Erd\H{o}s--R\'enyi graphs
$G_{ER}(n,s_n')$ and $G_{ER}(n,s_n'')$, if $s_n ' \geq s_n''$, then
$G_{ER}(n,s_n')\succeq G_{ER}(n,s_n'')$. Thus, by (\ref{gercp})
(\ref{GerGb-ton-tac11}) and \cite[Fact 3]{2013arXiv1301.0466R} on the
transitivity of graph coupling, we can set $s_n   =
\textstyle{\frac{(P_n{x_n}^2)^q}{q!}} \cdot \big[1 - o\big(\frac{1}{\ln n}\big)\big]$ to have
$ H_q(n,P_n,x_n) \succeq_{1-o(1)}G_{ER}(n,s_n) $, so  that Lemma \ref{er_brig} is proved. \pfe

\subsection{Proof of Lemma \ref{lem-mnd}} \label{app-mnd-thm}

\textbf{Lemma \ref{lem-mnd} on minimum   degree in $q$-composite random key graphs (Restated). }
For a $q$-composite random key graph $G_q(n,K_n,P_n)$, if there is a sequence $\phi_n$ with $\lim_{n \to \infty}{\phi_n} \in [-\infty, \infty]$
such that
\begin{align}
 {\frac{1}{q!} \cdot \frac{{K_n}^{2q}}{{P_n}^{q}}   = \frac{\ln  n + {(k-1)} \ln \ln n + {\phi_n}}{n},} \label{lem-mnd:pe-restated}
\end{align}
then  under \begin{align}
P_n  &= \begin{cases} \, \Omega(n), &\text{for } q=1,\\
\, \omega\big(n^{2-\frac{1}{q}}(\ln n)^{2+\frac{1}{q}}\big), &\text{for } q \geq 2 ,
\end{cases}  \label{scalingP-strongercopyvvs}
\end{align} we have
 \begin{align}
& \lim_{n \to \infty}  \mathbb{P}[\hspace{2pt}G_q(n,K_n,P_n)\text{ has a minimum   degree at least $k$.}\hspace{2pt}] \nonumber \\ &=  e^{-
\frac{e^{- \iffalse \lim\limits_ \fi \lim_{n \to \infty}{\phi_n}}}{(k-1)!}} \label{expr-mnd-all}
 \end{align}
 \begin{subnumcases}{\hspace{-35pt}=} 0,&\text{\hspace{-4pt}if  }$ \lim_{n \to \infty}{\phi_n} =- \infty$, \label{expr-mnd-0} \\
1,&\text{\hspace{-4pt}if  }$ \lim_{n \to \infty}{\phi_n} = \infty$, \label{expr-mnd-1} \\ e^{-
\frac{e^{-\phi^{*}}}{(k-1)!}}  ,&\text{\hspace{-4pt}if  }$ \lim_{n \to \infty}{\phi_n} = \phi^{*}\in (-\infty, \infty)$.\label{expr-mnd-exact-restated}  \end{subnumcases}

We will use Lemma \ref{inter-lem-mnd} below to establish Lemma \ref{lem-mnd}.

\begin{lem}[\textbf{Minimum   degree in the intersection of a $q$-composite random key graph $G_q(n,K_n,P_n)$ and an Erd\H{o}s--R\'{e}nyi graph $G_{ER}(n,s_n)$, a result presented in our   work \cite{zhao2017topological}}] \label{inter-lem-mnd}
For $G_q(n,K_n,P_n) \bcap G_{ER}(n,s_n)$ being the intersection of a $q$-composite random key graph $G_q(n,K_n,P_n)$ and an Erd\H{o}s--R\'{e}nyi graph $G_{ER}(n,s_n)$, with $b_{q,n}$ denoting the edge probability of a $q$-composite random key graph $G_q(n,K_n,P_n)$ so that $ b_{q,n}\times s_n $ is the edge probability of $G_q(n,K_n,P_n) \bcap G_{ER}(n,s_n)$, if there is a sequence $\varphi_n$ with $\lim_{n \to \infty}{\varphi_n} \in [-\infty, \infty]$
such that %
\begin{align}
 {b_{q,n}\times s_n   = \frac{\ln  n + {(k-1)} \ln \ln n + {\varphi_n}}{n},} \label{inter-lem-mnd:pe}
\end{align}
then it holds under $ K_n =
\omega(1)$ and $\frac{{K_n}^2}{P_n} = o(1)$ that
 \begin{align}
& \lim_{n \to \infty}  \mathbb{P}\left[\begin{array}{l}
	G_q(n,K_n,P_n) \bcap G_{ER}(n,s_n)\\ \text{has a minimum degree at least $k$.}
\end{array}\right] \nonumber \\ &=  e^{-
\frac{e^{- \iffalse \lim\limits_ \fi \lim_{n \to \infty}{\varphi_n}}}{(k-1)!}} \label{inter-expr-mnd-all}
 \end{align}
 \begin{subnumcases}{\hspace{-35pt}=} 0,&\text{\hspace{-4pt}if  }$ \lim_{n \to \infty}{\varphi_n} =- \infty$, \label{inter-expr-mnd-0} \\
1,&\text{\hspace{-4pt}if  }$ \lim_{n \to \infty}{\varphi_n} = \infty$, \label{inter-expr-mnd-1} \\ e^{-
\frac{e^{-\varphi^{*}}}{(k-1)!}}  ,&\text{\hspace{-4pt}if  }$ \lim_{n \to \infty}{\varphi_n} = \varphi^{*}\in (-\infty, \infty)$.\label{inter-expr-mnd-exact}  \end{subnumcases}
\end{lem}

Lemma \ref{inter-lem-mnd} is Theorem 1 of in our   work \cite{zhao2017topological}. Setting $s_n = 1$, we have  $G_q(n,K_n,P_n) \bcap G_{ER}(n,s_n) = G_q(n,K_n,P_n) $ and use Lemma  \ref{inter-lem-mnd} to obtain the following Lemma  \ref{inter-lem-mndcp}.

\begin{lem}[\textbf{Minimum   degree in a $q$-composite random key graph $G_q(n,K_n,P_n)$}] \label{inter-lem-mndcp}
With $b_{q,n}$ denoting the edge probability of a $q$-composite random key graph $G_q(n,K_n,P_n)$, if there is a sequence $\varphi_n$ with $\lim_{n \to \infty}{\varphi_n} \in [-\infty, \infty]$
such that %
\begin{align}
 {b_{q,n}  = \frac{\ln  n + {(k-1)} \ln \ln n + {\varphi_n}}{n},} \label{inter-lem-mnd:pecp}
\end{align}
then it holds under $ K_n =
\omega(1)$ and $\frac{{K_n}^2}{P_n} = o(1)$ that
 \begin{align}
& \lim_{n \to \infty}  \mathbb{P}[\hspace{2pt}G_q(n,K_n,P_n)\text{ has a minimum   degree at least $k$.}\hspace{2pt}] \nonumber \\ &=  e^{-
\frac{e^{- \iffalse \lim\limits_ \fi \lim_{n \to \infty}{\varphi_n}}}{(k-1)!}} \label{varpexpr-mnd-all}
 \end{align}
 \begin{subnumcases}{\hspace{-35pt}=} 0,&\text{\hspace{-4pt}if  }$ \lim_{n \to \infty}{\varphi_n} =- \infty$, \label{varpexpr-mnd-0} \\
1,&\text{\hspace{-4pt}if  }$ \lim_{n \to \infty}{\varphi_n} = \infty$, \label{varpexpr-mnd-1} \\ e^{-
\frac{e^{-\varphi^{*}}}{(k-1)!}}  ,&\text{\hspace{-4pt}if  }$ \lim_{n \to \infty}{\varphi_n} = \varphi^{*}\in (-\infty, \infty)$.\label{varpexpr-mnd-exact-restated}  \end{subnumcases}
\end{lem}

Note that the property of minimum degree being at least $k$ is  monotone increasing.
For any monotone increasing property $\mathcal {I}$, the probability that a spanning subgraph
(resp., supergraph) of graph $G$ has $\mathcal {I}$ is at most (resp., at least) the probability of $G$ having $\mathcal {I}$. Therefore, from Lemma \ref{graph_Gs_cpl} on Page \pageref{graph_Gs_cpl}, we can introduce an extra condition $|\phi_n |= o(\ln n)$ to prove Lemma \ref{lem-mnd}. Hence, to use Lemma  \ref{inter-lem-mndcp} for proving Lemma \ref{lem-mnd}, we only need to show that under the conditions of Lemma \ref{lem-mnd} along with the extra condition $|\phi_n |= o(\ln n)$, then the conditions of Lemma  \ref{inter-lem-mndcp} all hold and $\lim_{n \to \infty}{\varphi_n} = \lim_{n \to \infty}{\phi_n}$. Specifically, we only need to show under (\ref{lem-mnd:pe-restated}) (\ref{scalingP-strongercopyvvs}), $\lim_{n \to \infty}{\phi_n} \in [-\infty, \infty]$ and $|\phi_n |= o(\ln n)$, we have $ K_n =
\omega(1)$, $\frac{{K_n}^2}{P_n} = o(1)$, and the sequence ${\varphi_n}$ defined by (\ref{varpexpr-mnd-all}) satisfies $|\varphi_n- {\phi_n} | = o(1)$ so that whenever $\lim_{n \to \infty}{\phi_n}$ exists, $\lim_{n \to \infty}{\varphi_n} $ also exists and $\lim_{n \to \infty}{\varphi_n} = \lim_{n \to \infty}{\phi_n}$.
The rest of the proof is straightforward. Specifically, we use  (\ref{lem-mnd:pe-restated}) and $|\phi_n |= o(\ln n)$ to have $\frac{1}{q!} \cdot \frac{{K_n}^{2q}}{{P_n}^{q}}   = \frac{\ln  n + {(k-1)} \ln \ln n + {\phi_n}}{n} = \frac{\ln  n + {(k-1)} \ln \ln n \pm o(\ln n)}{n} \sim \frac{\ln n}{n}$, implying $\frac{{K_n}^2}{P_n} = \Theta\big( (\frac{\ln n}{n})^{1/q}\big) $, which along with (\ref{scalingP-strongercopyvvs}) further implies $ \begin{cases} K_n = \Omega(\sqrt{\ln n}\,) =
\omega(1) , & \text{for } q=1, \\  K_n = \omega\big(n^{1-\frac{1}{q}}(\ln n)^{1+\frac{1}{q}}\big) =
\omega(1) , & \text{for } q \geq 2, \end{cases} $. Then under the just proved $ \begin{cases} K_n = \Omega(\sqrt{\ln n}\,) =
\omega(1) , & \text{for } q=1, \\  K_n = \omega\big(n^{1-\frac{1}{q}}(\ln n)^{1+\frac{1}{q}}\big) =
\omega(\ln n) , & \text{for } q \geq 2, \end{cases} $ and $\frac{{K_n}^2}{P_n} =  \Theta\big( (\frac{\ln n}{n})^{1/q}\big) = o\big( \frac{1}{\ln n} \big)$, we use Property (ii) of Lemma \ref{lem_eval_psq} below to obtain $b_{q,n}
= \frac{1}{q!} \big( \frac{{K_n}^2}{P_n} \big)^{q} \times [1\pm o\big(\frac{1}{\ln n}\big)]$, \vspace{2pt} which along with the condition $\frac{1}{q!} \cdot \frac{{K_n}^{2q}}{{P_n}^{q}}   = \frac{\ln  n + {(k-1)} \ln \ln n \pm o(\ln n)}{n} $ implies that the sequence ${\varphi_n}$ defined by (\ref{varpexpr-mnd-all}) satisfies $|\varphi_n- {\phi_n} | = o(1)$. This further means that whenever $\lim_{n \to \infty}{\phi_n}$ exists, $\lim_{n \to \infty}{\varphi_n} $ also exists and $\lim_{n \to \infty}{\varphi_n} = \lim_{n \to \infty}{\phi_n}$. Thus, we have shown that under the conditions of Lemma \ref{lem-mnd} along with the extra condition $|\phi_n |= o(\ln n)$, then the conditions of Lemma  \ref{inter-lem-mndcp} all hold and $\lim_{n \to \infty}{\varphi_n} = \lim_{n \to \infty}{\phi_n}$. Then we can use Lemma  \ref{inter-lem-mndcp} to obtain Lemma \ref{lem-mnd} with the extra condition $|\phi_n |= o(\ln n)$. From Lemma \ref{graph_Gs_cpl} on Page \pageref{graph_Gs_cpl}, we further establish Lemma \ref{lem-mnd} regardless of $|\phi_n |= o(\ln n)$.

\subsection{An asymptotic expression for the edge probability of a $q$-composite random key graph $G_q(n,K_n,P_n)$} \label{app-asymptotic-edge-prob}

We present Lemma \ref{lem_eval_psq} below, which provides asymptotic expressions of the edge probability $b_{q,n} $ of a $q$-composite random key graph $G_q(n, K_n,P_n)$.

Recall that a $q$-composite random key graph $G_q(n, K_n,P_n)$ models  the topology of a secure sensor network with $n$ nodes working under the $q$-composite scheme. Let $\mathcal{V}_n =
\{v_1, v_2, \ldots, v_n \}$  represent the $n$ nodes. In the $q$-composite scheme, each node $v_i$ selects $K_n$ {distinct} cryptographic keys
uniformly at {random} from the same pool $\mathcal{P}_n$ consisting of $P_n$ keys, and two nodes can establish a secure link only if they have at least $q$ key(s) in common. For each node $v_i$, the set of its $K_n$ different keys is denoted by $S_i$, and is referred to as the \emph{key ring} of node $v_i$. Then graph $G_q(n,K_n,P_n)$ to model the network topology is defined on the
node set $\mathcal{V}_n$ such that any two different nodes $v_i$ and
$v_j$ possessing at least $q$ key(s) in common (such event is denoted by
$\Gamma_{ij}$) have an edge in between. With $S_{ij} $ defining as $S_{i} \cap
S_{j}$, event $\Gamma_{ij}$  equals $\big[ |S_{ij}| \geq q \big]$,
where $|A|$ with $A$ as a set means the cardinality of $A$. With $b_{q,n} $ denoting the edge probability of a $q$-composite random key graph $G_q(n, K_n,P_n)$, we have $b_{q,n}   = \mathbb{P}[\Gamma_{ij}] = \mathbb{P}[|S_{ij}| \geq q]= \sum_{u=q}^{K_n}
\mathbb{P}[|S_{ij}| = u]$.


\begin{lem} \label{lem_eval_psq}
The following two properties hold, where $b_{q,n} $ denotes the edge probability of a $q$-composite random key graph $G_q(n, K_n,P_n)$:
\begin{itemize}
\item[(i)] If $K_n = \omega(1)$ and $\frac{{K_n}^2}{P_n} = o(1)$, then\\$b_{q,n}
= \frac{1}{q!} \big( \frac{{K_n}^2}{P_n} \big)^{q} \times [1\pm o(1)]$; i.e., $b_{q,n}
\sim \frac{1}{q!} \big( \frac{{K_n}^2}{P_n} \big)^{q}$.\vspace{3pt}
\item[(ii)] If $ \begin{cases} K_n =
\omega(1) , & \text{for } q=1, \\  K_n =
\omega(\ln n) , & \text{for } q \geq 2, \end{cases} $ and $\frac{{K_n}^2}{P_n} = o\big(\frac{1}{\ln n}\big)$, then $b_{q,n}
= \frac{1}{q!} \big( \frac{{K_n}^2}{P_n} \big)^{q} \times [1\pm o\big(\frac{1}{\ln n}\big)]$.
\end{itemize}
\end{lem}

\textbf{Proof of Lemma \ref{lem_eval_psq}:}

\subsubsection{Proving Property (i) of Lemma \ref{lem_eval_psq}}~

We prove Property (i) of Lemma \ref{lem_eval_psq} below. We simplify
$S_{i} \cap S_{j}$ by writing it as $S_{ij}$. Clearly, $P_n \geq
2K_n$ for all $n$ sufficiently large, due to $ \frac{{K_n}^2}{P_n} =
o(1)$. Given $b_{q,n}  = \sum_{u=q}^{K_n}
\mathbb{P}[|S_{ij}| = u]$, Property (i) of Lemma \ref{lem_eval_psq} holds once we establish the following
(\ref{eq_psijq}) and (\ref{eq_psiju}):
\begin{align}
\mathbb{P}[|S_{ij}| = q]  & \sim  (q!)^{-1} \big( {{K_n}^2}/{P_n}
\big)^{q}, \label{eq_psijq}
\end{align}
and
\begin{align}
\mathbb{P}[|S_{ij}| = q] & \sim \sum_{u=q}^{K_n} \mathbb{P}[|S_{i}
\cap S_{j}| = u]. \label{eq_psiju}
\end{align}

We will first establish (\ref{eq_psijq}) by providing an upper bound
and a lower bound for $\mathbb{P}[|S_{ij}| = q]$, respectively.

Given $P_n \geq 2K_n $ (which holds for all $n$ sufficiently large given the condition $\frac{{K_n}^2}{P_n} = o(1)$), we derive that for $u = 0, 1, \ldots, K_n$,
\begin{align}
 \mathbb{P}[|S_{ij}| =
u]&  =
{\binom{K_n}{u}\binom{P_n-K_n}{K_n-u}}\Big/{\binom{P_n}{K_n}}.
\label{psiju}
\end{align}
Setting $u$ as $q$ in (\ref{psiju}), it is clear that
\begin{align}
  \mathbb{P}[|S_{ij}| \hspace{-2pt} = \hspace{-2pt} q]
& \hspace{-1pt}
 = \hspace{-1pt}
\frac{1}{q!}  \bigg[\frac{K_n!}{(K_n-q)!}\bigg]^2 \hspace{-3pt}
\cdot \hspace{-3pt} \frac{(P_n-K_n)!}{(P_n-2K_n+q)!} \hspace{-2pt}
\cdot \hspace{-3pt} \frac{(P_n-K_n)!}{P_n!}. \label{psijq}
\end{align}
 For the
upper bound on $\mathbb{P}[|S_{ij}| = q]$, using (\ref{psijq}) and
$\frac{{K_n}^2}{P_n-K_n} = o(1)$ which holds from $
\frac{{K_n}^2}{P_n} = o(1)$, and applying the fact that $1+x \leq
e^x$ for any real $x$, we have
\begin{align}
 & \mathbb{P}[|S_{ij}| = q]  \nonumber  \\ & \quad \leq
(q!)^{-1} {K_n}^{2d}
 {P_n}^{K_n-q} (P_n-K_n)^{-K_n}
  \nonumber  \\ & \quad = (q!)^{-1} \big( {{K_n}^2}/{P_n}
\big)^{q} \big[1+ {K_n}/(P_n-K_n)\big]^{K_n}  \nonumber
\\ & \quad \leq  (q!)^{-1} \big({{K_n}^2}/{P_n}
\big)^{q} e^{\frac{{K_n}^2}{P_n-K_n}}  \label{psijq_up-oldii}
\\ & \quad \leq  (q!)^{-1} \big( {{K_n}^2}/{P_n}
\big)^{q} \cdot [1+o(1)]. \label{psijq_up}
\end{align}
 For the part of finding the lower bound, we employ (\ref{psijq}), $ \frac{{K_n}^2}{P_n} =
 o(1)$ and $\big(1-\frac{2K_n}{P_n}\big)^{K_n} \to
 1$ as $n \to \infty$ which follows by $ \frac{{K_n}^2}{P_n} =
 o(1)$ and \cite[Fact 3]{ZhaoYaganGligor}. We also use
 $\frac{{(K_n-q)}^2}{P_n-2K_n} \sim \frac{{K_n}^2}{P_n}$ due to $K_n =
 \omega(q)$ by $K_n = \omega(1)$, and $P_n =
 \omega(K_n)$ by $ \frac{{K_n}^2}{P_n} =
 o(1)$. Therefore,
\begin{align}
 & \mathbb{P}[|S_{ij}| = q]  \nonumber  \\ & \quad \geq
(q!)^{-1} {(K_n-q)}^{2d}
 {(P_n-2K_n)}^{K_n-q}  {P_n}^{-K_n}
  \nonumber  \\ & \quad = (q!)^{-1}
  \big[ {{(K_n-q)}^2}/{(P_n-2K_n)}
\big]^{q} \cdot \big(1-{2K_n}/{P_n}\big)^{K_n}    \label{psijq_low-oldii}
\\ & \quad \sim (q!)^{-1} \big( {{K_n}^2}/{P_n}
\big)^{q} ;  \label{psijq_low}
\end{align}
i.e., $(q!)^{-1} \big( {{K_n}^2}/{P_n} \big)^{q} \cdot [1-o(1)]$ is
a lower bound for $\mathbb{P}[|S_{ij}| = q]$. Then (\ref{eq_psijq})
follows from (\ref{psijq_up}) and (\ref{psijq_low}).

Below we focus on proving (\ref{eq_psiju}). From (\ref{psiju}), for
$u \geq q$,
\begin{align}
  &  \mathbb{P}[|S_{ij}| = u]/{ \mathbb{P}[|S_{ij}| = q]
} \nonumber%
    \\ &  = \hspace{-2pt} q! (u!)^{-1}
  \hspace{-2pt} \bigg[\hspace{-2pt}\prod_{r=0}^{u-q-1}\hspace{-2pt}
   (K_n-q-r)\hspace{-1pt}\bigg]
   \hspace{-2pt} \bigg/ \hspace{-2pt} \bigg[\hspace{-2pt}
   \prod_{r=0}^{u-q-1} \hspace{-2pt}(P_n-2K_n+u-r)\hspace{-1pt}\bigg]
\nonumber \\
&  \leq \hspace{-2pt} [(u-q)!]^{-1} \big( {{K_n}^2}/{P_n}
\big)^{u-q}. \nonumber
\end{align}
Setting $t:=u-q$ and using $ \frac{{K_n}^2}{P_n} =
 o(1)$, we obtain (\ref{eq_psiju}) by
\begin{align}
&  \bigg\{\sum_{u=q}^{K_n} \mathbb{P}[|S_{ij}| = q]\bigg\} \bigg/
\mathbb{P}[|S_{ij}| = q] \nonumber \\ 
& \quad \leq
  \sum_{t=0}^{\infty}
 \big[ {t!}^{-1} \big( {{K_n}^2}/{P_n} \big)^t \big]
 = e^{{{K_n}^2}/{P_n}} \to 1,\textrm{ as }n \to \infty. \label{eqn-oldii}
\end{align}

Property (i) of Lemma \ref{lem_eval_psq} is completed with (\ref{eq_psijq}) and
(\ref{eq_psiju}).

\subsubsection{Proving Property (ii) of Lemma \ref{lem_eval_psq}}~

We prove Property (ii) of Lemma \ref{lem_eval_psq} below. We only need to consider $q \geq 2$ here since the case of $q = 1$ is already proved by Lemma 8-Property (a) in our work \cite{ZhaoYaganGligor}.

We simplify
$S_{i} \cap S_{j}$ by writing it as $S_{ij}$. Clearly, $P_n \geq
2K_n$ for all $n$ sufficiently large, due to $ \frac{{K_n}^2}{P_n} = o\big(\frac{1}{\ln n}\big) =
o(1)$. We will use $b_{q,n}  = \sum_{u=q}^{K_n}
\mathbb{P}[|S_{ij}| = u]$.

From (\ref{psijq_up-oldii}), it holds that
\begin{align}
 & \mathbb{P}[|S_{ij}| = q] \leq  (q!)^{-1} \big({{K_n}^2}/{P_n}
\big)^{q} e^{\frac{{K_n}^2}{P_n-K_n}} . \label{psijq_up-oldii1}
\end{align}
From (\ref{eqn-oldii}), it holds that
\begin{align}
 & \sum_{u=q}^{K_n} \mathbb{P}[|S_{ij}| = q]
 \leq  \mathbb{P}[|S_{ij}| = q] \times e^{{{K_n}^2}/{P_n}}  . \label{psijq_up-oldii1sb}
\end{align}
Combining (\ref{psijq_up-oldii1}) and (\ref{psijq_up-oldii1sb}), we have
\begin{align}
 b_{q,n}  &   \leq  (q!)^{-1} \big({{K_n}^2}/{P_n}
\big)^{q} e^{(\frac{{K_n}^2}{P_n-K_n}+{{K_n}^2}/{P_n})} \nonumber \\ & =  (q!)^{-1} \big({{K_n}^2}/{P_n}
\big)^{q} e^{2\frac{{K_n}^2}{P_n-K_n} } . \label{psijq_up-oldii1sb2}
\end{align}

From $ \frac{{K_n}^2}{P_n} = o\big(\frac{1}{\ln n}\big) $, we have   $ 2{\frac{{K_n}^2}{P_n-K_n}}  = o\big(\frac{1}{\ln n}\big) $ by considering for all $n$ sufficiently large that $2{\frac{{K_n}^2}{P_n-K_n}} \leq \frac{4{K_n}^2}{P_n}$ from $K_n \leq \frac{1}{2} P_n$.
We can easily prove $ e^{x}\leq  1 + 2 x $ for $0\leq x \leq 1$ by taking the derivative of  $ e^{x} - 1 - 2x$ to investigate its monotonicity. This implies that for a sequence $x_n = o\big(\frac{1}{\ln n}\big)$, we have $e^{x_n} = 1+ o\big(\frac{1}{\ln n}\big)$. Given the above, we obtain $e^{2\frac{{K_n}^2}{P_n-K_n}} = 1+o\big(\frac{1}{\ln n}\big)$. Using this in (\ref{psijq_up-oldii1sb2}), we have
\begin{align}
 &  b_{q,n}  \leq  (q!)^{-1} \big({{K_n}^2}/{P_n}
\big)^{q} \times \bigg[1+ o\bigg(\frac{1}{\ln n}\bigg)\bigg]  . \label{psijq_up-oldii2}
\end{align}


We can easily prove $1- x \geq e^{-2x}$ for $0\leq x  < \frac{1}{2}$ by taking the derivative of  $1- x - e^{-2x}$ to investigate its monotonicity. Given $\frac{{K_n}}{P_n} \leq \frac{{K_n}^2}{P_n} = o\big(\frac{1}{\ln n}\big)$, we have $\frac{{K_n}}{P_n} < \frac{1}{2}$ for all $n$ sufficiently large, which implies
\begin{align}
 & \big(1-{2K_n}/{P_n}\big)^{K_n} \geq \big(e^{-2\times {2K_n}/{P_n}}\big)^{K_n} \nonumber \\ & = e^{-4{K_n}^2/P_n} \geq 1- 4{K_n}^2/P_n = 1 -o\bigg(\frac{1}{\ln n}\bigg)   . \label{psijq_up-oldii3}
\end{align}
To use (\ref{psijq_up-oldii3}) in (\ref{psijq_low-oldii}), we further evaluate $ {{(K_n-q)}^{2d}}/{(P_n-2K_n)^{q}} $. Recall that we only need to consider $q \geq 2$ here since the case of $q = 1$ is already proved by Lemma 8-Property (a) in our work \cite{ZhaoYaganGligor}. We have the condition $K_n = \omega(\ln n)$ for $q \geq 2$. Thus, it holds that   $K_n > q$ for all $n$ sufficiently large. Then using \cite[Fact 2]{ZhaoYaganGligor}, we have $1 - \frac{q}{K_n} \times 2d \leq (1-\frac{q}{K_n})^{2d} \leq 1 - \frac{q}{K_n} \times 2d + \frac{1}{2} \times  \big(\frac{q}{K_n}\big)^2 \times (2d)^2 $, which along with $K_n = \omega(\ln n)$   implies
\begin{align}
 & \bigg(1-\frac{q}{K_n}\bigg)^{2d} = 1 -o\bigg(\frac{1}{\ln n}\bigg)   . \label{psijq_up-oldii4}
\end{align}
Given $\frac{{K_n}}{P_n} \leq \frac{{K_n}^2}{P_n} = o\big(\frac{1}{\ln n}\big)$, we have $\frac{2{K_n}}{P_n} < 1$ for all $n$ sufficiently large. Then using \cite[Fact 2]{ZhaoYaganGligor}, we have $1 - \frac{2K_n}{P_n} \times q \leq (1-\frac{2K_n}{P_n})^{q} \leq 1 - \frac{2K_n}{P_n} \times q + \frac{1}{2} \times  \big(\frac{2K_n}{P_n}\big)^2 \times q^2 $, which along with $\frac{{K_n}}{P_n} \leq \frac{{K_n}^2}{P_n} = o\big(\frac{1}{\ln n}\big)$ implies
\begin{align}
 & \bigg(1-\frac{2K_n}{P_n}\bigg)^{q} = 1 -o\bigg(\frac{1}{\ln n}\bigg)   . \label{psijq_up-oldii5}
\end{align}
From (\ref{psijq_up-oldii4}) and (\ref{psijq_up-oldii5}), we obtain
\begin{align}
 & \frac{\big(1-\frac{q}{K_n}\big)^{2d}}{\big(1-\frac{2K_n}{P_n}\big)^{q}}  = 1 \pm o\bigg(\frac{1}{\ln n}\bigg)   . \label{psijq_up-oldii6}
\end{align}
The reason is that for two sequences $x_n$ and $y_n$ satisfying $x_n = o\big(\frac{1}{\ln n}\big)$ and $y_n = o\big(\frac{1}{\ln n}\big)$, it holds that $\frac{1-x_n}{1-y_n} = 1 \pm  o\big(\frac{1}{\ln n}\big)$. To see this, we have $\frac{1-x_n}{1-y_n} -1 = \frac{y_n - x_n}{1- y_n} = \pm  o\big(\frac{1}{\ln n}\big)$ given $y_n - x_n\pm  o\big(\frac{1}{\ln n}\big)$ and $\lim_{n \to \infty}(1- y_n)=1$.

The left hand side of (\ref{psijq_up-oldii6}) can be written as $\big[ {{(K_n-q)}^2}/{(P_n-2K_n)}
\big]^{q} \big/ \big[\big({{K_n}^2}/{P_n}
\big)^{q}\big]$. Hence, (\ref{psijq_up-oldii6}) implies
\begin{align}
 & \big[ {{(K_n-q)}^2}/{(P_n-2K_n)}
\big]^{q}  = \big({{K_n}^2}/{P_n}
\big)^{q} \times \bigg[ 1 \pm o\bigg(\frac{1}{\ln n}\bigg)\bigg]   . \label{psijq_up-oldii7}
\end{align}
Using (\ref{psijq_up-oldii3}) and (\ref{psijq_up-oldii7}) in (\ref{psijq_low-oldii}), and noting that $\big[1 \pm o\big(\frac{1}{\ln n}\big)\big]\times \big[1 \pm o\big(\frac{1}{\ln n}\big)\big]$ can also be written as $\big[1 \pm o\big(\frac{1}{\ln n}\big)\big]$, we obtain
\begin{align}
 &  b_{q,n}  \geq   \mathbb{P}[|S_{ij}| = q] \geq  (q!)^{-1} \big({{K_n}^2}/{P_n}
\big)^{q} \times \bigg[1- o\bigg(\frac{1}{\ln n}\bigg)\bigg]  . \label{psijq_up-oldii8}
\end{align}

Property (ii) of Lemma \ref{lem_eval_psq} is completed with (\ref{psijq_up-oldii2}) and (\ref{psijq_up-oldii8}).

\end{document}